\documentclass[a4paper]{article}

\usepackage[english]{babel}
\usepackage[utf8]{inputenc}
\usepackage{amsmath}
\usepackage{fixmath}
\usepackage{amsfonts}
\usepackage{amssymb}
\usepackage{tabularx}
\usepackage{graphicx}
\usepackage{hyperref}
\usepackage{bbm}
\usepackage[colorinlistoftodos]{todonotes}
\usepackage{geometry}
\usepackage{booktabs}
\usepackage{orcidlink}
\usepackage[normalem]{ulem} 

\usepackage{color}

\newcommand{\anderson}[1]{\textcolor{blue}{#1}}

\newcommand{\snr}{{\mathrm{SNR}}}

\usepackage{authblk}

\title{KSZ Velocity Reconstruction with ACT and DESI-LS using a Tomographic QML Power Spectrum Estimator}

\author[1]{Anderson Lai\, \orcidlink{0000-0003-2741-4556}} 
\author[1]{Yurii Kvasiuk\, \orcidlink{0009-0002-4720-1320}} 
\author[1,2]{Moritz M\"unchmeyer\, \orcidlink{0000-0002-3777-7791}} 

\affil[1]{Department of Physics, University of Wisconsin-Madison, Madison, WI 53706, USA}
\affil[2]{NSF-Simons AI Institute for the Sky (SkAI), Chicago, IL 60611, USA}

\date{\today}

\begin{document}
\maketitle

\begin{abstract}
We perform kinetic Sunyaev-Zel'dovich (kSZ) velocity reconstruction on data from ACT DR6 and DESI-LS DR9. To estimate the cross-power between kSZ velocity reconstruction and galaxy density, we make use of a novel quadratic maximum likelihood (QML) power spectrum estimator in redshift binned spherical coordinates. We find a detection of the kSZ signal from the cross-correlation between the estimated velocity field and the large-scale galaxy field of $10 \sigma$. We estimate an amplitude $A=0.42 \pm 0.042$ of the kSZ signal relative to a halo model prediction, possibly indicating strong feedback in massive halos, in agreement with previous studies. We compare the result to the pseudo-$C_{\ell}$ analysis on the same dataset, finding a $\sim$20\% improvement in the SNR. Our work demonstrates the feasibility of an optimal QML analysis at the required resolution, and will be a powerful tool for kSZ cosmology with upcoming high-resolution surveys.
\end{abstract}

\tableofcontents

\section{Introduction}

Current and upcoming CMB experiments measure CMB secondary anisotropies with unprecedented precision, making them a powerful probe of astrophysics and cosmology. On intermediate scales these anisotropies are dominated by CMB lensing \cite{Lewis:2006fu}. On even smaller scales, the dominant contribution to the CMB blackbody anisotropy is the kinetic Sunyaev-Zeldovich (kSZ) effect \cite{Sunyaev1980}, the rescattering of CMB photons on the bulk movement of free electrons in the universe. The kSZ has been detected by cross-correlation with galaxies in \cite{Hand:2012ui,Ade:2015lza,Schaan:2015uaa,Soergel:2016mce,Hill:2016dta,DeBernardis:2016pdv,AtacamaCosmologyTelescope:2020wtv}. 
The strongest detection to date comes from the recent paper \cite{Hadzhiyska:2024qsl} which found 13$\sigma$ using ACT data \anderson{\cite{Henderson:2015nzj, AtacamaCosmologyTelescope:2025vnj}} in cross-correlation with photometric galaxy data (DESI-LS) \anderson{\cite{DESI:2018ymu, 2021AAS...23723503S,Zhou:2023DESILS_Cggsample2}}. However, for upcoming experiments the expected signal-to-noise (SNR) is much higher, for example Simons Observatory (SO) \cite{SO_sciencebook:2018koc, SO_LAT:2025wwn} and Rubin Observatory \cite{LSSTSciBook} will reach $\snr \simeq 200$ \cite{Smith:2018bpn}. The kSZ has both astrophysical and cosmological applications. In astrophysics, the kSZ has been used to measure the gas profile of galaxy clusters \cite{Ho:2009iw,Schaan:2015uaa,AtacamaCosmologyTelescope:2020wtv}, which contains information about astrophysical galactic processes such as AGN feedback \cite{Battaglia:2016_battaglia_profile}. However, the kSZ is also a powerful probe of cosmology (see e.g. \cite{DeDeo:2005yr,HernandezMonteagudo:2005ys,Bhattacharya:2007sk}) and can be used to constrain, for example, modified gravity \cite{Mueller:2014nsa,Bianchini:2015iaa}, and
neutrino masses \cite{Mueller:2014dba}. Recent kSZ analyses using ACT data with other methods than the one presented here include \cite{AtacamaCosmologyTelescope:2020wtv,MacCrann:2024PNG_intro,Calafut:2021wkx,Amodeo:2020mmu, Hadzhiyska:2025gti}.

In the present work we apply a method called \emph{kSZ velocity reconstruction}, which uses the kSZ to reconstruct the large-scale velocity field of the universe. This velocity field could then be used to probe cosmology, in a similar way to the CMB lensing potential (both methods probe the total matter distribution on large scales), but we defer a cosmological parameter analysis to future work. kSZ velocity reconstruction was first proposed in \cite{Terrana2016, Deutsch:2017_2Dksz_estim1} (there called remote dipole reconstruction) and further developed in \cite{Smith:2018bpn,Cayuso:2021ljq}. kSZ velocity reconstruction has been applied to data for the first time in 2024 in several different analyses using various data combinations \cite{McCarthy:2024_ACT_DESI_ksz,Bloch:2024PlanckxUnwise,Lague:2024czc_7.2sigma,Krywonos:2024mpb}. References \cite{Bloch:2024PlanckxUnwise,Krywonos:2024mpb} used data from Planck \cite{Planck:2018nkj} and unWISE \cite{Schlafly_2019} to constrain the optical depth bias as well as primordial non-Gaussianity, while \cite{McCarthy:2024_ACT_DESI_ksz,Lague:2024czc_7.2sigma} used ACT data in combination with DESI-LS and SDSS \cite{eBOSS:2017dtq} to detect the kSZ signal at $3.8\sigma$ and $7.2\sigma$, respectively. A number of authors have proposed possible applications of the kSZ velocity reconstruction method to constrain fundamental parameters of the universe, including modified gravity \cite{Pan:2019dax}, isocurvature perturbations \cite{Hotinli:2019wdp} and statistical anisotropies \cite{Cayuso:2019hen}. A perhaps particularly promising application is the measurement of the local non-Gaussianity parameter $f_\mathrm{NL}$, which was proposed in \cite{Munchmeyer:2018eey}. The kSZ velocity reconstruction method was also studied with simulations in \cite{Cayuso:2018lhv,Giri:2020pkk}. 

In this work, we apply kSZ velocity reconstruction to ACT DR6 data and photometric galaxy data from DESI Legacy Survey DR9. Two previous works have applied this method to ACT data \cite{McCarthy:2024_ACT_DESI_ksz,Lague:2024czc_7.2sigma}\footnote{We note that additional works on kSZ velocity reconstruction, using similar datasets, have appeared during the review process of this manuscript. For example,~\cite{McCarthy:2025brx} reported a $\sim 10\sigma$ detection using DESI-LS DR9 + ACT DR6 for kSZ velocity reconstruction, and DESI-LS DR10 galaxies for cross-correlation. A follow-up work from~\cite{EmbilVillagra:2026fnh} reported a $\sim 11\sigma$ detection using the same dataset with a parity-odd estimator.}. Our analysis differs from these works in the following ways: 
\begin{itemize}
    \item We use the DESI Legacy Survey (DESI-LS) both to construct the quadratic velocity estimator (which uses the small-scale galaxy distribution), and to cross-correlate the large-scale kSZ velocity reconstruction with the large-scale galaxy field. In contrast, previous analyses used spectroscopic galaxy data from SDSS to generate a large-scale velocity template from the galaxy field for cross-correlation with the kSZ velocity reconstruction. We also use finer redshift bins than the ACT DR6 analysis in \cite{McCarthy:2024_ACT_DESI_ksz}, which allows us to increase the SNR.
    \item We develop a novel implementation of the optimal quadratic maximum likelihood (QML) power spectrum estimator, to improve over the common but suboptimal pseudo-$C_\ell$ method. This estimator uses a pixel-wise covariance matrix, which we make computationally tractable for the required large-scale analysis ($\ell_{max}=60$, Healpix $N_{\rm side}=32$ and 20 redshift bins for both galaxy and velocity fields). More details about our novel QML estimator are provided in our companion paper~\cite{Kvasiuk:qmlpaper}, with the associated Python code \texttt{QML-FAST} publicly available at~\url{https://github.com/ykvasiuk/qmlfast}. We find that this estimator provides a 20\% improvement in the SNR compared to the more standard pseudo-$C_\ell$ approach with identical analysis choices. An interesting aspect of our implementation is that we calculate the full covariance matrix (including mask effects) analytically, rather than having to run a large ensemble of (Gaussian) Monte Carlo simulations. 
\end{itemize}
In combination, these improvements allow us to reach an SNR of 10 $\sigma$, which is significantly higher than the SNR reported in~\cite{McCarthy:2024_ACT_DESI_ksz}, and almost equal to the SNR reported in~\cite{SelimKendrickPaper} (pre-prints published at the same time). While in the present paper we only reconstruct the velocity field and do not perform a cosmological parameter analysis, our novel QML pipeline is well suited for an optimal cosmology analysis on large-scales, which we will provide in follow-up work. 

This paper is similar in scope to \cite{SelimKendrickPaper}, submitted to arXiv at the same time, which also does kSZ velocity reconstruction with ACT × (DESI-LS LRGs). Our pipeline is 2+1 dimensional, while \cite{SelimKendrickPaper} has a 3-dimensional Cartesian representation of the data. We also make different analysis choices. In particular, \cite{SelimKendrickPaper} uses only the DESI-LS northern galactic hemisphere area to reduce image systematics (while still retaining a large part of the SNR). The two papers are substantially different in their pipeline and analysis choices but reach very similar results in terms of SNR and kSZ velocity bias $A \sim 0.44$. The paper \cite{SelimKendrickPaper} also includes a measurement of local primordial non-Gaussianity, which we defer to future work.

The paper is organized as follows. In Sec. \ref{sec:analysispipeline} we describe the input data and the analysis pipeline, including the novel QML estimator implementation. In Sec. \ref{sec:results} we apply the pipeline to data and provide a number of consistency checks. In Sec. \ref{sec:linearrec} we make a linear reconstruction of the velocity field from DESI-LS, which we can then use to compare to the kSZ-derived velocity reconstruction. We conclude in Sec. \ref{sec:conclusion}. A number of appendices provide more details for our analysis. 

\section{Analysis Pipeline}

In this section we describe the dataset, review the quadratic estimator velocity reconstruction, describe our novel QML power spectrum implementation, and evaluate signal and noise covariance matrices.

\label{sec:analysispipeline}

\subsection{Data set}
\subsubsection{ACT DR6 temperature maps}\label{sec:ACTDR6data}
\begin{figure}[h!]
    \centering
    \includegraphics[width=0.9\columnwidth]{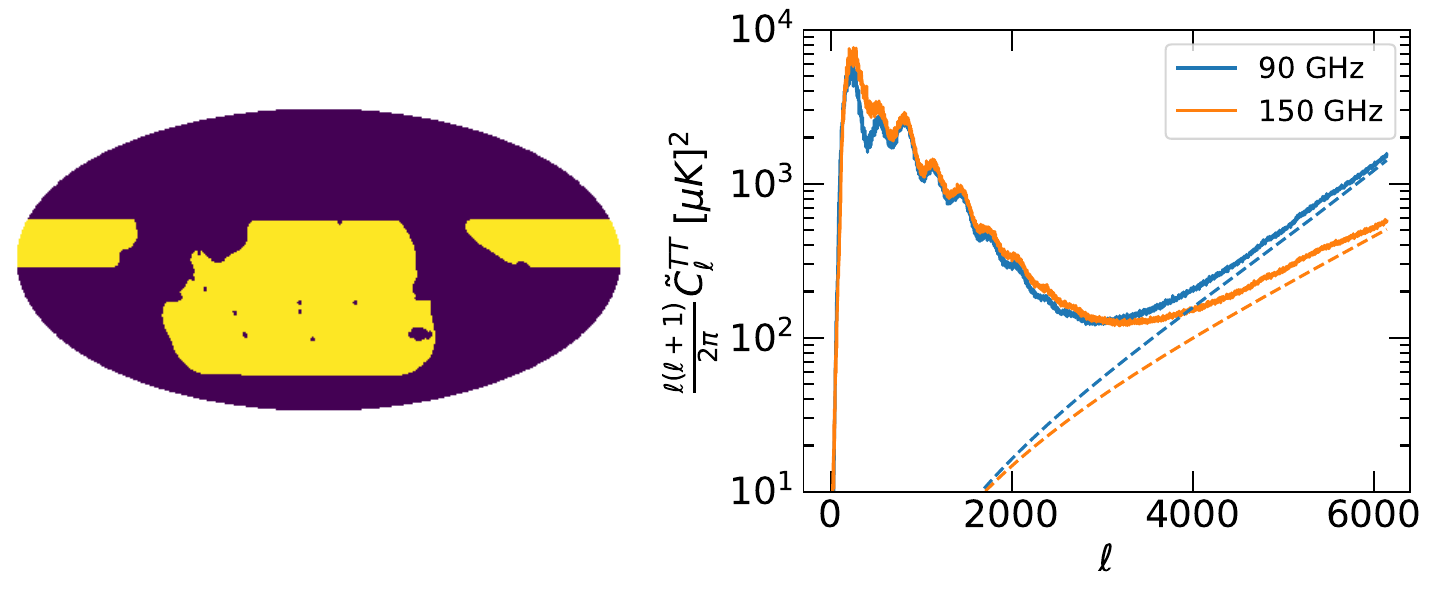}
    \caption{Left: A binary mask created from the apodized DR6 NILC analysis mask in~\cite{ACT:2023ilc_mask_ref}. Right: The observed temperature power spectrum $\tilde{C}^{TT}_{\ell}$ multiplied by $\frac{\ell(\ell+1)}{2\pi}$. It is calculated from the preprocessed, co-added temperature maps. The blue (orange) curve corresponds to the 90 (150) GHz channel. The temperature power spectrum is presented in the unit of $(\mu\text{K})^2$. The blue and orange dashed curves are included in the plot to illustrate the asymptotic behavior at small scales, due to beam deconvolution. The amplitudes of the dashed curves are obtained by fitting a beam-deconvolved, flat power spectrum to the beam-deconvolved data at large $\ell$. They are included for illustration only.}
    \label{fig:ACTDR6 Summary}
\end{figure}

We use the co-added temperature maps from ACT Data Release 6 (DR6), which are the latest dataset available to the public~\cite{AtacamaCosmologyTelescope:2025vnj}. The ACT DR6 maps preserve a relatively large footprint ($\approx43\%$), while offering a lower noise level beyond $\ell > 1000$ compared to the Planck satellite, which has a more complete sky coverage. To avoid strong foregrounds due to thermal dust emission from unresolved infrared sources, known as the cosmic infrared background (CIB)~\cite{Puget:1996fx, Lagache:2005sw}, we perform the kSZ reconstruction using only the 90 GHz and 150 GHz channel maps.
\begin{figure}[h!]
    \centering
    \includegraphics[width=\columnwidth]{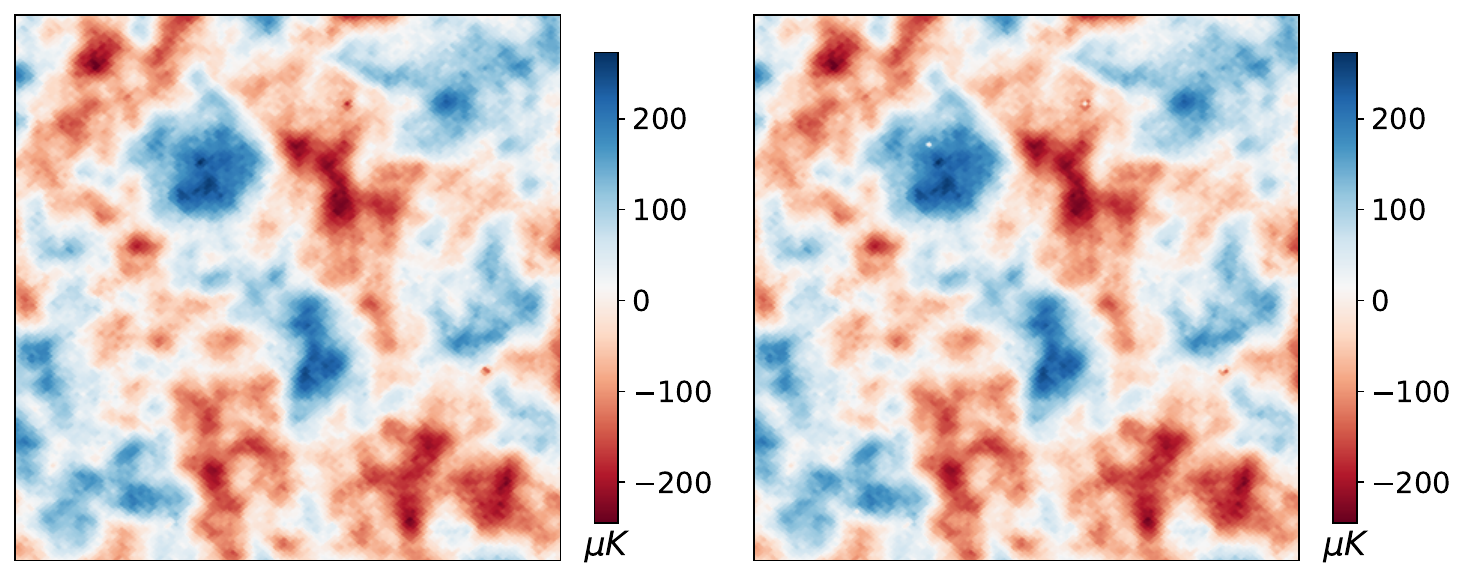}
    \caption{A figure illustrating the ACT DR6 90 GHz temperature map before and after the SZ cluster masking over a $4^\circ\times4^\circ$ region. The SZ clusters appear as dense cold spots at the 90 GHz channel, which can be seen in the middle-top and bottom-right regions of the left plot. The SZ clusters are then masked according to the catalog from~\cite{ACT:2020tSZ_mask} as shown in the right figure. We create the SZ cluster mask by zeroing a disc-like region of size $\sim 2.7'$ at the position of every cluster with SNR $>5$ before beam-deconvolving the maps. The SZ cluster mask could potentially mitigate anisotropic foregrounds that may bias the kSZ reconstruction. In Sec.~\ref{subsec: SNR result} we find that masking the SZ clusters changes the central value of the kSZ amplitude $A$ and improves goodness of fit for the galaxy-velocity cross-power.}
    \label{fig:tszmask_illustration}
\end{figure}
We prepare the temperature maps on individual frequencies. To avoid any bright point sources that may induce excessive power at small scales, we start with source-free temperature maps\footnote{The coadded, point-source-free temperature maps are provided as part of the ACT DR6 data products. Point sources are removed by subtracting the best-fit flux at locations with a matched-filtered point-source detection above $5\sigma$~\cite{Naess:2020ACTDR5coadd}. We do not apply any additional masking to point sources.}. Another known source of foreground that can bias the kSZ signal is the thermal SZ (tSZ) emission due to the scattering between the CMB photons and the thermal electrons located in galaxy clusters~\cite{Sunyaev:1970bma, Zeldovich:1969ff}. To mitigate the anisotropic tSZ foreground, we create an SZ cluster mask from the catalog given in ACT DR5~\cite{ACT:2020tSZ_mask} such that clusters with SNR $>$ 5 are masked with a binary disc of size $\sim 2.7'$\footnote{Specifically, we mask the central pixel, the neighboring pixels and the next-nearest-neighboring pixels for every cluster with SNR $>$ 5 on $N_{\rm side} = 4096$ temperature maps.}. The SZ cluster mask is then applied to the temperature maps. We study the impact of the SZ cluster mask used in this analysis in Sec.~\ref{subsec: SNR result} and defer a full investigation of the optimal choice of SZ cluster mask to future work. Fig.~\ref{fig:tszmask_illustration} illustrates a square region from the 90 GHz temperature map before and after the tSZ masking. The maps are then deconvolved using the equivalent beam profile as prescribed in~\cite{Naess:2020ACTDR5coadd} and overlaid with a sky mask from the DR6 internal-linear combination analysis~\cite{ACT:2023ilc_mask_ref} to remove regions that are near the galactic plane. The left plot in Fig.~\ref{fig:ACTDR6 Summary} shows the combined mask that we use for the temperature maps. The remaining sky coverage is approximately $23.5\%$. We present the temperature power spectrum $\tilde{C}^{TT}_{\ell}$'s computed from the processed temperature maps in the right plot of Fig.~\ref{fig:ACTDR6 Summary}. The source-free temperature maps, equivalent beam profiles, SZ cluster catalog and mask are all publicly available from the ACT repository. 

\subsubsection{DESI-LS DR9 extended LRG sample and calibration}\label{subsubsec: DESI LS data}
\paragraph{Data selection.}
We use the DESI-LS (Legacy Survey) DR9 extended LRG (Luminous Red Galaxy) sample from~\cite{DESI:202DESILS_Cggsample1, Zhou:2023DESILS_Cggsample2}. The full DR9 extended LRG catalog has a footprint of nearly $44\%$ of the full sky and contains approximately $3.4 \times10^7$ objects. We apply the veto mask in~\cite{Zhou:2023DESILS_Cggsample2} to remove objects that do not match the selection criteria. This reduces the number of objects to $2.8 \times 10^7$. The reduced sample doubles the number of galaxies compared to the galaxy sample used in the kSZ analysis in~\cite{McCarthy:2024_ACT_DESI_ksz} by extending the analysis to redshifts up to $z = 1.0$. 

\paragraph{Photo-z weighting.}
The galaxy catalog also provides the photo-z error $\sigma_z$ for each object. As mentioned in~\cite{SelimKendrickPaper}, the photo-z error distribution of the observed galaxies is non-Gaussian because of galaxies with extremely large photo-z errors ($\sigma_z(1+z) > 0.08$). We follow the approach in~\cite{SelimKendrickPaper} by assigning a weight $w_j$ to each observed galaxy:
\begin{align}\label{eq:per-object-weight}
    w_j(\sigma_z, z) &= \exp{\left(-\frac{\sigma^2_{z, j}}{0.0025(1+z_j)^2}\right)} \, .
\end{align}
Here the index $j$ runs over every galaxy, and $z_j$ is the photometric redshift of each galaxy. This mitigates the contribution from outlier galaxies \footnote{We note that the down-weighting procedure is newly introduced in the version 2 of this paper. The weights improve the goodness of fit for the QML estimates with respect to the fiducial theory, especially on large scales. As we will see in Sec.~\ref{subsec:Cvg result}, the reduced chi-square value of the galaxy-velocity cross-powers went down from 1.5 in the older version of the paper to $1.02$ (tSZ-masked, second iteration) in the present version. The photo-z weighting is the main source of this improvement.}.

\paragraph{Tomographic galaxy density maps.} 
We define a set of 20 photometric redshift bins from $z = 0.4$ to $z = 1.0$ of uniform width $\Delta z = 0.03$. For each redshift bin, we determine the position of the galaxies with their RA and Dec and then project them onto a \texttt{Healpix} map~\cite{Zonca2019:Healpix1, Gorski:Healpix2}, forming a weighted galaxy density map $n_g^\alpha$:
\begin{align}
    n_g^\alpha(\hat{n}_p) &= \sum_{j \in \alpha}w_j\,\,\delta_{p_j,\, p} \, .
\end{align}
Here $\alpha$ is the redshift bin index. $p_j$ is the pixel index for the $j$-th galaxy. We prepare the maps at three target resolutions: a set of high-resolution maps at $N_{\rm side} = 4096$ for the kSZ quadratic estimator, a set of low-resolution maps at $N_{\rm side} = 256$ for imaging calibration, and a set of $N_{\rm side} = 32$ maps, which is downgraded from $N_{\rm side} = 256$ maps and used for estimating the large-scale power spectra with QML.

\paragraph{High-resolution galaxy overdensity maps.} 
We use the high-resolution $N_{\rm side} = 4096$ galaxy density maps as input to the kSZ quadratic estimator, since the filter of the quadratic estimator peaks around $\ell = 2000$-$4000$. We obtain the galaxy overdensity map $\delta_g^\alpha$ from the following equation:
\begin{align}\label{eq: delta_g small scale}
    \delta_g^\alpha(\hat{n}_p) &= \frac{n_g^\alpha(\hat{n}_p)}{\bar{n}^\alpha_g} - 1 \, .
\end{align}
Here $\bar{n}^\alpha_g$ is a pixel-independent number defined as the average of the weighted galaxy number density over the unmasked region for the $\alpha$-th redshift bin. Note that $\delta_g^\alpha(\hat{n}_p)$ is defined over the unmasked region and set to zero elsewhere. Eq.~\ref{eq: delta_g small scale} is a good approximation to the observed galaxy overdensity field when the scales of the spatial variation of survey depth are much larger than the scales of interest.

\paragraph{Random imaging maps.} 
We use $N_{\rm side} = 256$ galaxy density maps to compute the galaxy overdensity field for the large-scale cross-correlation between the reconstructed kSZ velocities and the galaxy field. On these scales, survey and imaging properties, such as completeness and galaxy depth, can correlate with the cosmological signal~\cite{DESI:202DESILS_Cggsample1, Rezaie:2023_gmap_calibration}. Therefore, we create random maps from the random catalog associated with the data product. The random catalog accurately describes the variation of survey completeness. The size of the random catalog used for this analysis is about 60 times the number of observed galaxies. Each object from the random catalog contains position information and a set of 11 imaging properties. There is, however, no redshift information. For each random object we therefore draw a pair of photo-z and photo-z error from the observed distributions. This is achieved by the \texttt{RegulatedDeconvolver} class in \texttt{kszx}~\cite{SelimKendrickPaper}. Then, for each redshift bin $\alpha$ and imaging property $\gamma$, we define a random imaging map $r^{\alpha, \gamma}$:
\begin{align}
    r^{\alpha, \gamma}(\hat{n}_p) &= \sum_{j \in \alpha}w^r_j\,U^\gamma_j\,\,\delta_{p_j,\, p} \, .
\end{align}
Here $U^\gamma_j$ is the value of the $\gamma$-th imaging property for the $j$-th random object within the redshift bin $\alpha$. Note that the random points are also weighted by their own photo-z error weights $w^r_j$.

\paragraph{Imaging calibration.} 
We perform linear regression on the random imaging maps and construct the best-fit random galaxy density template $n^\alpha_r$~\cite{DESI:202DESILS_Cggsample1, Rezaie:2023_gmap_calibration}:
\begin{align}
    n^\alpha_r(\hat{n}_p) &= c^\alpha_0 + \sum_{\gamma=1}^{11}\,c^{\alpha}_{\gamma}\,r^{\alpha, \gamma}(\hat{n}_p) \, .
\end{align}
The best-fit intercept $c^\alpha_0$ and coefficients $c^\alpha_\gamma$ are determined by minimizing the chi-square difference ${\chi^\alpha}^2_{\rm random}$ between the random template and the observed galaxy density map:
\begin{align}
    {\chi^\alpha}^2_{\rm random} = \sum_p\left(n_g^\alpha(\hat{n}_p) - n^\alpha_r(\hat{n}_p)\right)^2 \, .
\end{align}
Note that we calculate the coefficients for every redshift bin $\alpha$. We show the imaging coefficients used in this analysis in App.~\ref{app:calibration}.

\begin{figure}[h!]
    \centering
    \includegraphics[width=0.9\columnwidth]{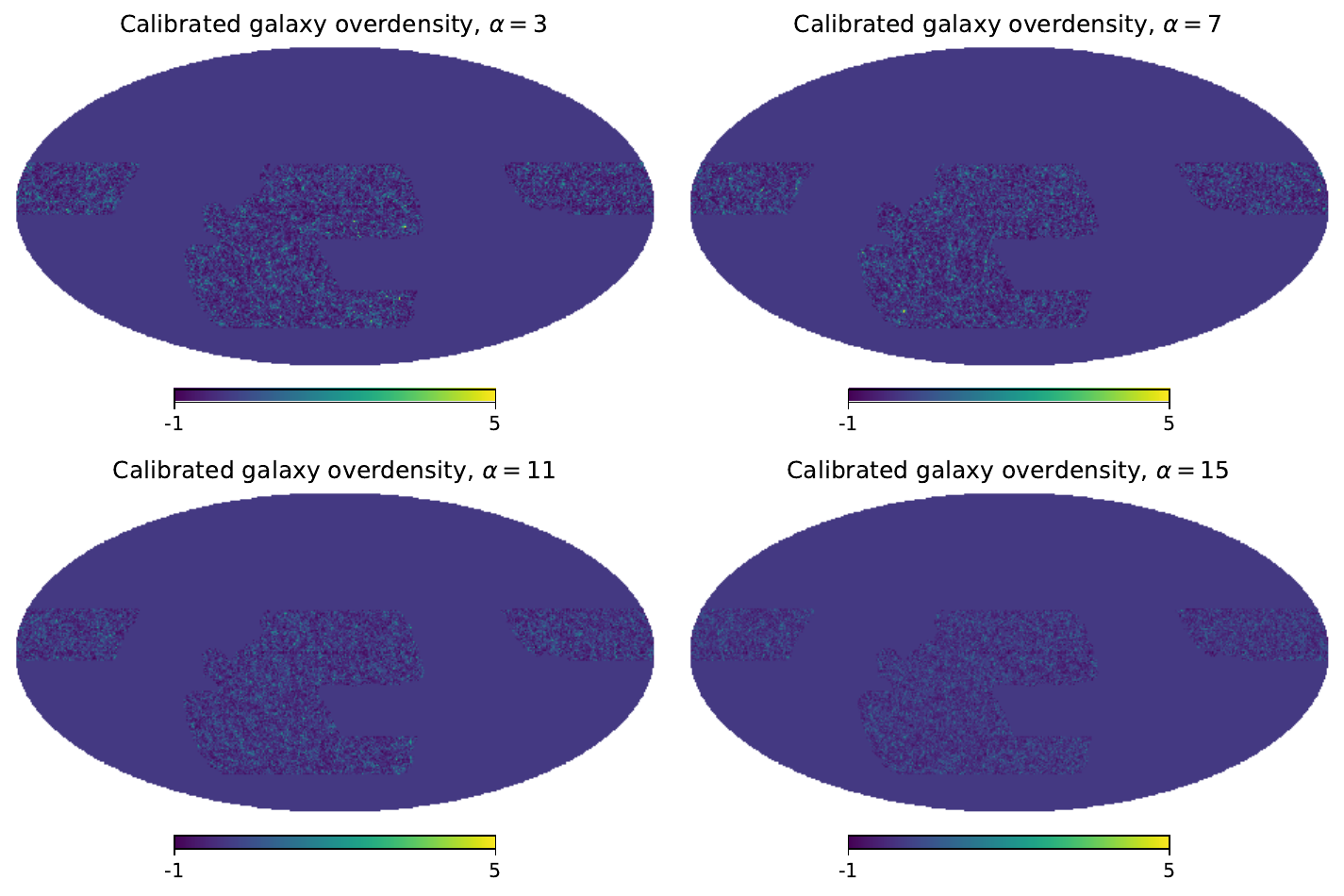}
    \caption{A selection of DESI LRG galaxy overdensity maps used in this analysis. The overdensity maps are calibrated using a regression model of 11 imaging systematic maps over the DESI footprint. The maximum overdensity is clipped at 5 for the purpose of illustration.}
    \label{fig:desi_imgfix_odmaps}
\end{figure}

\paragraph{Low-resolution galaxy overdensity maps.}
The $N_{\rm side} = 256$ calibrated galaxy overdensity maps can be obtained in same way as the $N_{\rm side} = 4096$ case (Eq.~\ref{eq: delta_g small scale}), with a change from the mean galaxy number density $\bar{n}^\alpha_g$ to the spatially-varying random template $n^\alpha_r$. Again, we compute the overdensity only within the unmasked footprint. We downgrade these maps to $N_{\rm side} = 32$ for the power spectrum estimation in Sec.~\ref{sec:results} using a three-step approach: transform the overdensity maps to spherical harmonic coefficients, zero out any coefficients with $\ell > 3 N_{\rm side} - 1=95$, and then transform back to a map at $N_{\rm side} = 32$. We show a selection of calibrated galaxy overdensity maps in Fig.~\ref{fig:desi_imgfix_odmaps}. 

\paragraph{Radial window functions.} 
We also define a set of radial window functions $W^\alpha(z)$ for calculating the fiducial angular power spectrum in Sec.~\ref{subsec: fiducial powers}. Following the method in~\cite{McCarthy:2024_ACT_DESI_ksz}, we create a bin-wise histogram for the galaxy catalog over a finer redshift grid (by uniformly subdividing each redshift bin $\alpha$ into 20 smaller parts), with each galaxy weighted by $w_j$. Then we convolve the histogram with a Gaussian kernel of width $\bar{\sigma}_z (1+z)$, where $\bar{\sigma}_z$ is the weighted photo-z error\footnote{In practice we use the uniformly-weighted window function throughout this analysis. We find that the photo-z error weights $w_j$ do not significantly change the shape of the window function.}. We normalize the convolved histogram to obtain the radial window functions.

\begin{figure}[h!]
    \centering
    \includegraphics[width=0.8\columnwidth]{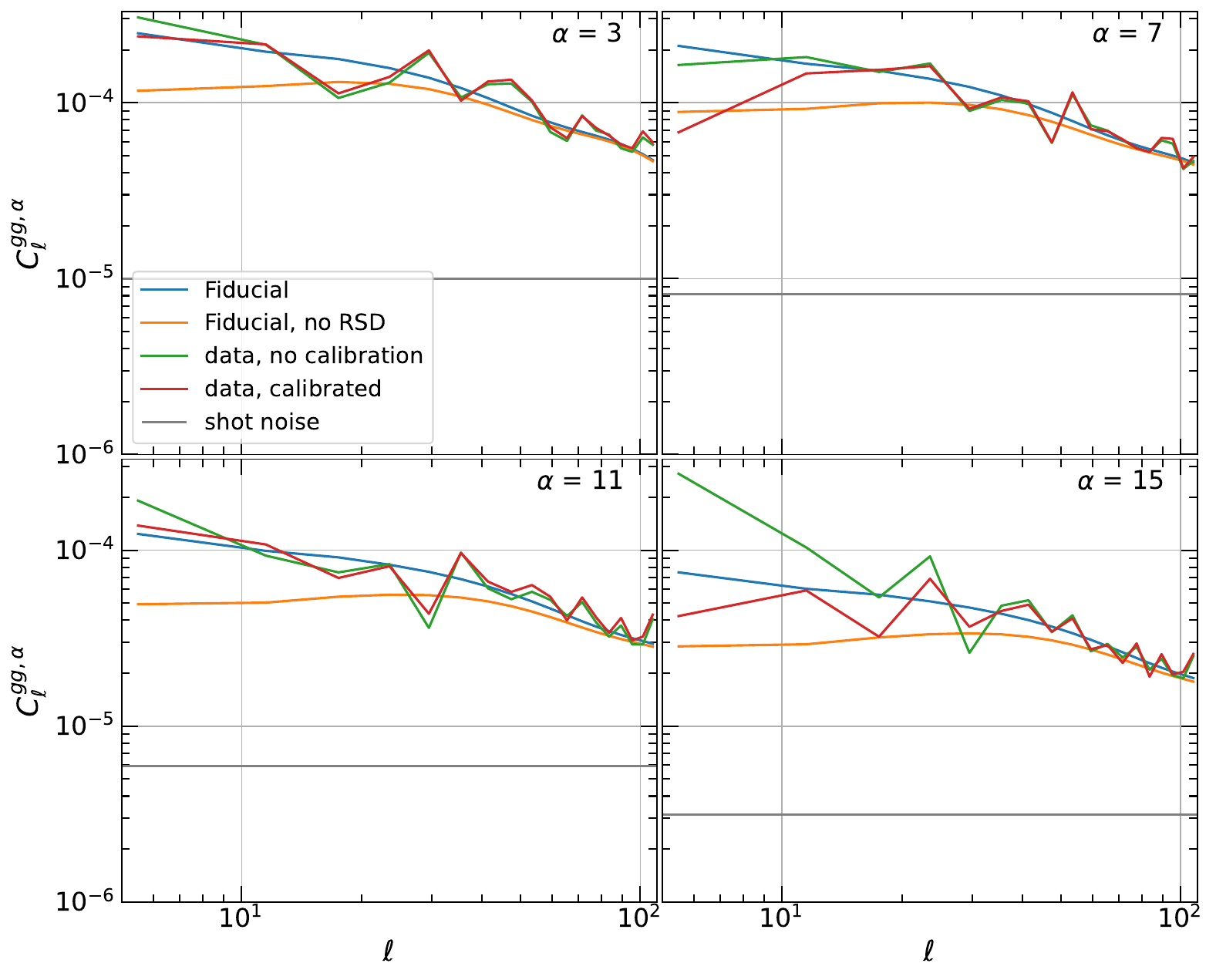}
    \caption{Galaxy auto-power spectrum in four redshift bins (selected from 20 redshift bins). The plot shows the influence of the imaging calibration on the observed large-scale galaxy auto-power spectrum. Green curve: galaxy auto-powers evaluated from overdensity maps without imaging calibration. Red curve: galaxy auto-powers evaluated from maps with calibrated imaging systematics. The calibration tends to reduce excess power on large scales. We also show in the blue curve that the RSD contribution to the fiducial power is not negligible, as compared to the orange curve, which has no RSD term. The fiducial powers are computed using $\texttt{PyCCL}$.}
    \label{fig:desils_dr9_extlrg_Cgg}
\end{figure}

\paragraph{Estimated galaxy power spectra.} Fig.~\ref{fig:desils_dr9_extlrg_Cgg} displays four of the 20-bin galaxy auto-powers. We compute the mask-decoupled pseudo-$C_{\ell}$ power spectrum from the $N_{\text{side}}=256$ maps using the Python package \texttt{PyMaster}~\cite{Alonso:2018pymaster, namaster}. The green curves show the power spectrum calculated from galaxy overdensity maps with just a random catalog subtraction. In contrast to the green curves, the red curves are computed from the calibrated galaxy overdensity maps, which mitigate the excess power on large scales $\ell < 20$ across all redshift intervals. The imaging calibration becomes almost negligible at $\ell > 50$ and the two curves converge. For reference, we include the fiducial galaxy auto-powers in the plot as the blue curves, which include the redshift space distortion (RSD) contribution, and the orange curves, which do not. The RSD contribution is non-negligible on large scales for this analysis.

\begin{figure}[h!]
    \centering
    \includegraphics[width=1.0\columnwidth]{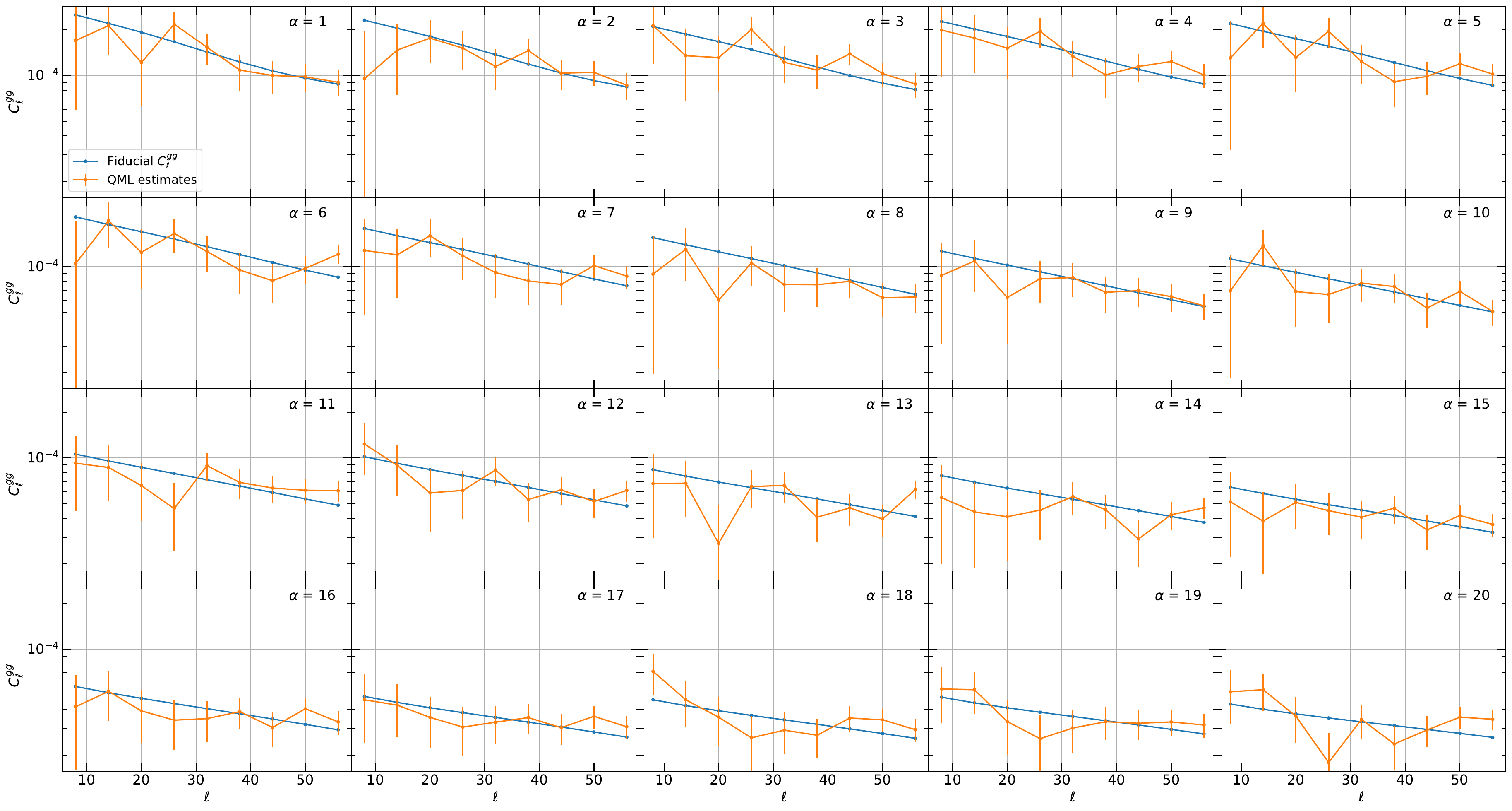}
    \caption{QML estimates on the large-scale galaxy auto-powers $C^{gg, \alpha}_{\ell}$ between $\ell=5$ and $\ell = 60$. We bandpower the estimates in 6 modes per band with a uniform boxcar for better visualization. The data points mark the center of each bandpower. The errorbars are obtained from the analytic covariance matrix presented in Sec.~\ref{subsec:Cvg result}.}
    \label{fig:Cgg_QML_estimates}
\end{figure}

We also present the bandpowered QML estimates for the 20 galaxy auto-powers in Fig.~\ref{fig:Cgg_QML_estimates} over a smaller range of modes $5 \leq \ell \leq 60$. We refer the reader to App.~\ref{app:bandpowering scheme} for the details of the bandpowering scheme employed in this work. The $N_{\text{side}}=32$, calibrated maps are used for the QML estimators.

\begin{figure}[h!]
    \centering
    \includegraphics[width=0.6\columnwidth]{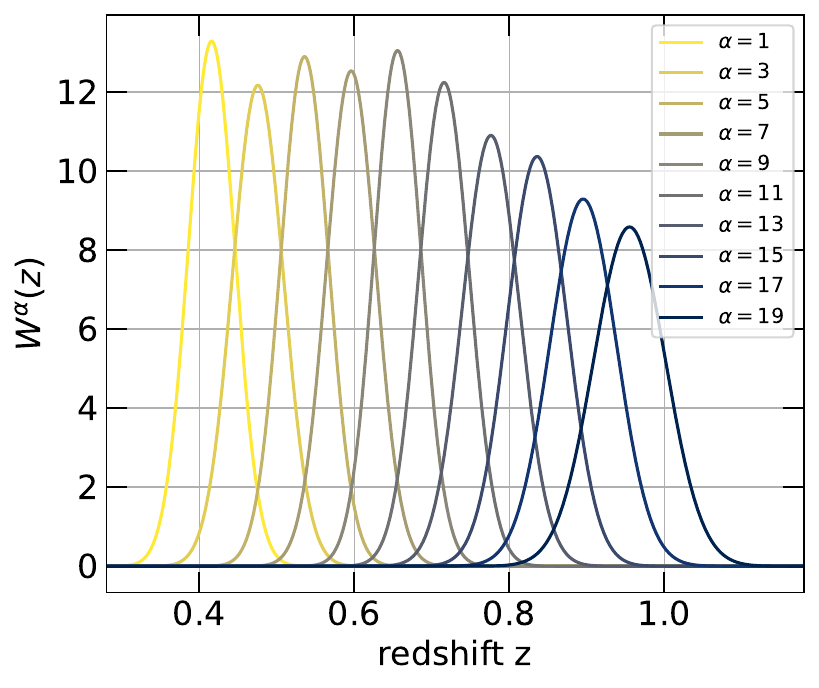}
    \caption{The normalized radial window functions $W^{\alpha}(z)$ of the 20 galaxy tomographic bins in an interval of 2 bins. Objects from $z = 0.4$ to $z = 1$ are used in this analysis. The redshift bins presented in this figure have been convolved with the photo-$z$ error.}
    \label{fig:desils_dr9_extlrg_ng}
\end{figure}

% This number reduces to $2.8 \times 10^7$ after applying the veto mask to remove objects that do not match the selection criteria as in~\cite{Zhou:2023DESILS_Cggsample2}. This sample doubles the number of galaxies that were used in~\cite{McCarthy:2024_ACT_DESI_ksz} by extending the analysis to a higher redshift up to 1.0, hence improving the constraining power on the baryonic feedback.

\subsection{Quadratic Estimator and Noise Covariance Estimation}
\label{subsec:velocity field QE}

We give a brief summary on the binned-2D quadratic velocity field estimator used for the kSZ reconstruction in this analysis.\footnote{The alternative optimal maximum a posteriori (MAP) estimator~\cite{Contreras:2022zdz} leads to an improvement in reconstruction noise only in the case of significantly lower noises in the CMB and galaxy data than present surveys provide \cite{Kvasiuk:2023_2Dksz_estim3}.} The velocity modes at a particular redshift bin $\hat{v}^{\alpha}_{\ell m}$ can be estimated from a quadratic combination of CMB anisotropy $a_{\ell m}$ and redshift-binned galaxy overdensities $\delta^{\alpha}_{g, \ell m}$:
\begin{align}\label{eq:v estim}
    \hat{v}_{lm}^{\alpha} &= \sum_{l_1m_1,l_2m_2}W_{lm, l_1m_1l_2m_2} a_{l_1m_1}\delta^{\alpha}_{g,l_2m_2} \, .
\end{align}
Here $W_{lm, l_1m_1l_2m_2}$ is a weight to be determined through the constraint of an unbiased estimator $\left<\hat{v}_{lm}^{\alpha}\right> = v_{lm}^{\alpha}$ and minimum variance $\left<\left(\hat{v}_{lm}^{\alpha} -  v_{lm}^{\alpha}\right)^2\right>$. As described in~\cite{Deutsch:2017_2Dksz_estim1, Smith:2018bpn, Kvasiuk:2023_2Dksz_estim3, Bloch:2024PlanckxUnwise}, by treating $a_{\ell m}$ and $\delta^{\alpha}_{g, \ell m}$ to be Gaussian to the leading order and by working in the regime where the kSZ signal is dominated by the small-scale power, the weight can be solved using the Lagrange multiplier, leading to the following result:
\begin{align}\label{eq:v estim result}
    \hat{v}_{lm}^{\alpha} &= N^{vv, \alpha}_{\ell}\sum_{\ell_1, \ell_2, m_1, m_2}
    (-1)^{m+1}\Gamma^{\alpha}_{\ell_1\ell_2\ell}
    \begin{pmatrix}
    \ell_1 & \ell_2 & \ell  \\
    m_1 & m_2 & -m  
    \end{pmatrix}\frac{a_{\ell_1m_1}\delta^{\alpha}_{g,\ell_2m_2}} {C^{TT}_{\ell_1}C^{gg, \alpha}_{\ell_2}} \, ,
\end{align}
where $C^{TT}_{\ell_1}$ is the full-sky CMB auto-power, $C^{gg, \alpha}_{\ell_2}$ is the full-sky galaxy power spectrum at redshift bin $\alpha$ and both include signal and noise. 

\begin{figure}[h!]
    \centering
    \includegraphics[width=0.5\columnwidth]{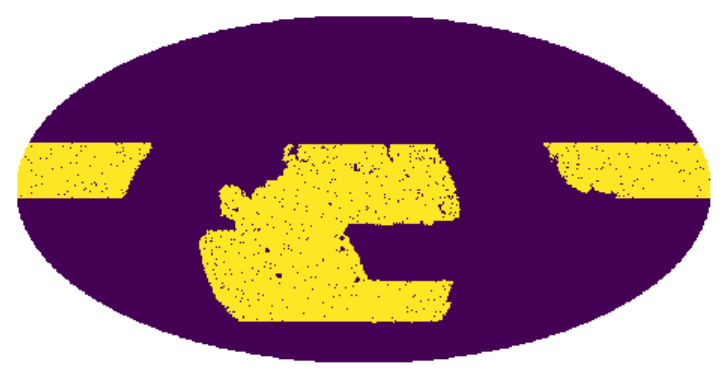}
    \caption{The mask constructed from the overlapping region between the ACT DR6 CMB map and DESI-LS DR9 survey. Dots with masked value of zero show the regions that do not pass the selection defined in the DESI LRG analysis~\cite{Zhou:2023DESILS_Cggsample2}. The mask is applied to the pixel-space velocity estimator $\hat{v}(\hat{n})$ on the estimator output data.}
    \label{fig:ACTxDESI overlap}
\end{figure}

 The kSZ coefficient $\Gamma^{\alpha}_{\ell_1\ell_2\ell}$ is defined as:
\begin{align}\label{eq:ksz coefficient}
    \Gamma^{\alpha}_{\ell_1\ell_2\ell} &= \sqrt{\frac{(2\ell_1+1)(2\ell_2+1)(2\ell+1)}{4\pi}}\begin{pmatrix}
    \ell_1 & \ell_2 & \ell \\
    0 & 0 & 0 
    \end{pmatrix}
    C^{\tau g, \alpha}_{\ell_2} \, ,
\end{align}
such that the Wigner 3-j symbols $\begin{pmatrix}
    \ell_1 & \ell_2 & \ell_3 \\
    m_1 & m_2 & m_3
\end{pmatrix}$
can be expressed through the angular integral over 3 spherical harmonics as follows:
\begin{align}
    \sqrt{\frac{(2\ell_1+1)(2\ell_2+1)(2\ell+1)}{4\pi}}\begin{pmatrix}
    \ell_1 & \ell_2 & \ell \\
    0 & 0 & 0 
    \end{pmatrix} \begin{pmatrix}
    \ell_1 & \ell_2 & \ell  \\
    m_1 & m_2 & -m  
    \end{pmatrix} &= \int d\mathrm{\Omega}(\hat{n}) \,Y_{\ell_1,m_1}(\hat{n})\,Y_{\ell_2, m_2}(\hat{n})\,Y_{\ell,\,-m}(\hat{n}) \, .
\end{align}
$C^{\tau g, \alpha}_{\ell_2}$ is the cross power between the electron density and galaxy at redshift bin $\alpha$. Note that Eq.~\ref{eq:v estim result} can be expressed and computed in a more compact form by transforming $\hat{v}_{\ell m}$ back to pixel space, such that~\cite{Bloch:2024PlanckxUnwise}:
\begin{align}\label{eq: v estimator pixel speace expression}
    v^{\alpha}(\hat{n}) &= -N^{vv, \alpha} a'(\hat{n})\,\delta'^{\alpha}_{g}(\hat{n}) \, ,
\end{align}
where we make use of the fact that the kSZ reconstruction noise $N^{vv, \alpha}$ can be approximated by its mean across the multipoles. The primed variables denote the inverse-variance-filtered quantities in pixel space, such that:
\begin{align}
    a'(\hat{n}) &= \sum_{\ell_1, m_1}\frac{1}{C^{TT}_{\ell_1}}\,a_{\ell_1 m_1}\,Y_{\ell_1, m_1}(\hat{n}) \, , \nonumber \\
    \delta'^{\alpha}_{g}(\hat{n}) &= \sum_{\ell_2, m_2}\frac{C^{\tau g, \alpha}_{\ell_2}}{C^{gg, \alpha}_{\ell_2}}\,\delta^{\alpha}_{g, \ell_2 m_2}\,Y_{\ell_2, m_2}(\hat{n}) \, .
\end{align}
The kSZ reconstruction noise $N^{vv, \alpha}_{\ell}$ can be obtained through the following expression:
\begin{align}\label{eq:ksz noise}
    \frac{1}{N^{vv, \alpha}_{\ell}} &= \frac{1}{2\ell+1}\sum_{\ell_1, \ell_2}\frac{\left(\Gamma^{\alpha}_{\ell_1\ell_2\ell}\right)^2}{\tilde{C}^{TT}_{\ell_1}\tilde{C}^{gg, \alpha}_{\ell_2}} \, .
\end{align}
%\moritz{clarify that Ct and Cg are estimated from the maps and include noise.}
In practice, the pixel-space expression Eq.~\ref{eq: v estimator pixel speace expression} is implemented to estimate the pixel-space velocity fields. We use observed CMB maps and galaxy overdensity maps of resolution $N_{\text{side}}=4096$, such that the kSZ component in CMB maps can be well captured. A mask in pixel space that accounts for the overlapping footprint between ACT DR6 (Sec.~\ref{sec:ACTDR6data}) and the DESI LRG survey (Sec.~\ref{subsubsec: DESI LS data}) is applied to the maps. The shape of the overlapping mask is shown in Fig.~\ref{fig:ACTxDESI overlap}. We also apply a filter in the spherical harmonic space to remove modes beyond $\ell = 10000$. The theoretical reconstruction noise is calculated separately in spherical harmonic space through Eq.~\ref{eq:ksz noise}. We include modes with $\ell_1, \ell_2 \in [300, 10000]$ within the summation. Extending the summation of modes below $\ell = 300$ only leads to sub-percent changes to the reconstruction noise. Hence, it can be neglected.

$\tilde{C}^{TT}_{\ell_1}$ and $\tilde{C}^{gg, \alpha}_{\ell_2}$ in Eq.~\ref{eq:ksz noise} denote the full-sky estimates of the CMB and galaxy auto-power spectra. They are derived from the observed data, such that $\tilde{C}^{TT}_{\ell_1}$ includes survey noise and foreground contamination from the tSZ effect and CIB. Similarly, $\tilde{C}^{gg, \alpha}_{\ell_2}$ carries the shot noise arising from the finite number of galaxies and uncertainties due to photo-z smearing. For this analysis, we estimate the full-sky $\tilde{C}^{TT}_{\ell_1}$ and $\tilde{C}^{gg, \alpha}_{\ell_2}$ from the unmasked region using the pseudo-$C_\ell$ routine in \texttt{PyMaster}. For these high-$\ell$ power spectra which enter the kSZ reconstruction (Eq.~\ref{eq:ksz noise}), the QML is neither suitable nor necessary.

There is currently no direct probe to the electron-galaxy cross-power $C^{\tau g, \alpha}_{\ell_2}$. In this analysis, we calculate $C^{\tau g, \alpha}_{\ell_2}$ as described in Sec.~\ref{subsec: fiducial powers}. The uncertainty on the exact amplitude of $C^{\tau g, \alpha}_{\ell_2}$ induces a linear bias on the overall amplitude of the velocity-galaxy cross-power $C^{vg}_{\ell}$, also known as the optical depth degeneracy~\cite{Smith:2018bpn, Bloch:2024PlanckxUnwise}. 

\paragraph{Velocity field monopole removal.}\label{subsec: mean-subtraction}
The kSZ quadratic estimator in Eq.~\ref{eq:v estim result} assumes that the two-point correlation between the galaxy field and observed CMB anisotropy, $\left<gT\right>$, is negligible. This is not always true because of residual foregrounds in the CMB maps. To lowest order, such a correlation could induce a bias in the monopole of the reconstructed velocity field. We thus remove the monopole from our kSZ velocity reconstruction by subtracting the mean of the reconstructed maps, where the mean is determined from the unmasked region. We note that~\cite{SelimKendrickPaper} and~\cite{McCarthy:2025brx} included similar procedures in their pipelines.

\begin{figure}[h!]
    \centering
    \includegraphics[width=0.4\columnwidth]{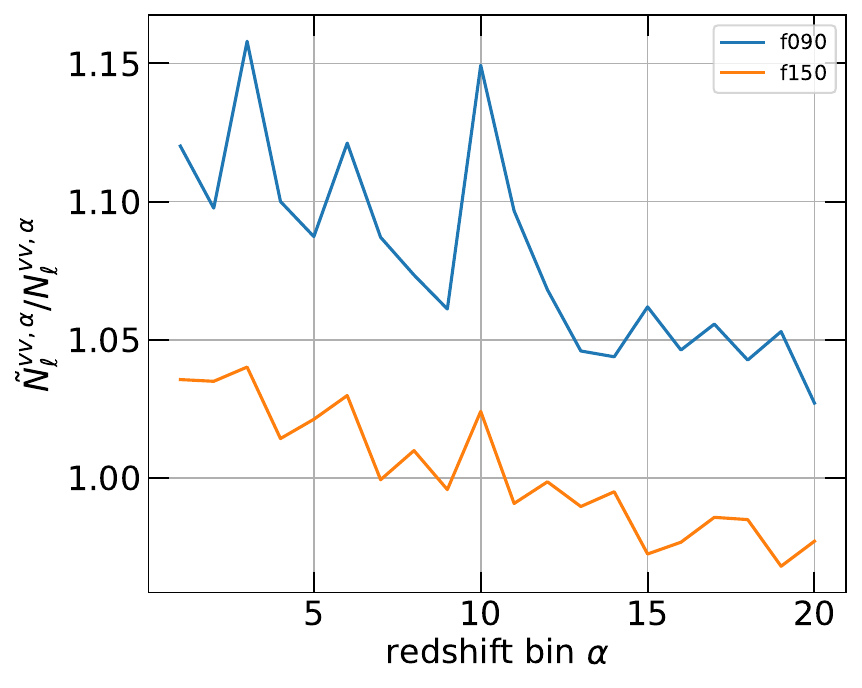}
    \caption{Figure comparing the theoretical kSZ reconstruction noise $N^{vv,\alpha}_{\ell}$ to the best-fit value $\tilde{N}^{vv,\alpha}_{\ell}$ estimated from the data. The theoretical calculations follow from Eq.~\ref{eq:ksz noise}. The best-fit estimated noises are obtained by taking the mean value between $\ell = 100$ to $\ell = 300$ in the pseudo-$C_{\ell}$ velocity auto-powers of the estimated velocity field, which is noise-dominated at these scales. The pseudo-$C_{\ell}$ power spectra are computed using \texttt{PyMaster} on $N_{\text{side}} =128$, mean-subtracted velocity maps.}
    \label{fig:Nvv theory vs fit}
\end{figure}

\paragraph{Channel-wise estimator noise.} The noise in each channel can be calculated theoretically using Eq. \eqref{eq:ksz noise}, assuming Gaussian fields (corrections to this noise model were studied in \cite{Giri:2020pkk}). We can also estimate the noise covariance of the kSZ estimator, in the $\ell$ range of interest ($\ell < 300$), directly from the data. To do this, we assume that the noise is Gaussian. We also assume that the noise is flat in $\ell$, which implies that noise in different pixels is uncorrelated (on large scales, i.e. on $N_{\rm side}$=32). With these assumptions, and knowing that at $\ell > 50$ the estimator is completely noise dominated, we can fit the noise amplitude to the estimator response. Alternatively, one could run Gaussian simulations to determine the noise level. However, it is not trivial to make the simulations accurate at the small scales required here (including non-Gaussianity), and we believe that bootstrapping the noise covariance from the data is simpler and more reliable. We investigate the potential mismatch between the theoretical reconstruction noise $N^{vv, \alpha}_{\ell}$ and noise level given by the estimator response. In particular, we compute the pseudo-$C_\ell$ velocity auto-powers $C^{vv}_{\ell}$ from the reconstructed, masked velocity maps $\hat{v}(\hat{n})$ and take the mean value between $\ell = 100$ and $\ell = 300$ as the best-fit reconstruction noise $\tilde{N}^{vv, \alpha}_{\ell}$. We find a mismatch between the theoretical and the estimator noise level, which is displayed in Fig.~\ref{fig:Nvv theory vs fit}. The ratio of two noise levels, depending on the frequency channel and the redshift bin, can range from 0.98 to 1.15~\footnote{In the initial version of this paper, we found a larger mismatch between the theoretical and estimated noise. The earlier calculation of the theoretical noise was biased by an insufficient apodization of the data prior to power spectrum estimation.}. Note that a similar mismatch was also found in other kSZ reconstruction pipelines~\cite{McCarthy:2024_ACT_DESI_ksz, Bloch:2024PlanckxUnwise}. In~\cite{McCarthy:2024_ACT_DESI_ksz} they associated the remaining mismatch with the small-scale non-Gaussianity in the CMB maps. To avoid the possibility of over/underestimating the reconstruction noise, we adopt the best-fit values $\tilde{N}^{vv, \alpha}_{\ell}$ as our fiducial kSZ noise for the QML estimator. We drop the notation $\tilde{N}^{vv, \alpha}_{\ell}$; $N^{vv, \alpha}_{\ell}$ represents the best-fit reconstruction noise hereafter.

\paragraph{Cross-channel estimator noise.} Note that the kSZ reconstruction is performed over each CMB map, leading to individual estimates on the velocity field for each frequency channel. The CMB signal in different frequencies is strongly correlated. Therefore, one has to account for the kSZ reconstruction noise from the cross-channel velocity auto-powers. For the 90 GHz and 150 GHz CMB used in this analysis, we only consider the cross-channel noise $N^{vv, \alpha}_{\ell, 90\times150}$. Using Eq.~\ref{eq:v estim result}, one can compute the cross-channel noise as follows:
\begin{align}\label{eq:cross channel noise}
    N^{vv, \alpha}_{\ell, 90\times150} &= \left<\hat{v}_{lm, 90}^{\alpha}\;\hat{v}_{lm, 150}^{\alpha}\right> \, , \nonumber \\
    &= N^{vv, \alpha}_{\ell, 90}N^{vv, \alpha}_{\ell, 150}\sum_{\ell_1, \ell_2, m_1, m_2} \left(\Gamma^{\alpha}_{\ell_1\ell_2\ell}\right)^2
    \begin{pmatrix}
    \ell_1 & \ell_2 & \ell  \\
    m_1 & m_2 & -m  
    \end{pmatrix}^2\frac{\tilde{C}^{TT}_{\ell_1, 90\times150}} {\tilde{C}^{TT}_{\ell_1, 90}\tilde{C}^{TT}_{\ell_1, 150}\tilde{C}^{gg, \alpha}_{\ell_2}} \, , \nonumber \\
    &= \frac{N^{vv, \alpha}_{\ell, 90}N^{vv, \alpha}_{\ell, 150}}{2\ell+1}\sum_{\ell_1, \ell_2} \frac{\left(\Gamma^{\alpha}_{\ell_1\ell_2\ell}\right)^2\;\tilde{C}^{TT}_{\ell_1, 90\times150}} {\tilde{C}^{TT}_{\ell_1, 90}\tilde{C}^{TT}_{\ell_1, 150}\tilde{C}^{gg, \alpha}_{\ell_2}} \, .
\end{align}
In practice, we estimate the cross-channel noise directly from the reconstructed velocity maps, following the same approach used for the channel-wise estimator noise.

\subsection{Velocity Map co-adding}\label{subsec:vmap coadd}

To maximize the efficiency of the power spectrum estimation from the set of galaxy and velocity maps, we first create a set of co-added velocity maps by defining a linear combination of individual velocity maps from the same redshift bins and with different frequency channels:
\begin{align}
    v^{\alpha}_{\text{coadd}} = Av^{\alpha}_{90}+Bv^{\alpha}_{150} \, .
\end{align}

$A$ and $B$ are coefficients to be determined. Taking the covariance of the maps and using the fact that $\left<v^{\alpha}_{90, 150, 90\times150}\right> = 0$, we arrive at the expression:
\begin{align}\label{eq: v coadded}
    C^{vv} + N^{vv}_{\text{coadd}} &= \left(A+B\right)^2 C^{vv} + A^2 N^{vv}_{90}+B^2 N^{vv}_{150} + 2ABN^{vv}_{90\times150} \, ,
\end{align}
where $C^{vv}$ is now the pixel covariance matrix and $N^{vv}_{\rm{coadd}}$ is the co-added noise covariance matrix in pixel space. Then one can enforce the normalization condition that $A+B = 1$ and minimize the expression with respect to $A$, giving:
\begin{align}\label{eq: v coadded weights}
    A = \frac{N^{vv}_{150} - N^{vv}_{90\times150}}{N^{vv}_{90}+N^{vv}_{150}-2N^{vv}_{90\times150}} \,\,\,\,,\,B = \frac{N^{vv}_{90} - N^{vv}_{90\times150}}{N^{vv}_{90}+N^{vv}_{150}-2N^{vv}_{90\times150}} \, ,
\end{align}
and the co-added noise covariance $N^{vv}_{\text{coadd}}$ in pixel space is given by the sum of the remaning terms on the right hand side of Eq.~\ref{eq: v coadded}. In general $N^{vv}_{\text{coadd}}$ can be anisotropic, so that the diagonal terms can be different from pixel to pixel. However, under the assumption that the co-added noise behaves as the single-channel noise, which is nearly isotropic, the conversion between pixel space $N^{vv}_{\text{coadd}}$ and the power spectrum $N^{vv}_{\ell, \text{coadd}}$ can be simplified:
\begin{align}
    N^{vv}_{\text{coadd}} &= \frac{12N^2_{\text{side}}}{4\pi}N^{vv}_{\ell, \text{coadd}} \, .
\end{align}
For simplicity, we omit the subscript `$\text{coadd}$' from the velocity field for the remaining text. Any quantity associated with the velocity estimation should be understood as co-added.

\subsection{Quadratic Maximum Likelihood Cross-Power Spectrum Estimator}\label{subsec:QML theory}
We apply the quadratic maximum likelihood (QML) method ~\cite{Tegmark:2001cmbqml, Vanneste:2018xqml, Bilbao-Ahedo:eclipse_qml} to the partial-sky velocity and galaxy overdensity data to reconstruct the full-sky velocity-galaxy cross-power $C^{v_\alpha g_\beta}_{\ell}$ and compare them to fiducial calculations. QML is an optimal estimator for the power spectrum of a Gaussian field with an arbitrary mask, and can extract more SNR than the common but sub-optimal pseudo-$C_\ell$ estimator. Our implementation of the estimator also features an analytically calculated covariance matrix, rather than one estimated from Monte Carlo simulations, which would require a large number of simulated maps. We make our QML pipeline publicly available at~\url{https://github.com/ykvasiuk/qmlfast}. More details including the formalism, optimization and various robustness tests can be found in our companion paper~\cite{Kvasiuk:qmlpaper}.

\paragraph{Data vector.} The data vector used for QML is constructed from the pixel-space galaxy overdensity maps and the co-added kSZ reconstructed velocity maps such that $\mathbf{d}^T = \left(\mathbf{g}^T, \mathbf{v}^{T}\right) = \left(\delta^1_g, \delta^2_g, ..., \delta^{20}_g, v^{1}, v^{2}, ...,v^{20}\right)$. Here each $\delta^\alpha_g$ represents the galaxy overdensity map in redshift bin $\alpha$ and each $v^\alpha$ represents the reconstructed velocity field in redshift bin $\alpha$. Note that the maps contain only unmasked pixels in both the galaxy and CMB survey. For this analysis with an overlapping sky fraction of around $23\%$, the number of pixels included for each map in $N_{\text{side}} = 32$ will be approximately $3000$ pixels. We note that this pixel resolution should be sufficient to obtain the full SNR, because the maximum multipole available from this resolution is $3 N_{\text{side}} -1 = 95$, and as we show in App.~\ref{app: fisher forecast qml} from Fisher forecast the SNR converges at around $\ell \simeq 50$.

\paragraph{Optimal Quadratic Estimator.} As described in~\cite{Tegmark:2001cmbqml, Vanneste:2018xqml, Bilbao-Ahedo:eclipse_qml, Kvasiuk:qmlpaper}, we construct an estimator for the power spectrum as follows:
\begin{align}
    \label{eq:qmlmaster}
    \hat{C}_{b} &= \mathbf{d}^T\mathbf{Q}_{b}\mathbf{d} \, .
\end{align}
Here $\hat{C}_{b}$'s are the QML estimated powers. The indexing of $b$ covers all the power spectrum modes constructed from the fields in the data vector. We compute the QML estimation up to $\ell_{max} = 60$ (note that $\ell_{max}$ is defined as the largest multipole output from the QML estimator in this analysis and is different from the maximum multipole provided by a Healpix map with a given $N_{\rm side}$) for all the power spectra constructed from a 40-field data vector, so that $b$ ranges from $b = 0$ to $b = 50019$ (820 power spectra $\times$ 61 modes). The matrix $\mathbf{Q}_{b}$ is computed for each mode and is determined by the following equation:
\begin{align}\label{eq:qml Q definition}
    \mathbf{Q}_{b} &= \frac{1}{2}\sum_{b'}\left(F^{-1}\right)_{bb'}\;\mathbf{C}^{-1}\mathbf{P}_{b'}\mathbf{C}^{-1} \, .
\end{align}
$F_{bb'}$ is a component in the Fisher matrix $\mathbf{F}$ that represents inverse-covariance between two different modes $b$ and $b'$. The matrix $\mathbf{C}$ is the pixel-space covariance matrix that includes both the signal computed from the fiducial power spectrum and the noise that depends on the survey properties, explicitly:
\begin{align}\label{eq: pixel covariance matrix explicit}
    \mathbf{C} &= \mathbf{S}+\mathbf{N} \, , \nonumber \\
    &=\begin{pmatrix}
        \mathbf{S}^{gg}, &\mathbf{S}^{gv} \\
        \mathbf{S}^{vg}, &\mathbf{S}^{vv} \\
    \end{pmatrix} +
    \begin{pmatrix}
        \mathbf{N}^{gg}, &0 \\
        0, &\mathbf{N}^{vv}
    \end{pmatrix}\, ,
\end{align}
with each term understood to be a smaller block matrix that contains the covariances from different redshift bins. The noise covariance matrix $\mathbf{N}$ is assumed to be uncorrelated between two different pixels and redshift bins. $\mathbf{P}_{b}$ is the Legendre matrix in pixel space for a particular mode $b$ and is purely geometric, which means that it depends only on the separation between two pixels.

\paragraph{Varying pixel-space noise covariance.} The kSZ noise covariance $N^{vv}$ can be position dependent due to the inhomogeneous noise and systematics in the ACT CMB temperature maps and DESI-LS galaxy maps. While in~\cite{SelimKendrickPaper} this is addressed by discarding the Southern sky region in the analysis, the dense-covariance QML introduced in this work can naturally incorporate such inhomogeneity in the noise covariance matrix due to its pixel-space construction. This is accomplished by first performing the kSZ reconstruction on the Northern sky and Southern sky individually and then modelling the QML noise covariance in Northern and Southern pixels according to their own kSZ reconstruction noise $N^{vv}_{\ell}$. In this work, we try both approaches (North + South kSZ reconstruction and North, South separately) and find a slight boost of SNR ($\sim 0.5$) using the latter pipeline, which we report in this work as the main result\footnote{The North + South configuration gives $A = 0.445 \pm 0.0467$ on the first QML iteration, which is consistent with the main result in this work within 1-$\sigma$ (see Table~\ref{Tb:QML summary}).}.

\paragraph{Fisher matrix.} The Fisher matrix components $F_{bb'}$ are also precomputed from the definition:
\begin{align}\label{eq:QML Fisher matrix}
    F_{bb'} &= \frac{1}{2}\text{Tr}(\mathbf{P}_{b}\mathbf{C}^{-1}\mathbf{P}_{b'}\mathbf{C}^{-1}) \, .
\end{align}
Eq.~\ref{eq:qml Q definition} implies that the QML estimator for mode $b$ is the sum of individual contributions from all modes $b'$, weighted by the corresponding inverse Fisher element. The detailed definitions of $\mathbf{P}_{b}$ and $(\mathbf{F}^{-1})_{bb'}$, in Eq.~\ref{eq:qml Q definition} can be found in~\cite{Tegmark:2001cmbqml, Vanneste:2018xqml, Bilbao-Ahedo:eclipse_qml, Kvasiuk:qmlpaper}. In general, each evaluation of the Fisher element requires an $O(N^3)$ operation, where $N \approx 120000$ is the number of unmasked pixels multiplied by the number of fields involved ($\sim 3000$ pixels $\times$ 40 fields). Therefore, a full calculation of the Fisher matrix can be a computationally challenging process. However, one can utilize the fact that $\mathbf{P}_{b}$ is very sparse in the scenario of a large number of input fields and simplify the calculation by breaking up a large matrix multiplication into smaller ones~\cite{Tegmark:2001cmbqml, Bilbao-Ahedo:eclipse_qml, Kvasiuk:qmlpaper}. Additionally, one can use the addition theorem of spherical harmonics to further reduce the number of operations needed to compute the trace. We refer the readers to our companion paper on the QML estimator, which is a \texttt{Python}-based, optimized implementation of the QML estimator over an arbitrary number of correlated fields ~\cite{Kvasiuk:qmlpaper}. Due to our extensive optimizations, the full power spectrum analysis, including calculating the Fisher matrix, at the resolution used in this paper, takes only about one hour on a computing node. We have tested the unbiasedness of our estimator and the correctness of its covariance matrix extensively on Gaussian simulations, confirming that it outperforms the pseudo-$C_\ell$ as expected. These tests are presented in the companion paper on our QML estimator~\cite{Kvasiuk:qmlpaper}.

\paragraph{Cross-power purification.} The QML in its general form will use all quadratic combinations of the data for an optimal estimate, i.e. estimates of the cross-power of galaxies and velocities would contain information also from auto-powers of these fields. In the present case, auto-powers are affected by calibration issues, and we thus want to exclusively use quadratic combinations of two different fields (galaxies and velocities). This issue is similar to the traditional QML method applied to the CMB analysis, where there is an "E-to-B mode leakage" issue.

Since the goal of our analysis is to establish the SNR of the cross correlation, the variance should not include the contribution from the auto powers.
To achieve this, we set up the cross-power-purified covariance matrix $\textbf{C}$ by manually zeroing out fiducial velocity-galaxy cross-powers, such that the QML estimators $\mathbf{Q}_i$, constructed from the new covariance matrix, acquire zero contributions from auto-correlations, i.e., $\mathbf{v}^{T}\mathbf{Q}_b\mathbf{v} = \mathbf{g}^{T}\mathbf{Q}_b\mathbf{g} = 0$, where $\mathbf{v}$ and $\mathbf{g}$ are the data vectors for the galaxy and velocity maps.

The cross-power-purified pixel-space covariance matrix is given by:
\begin{align}\label{eq:pixel-space covariance mode-purified}
    \mathbf{C}_{C^{vg}\text{-purified}} &= 
    \begin{pmatrix}
        \mathbf{S}^{gg}+\mathbf{N}^{gg}, &0 \\
        0 &\mathbf{S}^{vv}+\mathbf{N}^{vv} \\
    \end{pmatrix} \, .
\end{align}
This technique was already proposed in~\cite{Tegmark:2001cmbqml, Bilbao-Ahedo:eclipse_qml} in the CMB context. We point out that we indeed need to set the cross-powers to zero in the covariance, even though these are the powers we want to measure. These weights ensure that only combinations of form $(v,g)$ have non-zero weights, while $(g,g)$ and $(v,v)$ are nulled out. This is also consistent with the definition of the detection SNR as the significance of the rejection of the null hypothesis, because our null hypothesis is that the reconstructed velocities are not correlated with the galaxies.

\paragraph{Low $\ell$ mode-deprojection.}
In Sec.~\ref{subsec: mean-subtraction} we describe the mean-subtraction procedure that mitigates the bias in the kSZ velocity reconstruction. The treatment is exact only on the full sky. In the presence of a mask, these biases, including both the monopole subtraction itself and higher-order correlations between the CMB foreground and the galaxy field, can be coupled up to a few multipoles above $\ell = 0$. This is the ``mean-field bias'' problem known in the CMB lensing reconstruction literature~\cite{Namikawa:2012pe, Benoit-Levy:2013uxb}. These biases exist at the field level and cannot be simply removed by zeroing the entries in the power spectrum covariance matrix. In this work, we use the mode-deprojection method suggested in~\cite{Tegmark:2001cmbqml}, which is also a built-in feature in our estimator~\cite{Kvasiuk:qmlpaper}, to prevent field-level contributions from modes with $\ell < \ell_{\rm min} = 5$ from entering the QML estimator. The $\ell_{\rm min}$ cutoff also prevents survey systematics, which get worse on large scales, from contaminating the QML power spectrum estimators.

\paragraph{Iterating the QML fiducial power.} As described in~\cite{Tegmark:2001cmbqml, Bilbao-Ahedo:eclipse_qml, Kvasiuk:qmlpaper}, the QML estimator is only exactly optimal if the fiducial powers in the estimator match the true data distribution (which is to be measured). It is shown in~\cite{Vanneste:2018xqml, Kvasiuk:qmlpaper} that the choice of a fiducial power spectrum can affect the error bars in the signal-dominated regime, which is the case for the velocity-galaxy power spectrum at large scales. Therefore the QML estimator is sometimes used iteratively: One first estimates best-fit model parameters using some fiducial covariance, and then updates the fiducial covariance with the new model parameters to run the estimator again. This will converge to optimal measurements. In our case, we found that our fiducial value for the so-called optical depth degeneracy parameter $b_v$ is not very close to the inferred value. We therefore do a second round of QML estimation with updated parameters (we update an overall redshift independent amplitude $A$, not individual bin-wise $b_v$, since these are noisy). Indeed, as we will see, this improves the SNR of the cross-correlation (an improvement of $9.7\%$ (tSZ-masked configuration) for this work, see Table~\ref{Tb:QML summary}).

\subsection{Fiducial Powers}
\label{subsec: fiducial powers}
As shown in Eq.~\ref{eq:qml Q definition}, the construction of QML estimators relies on the fiducial power. This includes the velocity auto-powers $C^{v_\alpha v_\beta}_{\ell}$ and the galaxy auto-powers $C^{g^\alpha g^\beta}_{\ell}$ for any two redshift bins $(\alpha, \beta)$ selected from the 20 defined in Sec.~\ref{subsubsec: DESI LS data}. Although velocity-galaxy cross-powers $C^{v^\alpha g^\beta}_{\ell}$ are not required for QML estimators, they are still involved in the SNR calculations of the cross-power as described in Sec.~\ref{subsec:SNR calculation}. 
%The formulas are presented as various redshift-binned angular power spectrum $C_{\ell}$'s in terms of their three-dimensional analogue $P(k)$'s. 
For the large-scale power spectra we have compared our own implementation of the integrals below with the results from \texttt{PyCCL} and found consistency. Unless otherwise specified, we use the results from $\texttt{PyCCL}$~\cite{LSSTDarkEnergyScience:pyccl}. 

\paragraph{Galaxy power spectrum on large scales and galaxy bias.} We calculate the large-scale galaxy angular power spectrum between two redshift bins $(\alpha, \beta)$ as follows:
\begin{align}\label{eq: fiducial Cgg}
    C^{g^\alpha g^\beta}_{\ell} &= \frac{2c^2}{\pi} \int dz' dz'' W^{\alpha}(z')W^{\beta}(z'')\int dk \, k^2\,P^{gg}(z', z'', k) j_{\ell}(k\chi(z'))\,j_{\ell}(k\chi(z'')) \, ,
\end{align}
here $c$ is the speed of light. $W$ is the normalized redshift bin window function to account for the photo-z error as described in Sec.~\ref{subsubsec: DESI LS data}. Their implicit dependence on the redshifts is denoted by the superscript $\alpha$ and $\beta$. $P^{gg}(z', z'', k)$ gives the two-halo term of the three-dimensional linear galaxy cross power at redshifts $z'$ and $z''$, which are integrated. To the linear order $P^{gg}(z', z'', k)$ can be factored as $P^{gg}(z', z'', k) = \sqrt{P^{gg}(z', k)\,P^{gg}(z'', k)}$. The integral over the wavenumber $k$ includes the spherical Bessel functions $j_\ell(k\chi)$, where $\chi(z')$ is the comoving distance at $z'$.

The fiducial linear galaxy auto-powers are calculated from the matter power spectrum $P^{mm}(z', z'', k)$ using the following relation:
\begin{align}
    P^{gg}(z', k) &= (b_g(z')+f\mu^2)^2\, P^{mm}(z', k) \, ,
\end{align}
here $b_{g}(z')$ is the standard redshift-dependent galaxy bias, $f = d\log D(a)/d\log a$ is the logarithmic growth rate, $D(a)$ is the growth factor, and $\mu = \hat{k}\cdot\hat{r}$ is the line-of-sight projection factor. The term $f\mu^2$ accounts for the correction to the galaxy power spectrum due to redshift-space distortions.

\begin{figure}[h!]
    \centering
    \includegraphics[width=0.7\textwidth]{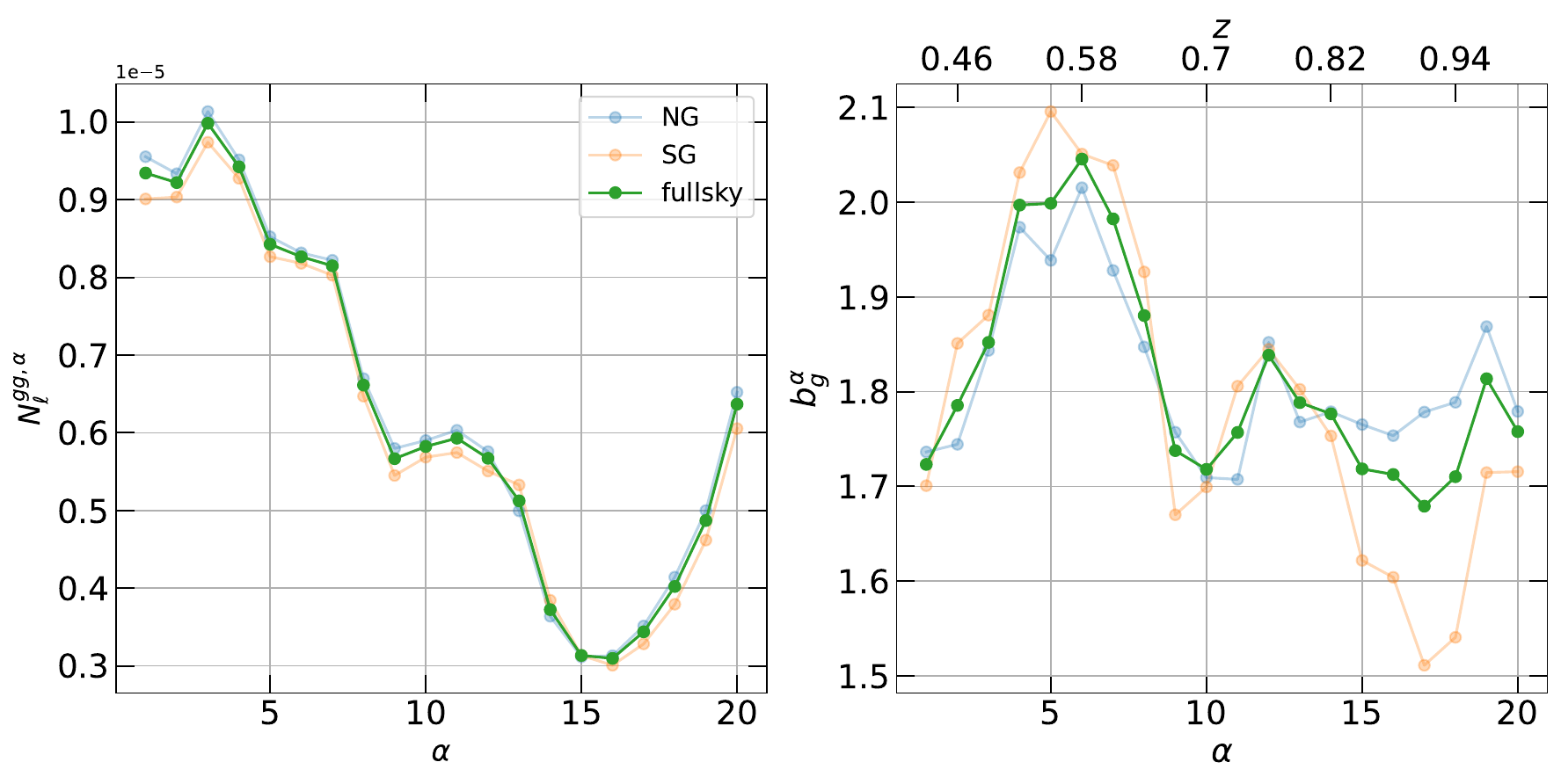}     
    \caption{Left: The galaxy shot noise $N^{gg, \alpha}_{\ell}$ for the 20 redshift bins $\alpha$, in units of steradians. For reference, we also include $N^{gg, \alpha}_{\ell}$ from just the Northern (NG, blue) and the Southern (SG, orange) galactic regions. They are computed from Eq.~\ref{eq: fiducial shot noise}, with the galaxy number density obtained from the DESI LRG extended sample. Right: A figure of the best-fit galaxy bias to the data across the redshift bins. The biases are defined with respect to the fiducial matter power spectrum $C^{mm}_{\ell}$, which is computed using \texttt{PyCCL}~\cite{LSSTDarkEnergyScience:pyccl}, such that on large scales $C^{gg,\alpha}_{\ell} = (b^\alpha_g)^2C^{mm}_{\ell}$.}
    \label{fig:galaxy shotnoise_bias}
\end{figure}

For narrow redshift windows, the galaxy biases $b_{g}(z')$ are well-approximated by their values at the redshift where the window function peaks, and can be pulled out from the integration in Eq.~\ref{eq: fiducial Cgg}. We determine redshift galaxy biases $b^\alpha_{g}$ by minimizing the chi-square difference between the observed galaxy auto-power spectrum $\tilde{C}^{g^\alpha g^\alpha}_\ell$ and the 2-halo angular matter power spectrum $C^{m^\alpha m^\alpha}_{\ell,\,\texttt{CCL}}$ over $\ell \in [138, 225]$:
\begin{align}
    \chi^2_\alpha = \sum_{\ell \in [138, 225]}\left( \tilde{C}^{g^\alpha g^\alpha}_\ell - ({b^\alpha_{g}})^2C^{m^\alpha m^\alpha}_{\ell,\,\texttt{CCL}}\right)^2 \, .
\end{align} 
The fiducial angular matter power spectrum is computed with \texttt{PyCCL}, which is a good approximation on large scales. We show the best-fit galaxy biases $b^\alpha_g$ in the right panel of Fig.~\ref{fig:galaxy shotnoise_bias}.

\paragraph{Galaxy shot noise.}  The QML estimator also requires a fiducial input for the noise of the galaxy auto-powers $N^{gg, \alpha}_{\ell}$. It can be directly translated from the galaxy number density $N^\alpha_{g}$:
\begin{align}\label{eq: fiducial shot noise}
    N^{gg, \alpha}_{\ell} = \frac{4\pi}{N^\alpha_{g}} \, ,
\end{align} 
which is uniform over $\ell$ at a given redshift bin $\alpha$. In fact, at the scales where the kSZ signal is significant ($\ell>2000$), all 20 bins are shot noise dominated. We also include a plot for the shot noise used for this analysis in the left panel of Fig.~\ref{fig:galaxy shotnoise_bias}.

\paragraph{Velocity power spectrum on large scales.} From the linear theory that relates the radial velocity mode and matter density modes, we write down the fiducial angular velocity auto-power and the fiducial angular cross power between velocity fields and galaxy density fields in a similar manner as Eq.~\ref{eq: fiducial Cgg}~\cite{Kvasiuk:2023_2Dksz_estim3}.
\begin{align}\label{eq: fiducial Cvv and Cvg}
    C^{v^\alpha v^\beta}_{\ell} &= \frac{2c^2}{\pi} \int dz' dz'' \left(faH\right)(z') \left(faH\right)(z'') W^{\alpha}(z')W^{\beta}(z'')\int dk \, P^{m m}(k, z', z'') j'_{\ell}(k\chi(z'))\,j'_{\ell}(k\chi(z'')) \, , \nonumber \\
    C^{v^\alpha g^\beta}_{\ell} &= \frac{2c^2}{\pi}\,b^{\beta}_{g} \int dz' dz'' \left(faH\right)(z') W^{\alpha}(z')W^{\beta}(z'')\int dk\, k \, P^{m m}(k, z', z'') j'_{\ell}(k\chi(z'))\,j_{\ell}(k\chi(z'')) \, .
\end{align}
Here $b^{\beta}_{g}$, $a(z')$ and $H(z')$ are the galaxy bias, the scale factor and the Hubble parameter, respectively.

\paragraph{Electron galaxy cross-power on small scales.} As mentioned in Sec.~\ref{subsec:velocity field QE}, the kSZ quadratic estimator requires a fiducial input of the electron-galaxy cross-power $C^{\tau g, \alpha}_{\ell}$. Unlike $C^{v^\alpha v^\beta}_{\ell}$, which only requires large-scale powers, $C^{\tau g, \alpha}_{\ell}$ has to be calculated up to $\ell > 3000$ in order to filter the kSZ signal in the CMB maps that are of the same scale. In practice, we follow the halo model description, which states that all the dark and baryonic matter are bound in halos with given density profiles~\cite{White:1977hm, Navarro:1995nfw, Cooray:2002hm}. We split the calculation of $C^{\tau g, \alpha}_{\ell}$ into the 1-halo term and 2-halo term; both are calculated using the Limber formula, which is a valid approximation to the small-scale power spectrum for two tracers with identical window functions~\cite{Limber:1953zz}. We note that the Limber approximation is implemented only in small-scale power spectrum calculations. For the 2-halo contribution:
\begin{align}
    C^{\tau g, \alpha}_{\ell, \, 2h} &= b^\alpha_{g}\,\sigma_T\,n_{e, 0} \int dz' W^{\alpha}(z')\left(\frac{1+z'}{\chi(z')}\right)^2\int dk \, P^{em}_{2h}(k, z') \, ,
\end{align}
here $\sigma_T$ is the Thomson cross-section, $n_{e,0}$ is the mean electron density at $z = 0$ such that $P^{eg}_{2h} \approx b^\alpha_{g}P^{em}_{2h}$ at redshift window $W^{\alpha}(z')$ on large scales. The 2-halo electron-matter power spectrum $P^{em}_{2h}$ is calculated using Python code \texttt{hmvec} with the Battaglia AGN-type electron profile as input~\cite{Smith:2018bpn, Battaglia:2016_battaglia_profile}. On small scales, the 1-halo term $C^{\tau g, \alpha}_{\ell, \, 1h}$ becomes important:
\begin{align}\label{eq:Cgtau 1 halo}
    C^{\tau g, \alpha}_{\ell, \, 1h} &= \sigma_T\,n_{e, 0} \int dz' W^{\alpha}(z')\left(\frac{1+z'}{\chi(z')}\right)^2\int dk \, P^{eg}_{1h}(k, z') \, .
\end{align}
We obtain $P^{eg}_{1h}$ directly from the $\texttt{hmvec}$ using the same electron profile. 
In addition, $P^{eg}_{1h}$ requires a halo occupation distribution model (HOD) that describes the relation between the number of galaxies and the halo mass. We adopt a 5-parameter vanilla model with the best-fit LRG parameters reported by~\cite{Yuan:2023_desi_one_percent} for the DESI one-percent survey. The sum $C^{\tau g, \alpha}_{\ell} = C^{\tau g, \alpha}_{\ell, \, 2h}+C^{\tau g, \alpha}_{\ell, \, 1h}$ is used as the input for kSZ reconstruction for our main result. 

We have also experimented with a simpler prescription for $C^{\tau g, \alpha}$ using a biasing model instead of an HOD:
\begin{align}\label{eq:Cgtau 1 halo, bias approximation}
    C^{\tau g, \alpha}_{\ell, \, 1h, \text{bias}} &\approx b^\alpha_{g}\,\sigma_T\,n_{e, 0} \int dz' W^{\alpha}(z')\left(\frac{1+z'}{\chi(z')}\right)^2\int dk \, P^{em}_{1h}(k, z') \, .
\end{align}

\begin{figure}[h!]
    \centering
    \includegraphics[width=0.8\textwidth]{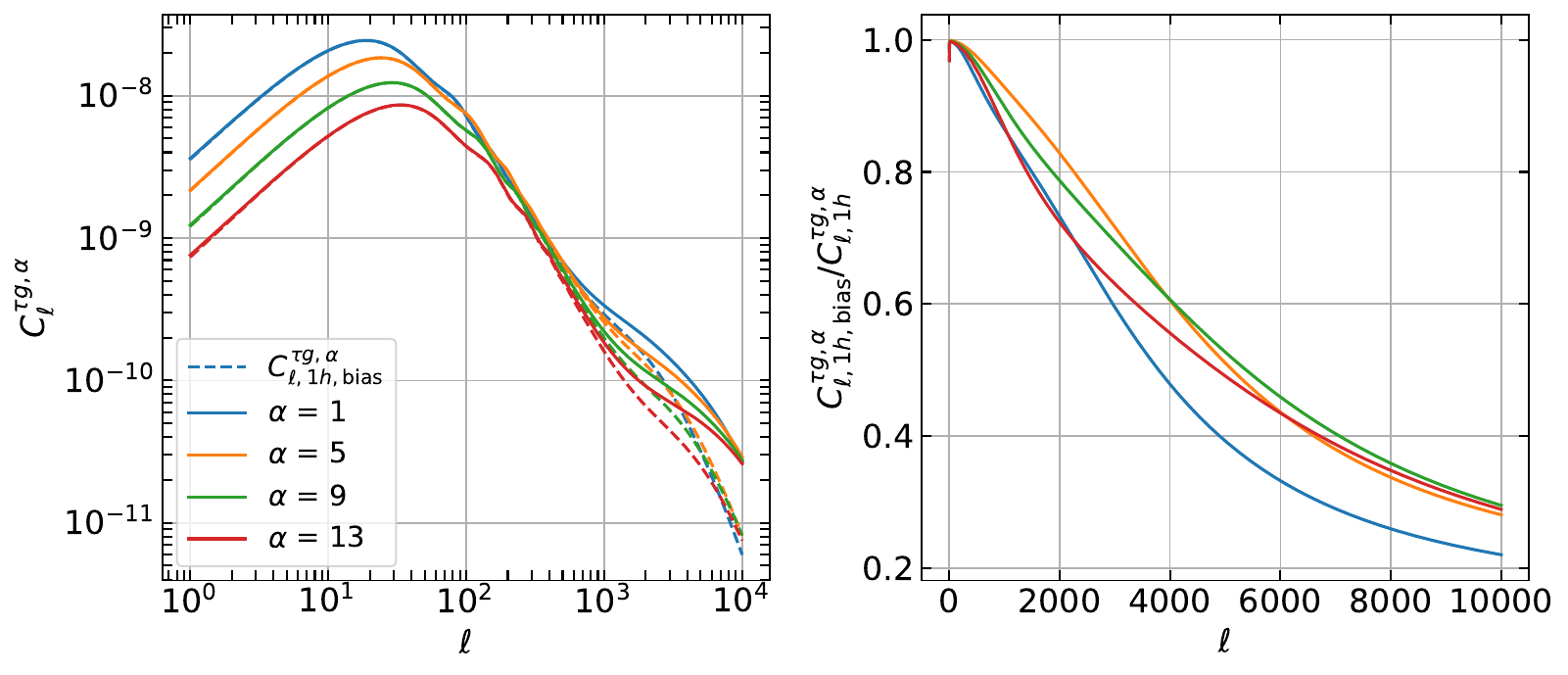}     
    \caption{The theoretical electron-galaxy cross-power spectrum $C^{\tau g, \alpha}_{\ell}$ for four redshift bins $\alpha = 1,5,9,13$. The power spectra involved in the calculation are obtained from \texttt{hmvec} using the DESI one-percent survey LRG parameters~\cite{Yuan:2023_desi_one_percent} for galaxy HOD and Battaglia 'AGN' model~\cite{Battaglia:2016_battaglia_profile} for the electron profile. We refer readers Sec.~\ref{subsec: fiducial powers} for the detail. Left: The total electron-galaxy cross-powers $C^{\tau g, \alpha}_{\ell}$ from summing the one-halo and the two-halo term, dashed line shows the same curve with the one-halo term replaced by the galaxy bias approximation $C^{\tau g, \alpha}_{\ell, \, 1h, \text{bias}}$. Right: the ratio of the one-halo term with galaxy bias approximation (Eq.~\ref{eq:Cgtau 1 halo, bias approximation}) to the one derived from the galaxy HOD (Eq.~\ref{eq:Cgtau 1 halo}).}
    \label{fig:Cgtau_my_vs_Cgtau_bias}
\end{figure}

We compare the overall $C^{\tau g, \alpha}_{\ell}$ in the left panel of Fig.~\ref{fig:Cgtau_my_vs_Cgtau_bias} and the ratio $C^{\tau g, \alpha}_{\ell, \, 1h, \text{bias}}/C^{\tau g, \alpha}_{\ell, \, 1h}$ in the right panel. In our main analysis, we use the HOD version of Eq.~\eqref{eq:Cgtau 1 halo} which is more physical, not the bias model.

\subsection{Cross-Correlation Signal-to-Noise}
\label{subsec:SNR calculation}

We will frame the discussion of the cross-correlation SNR in terms of the (redshift dependent) velocity bias. The velocity bias $b_v$ of the kSZ velocity reconstruction is the proportionality factor $v_{\mathrm{kSZ}}(\mathbf{k}) = b_v v({\mathbf{k}})$ between large-scale modes of the matter velocity field and large-scale modes of the kSZ reconstructed velocity field. This factor would be unity if the electron distribution in halos were precisely known, i.e. for a correct $P_{ge}$ the estimator would be unbiased by construction since we impose $\left<\hat{v}\right> = v$. The uncertainty on the electron distribution leads on large scales to the so-called kSZ optical depth degeneracy and can be parametrized by a linear bias factor. We refer to \cite{Smith:2018bpn} for a discussion of the optical depth degeneracy and a possible redshift dependence of $b_v$. Since $b_v$ is also the amplitude of the kSZ signal ($b_v=0$ would mean no kSZ), in each redshift bin $\alpha$ we will detect the kSZ individually with $\mathrm{SNR}_\alpha$ $=\frac{\hat{b}_{v}^\alpha}{\sigma_\alpha}$. Alternatively, we can fit an overall redshift independent amplitude so that $b^{\alpha}_{v} = A$ for all redshift bins. In this case, our overall significance is $\mathrm{SNR} = \frac{\hat{A}}{\sigma_A}$. We now discuss these definitions in more detail.

Our goal is to establish the significance of the detection of the velocity from kSZ reconstruction. Hence, we focus on the velocity-galaxy cross-powers. We define our signal as QML estimates of $vg$ cross spectra and define the total SNR in terms of the velocity biases $b^\alpha_v$ in each redshift bin. We consider the following likelihood function:
\begin{align}
    -2\text{ln}\,\mathcal{L}\left(\hat{\mathbf{C}}^{vg}, \mathbf{C}^{vg}, b^\alpha_v\right) &= \left(\hat{\mathbf{C}}^{vg} - \mathbf{C}^{vg}\left(b^{\alpha}_v\right)\right)^T\text{Cov}^{-1}\left(\mathbf{C}^{vg}\right)\left(\hat{\mathbf{C}}^{vg} -\mathbf{C}^{vg}\left(b^{\alpha}_v\right)\right) \, .
\end{align}
Here $\hat{\mathbf{C}}^{vg}$ is understood as a vectorized output of the QML estimator that contains all independent velocity-galaxy cross powers. It takes the following explicit form:
\begin{align}
    \hat{\mathbf{C}}_{vg} &= \begin{pmatrix}
    \hat{C}^{v^{1}g^{1}}_{\ell_{min}},&\hat{C}^{v^{1}g^{1}}_{\ell_{min}+1} &,...&, \hat{C}^{v^{20}g^{20}}_{\ell_{\text{max}}-1}&, \hat{C}^{v^{20}g^{20}}_{\ell_{\text{max}}}
    \end{pmatrix}^{T} \, .
\end{align}
The vectorized power spectrum contains the velocity biases as the modeling parameters, and is expressed as:
\begin{align}
    \mathbf{C}_{vg} = \begin{pmatrix}
    b^{1}_{v}C^{v^{1}g^{1}}_{\ell_{min}},&b^{1}_{v}C^{v^{1}g^{1}}_{\ell_{min}+1} &,...&, b^{20}_{v}C^{v^{20}g^{20}}_{\ell_{\text{max}}-1}&, b^{20}_{v}C^{v^{20}g^{20}}_{\ell_{\text{max}}}
    \end{pmatrix}^{T} \, .
\end{align}

We can introduce the matrix that assigns biases, so that
\begin{align}\label{eq:linear bias model}
    \mathbf{C}_{vg}(b^{\alpha}_{v}) &= \mathbf{L}\mathbf{b}_v \, .
\end{align}
Where $\mathbf{b}_v$:
\begin{align}
    \mathbf{b}_v &= \begin{pmatrix}
        b^1_{v}, & b^2_{v} ,& ... ,& b^{20}_{v}
    \end{pmatrix}^{T}\, ,
\end{align}is a vector of length $20$, and $\mathbf{L}$ is as follows:
\begin{align}
    \mathbf{L} &= \begin{pmatrix}
        C^{v^{1}g^{1}}_{\ell_{min}}, &...,&C^{v^{1}g^{20}}_{\ell_{\text{max}}},&0,&...,&0,&...,&0 \\
        0, &...,&0,&C^{v^{2}g^{1}}_{\ell_{min}},&...,&C^{v^{2}g^{20}}_{\ell_{\text{max}}},&...,&0 \\
        \vdots & \vdots&\vdots&\vdots&\vdots&\vdots&\vdots&\vdots \\
        0, & ...,&0,&0, & ...,&0,&...,&C^{v^{20}g^{20}}_{\ell_{\text{max}}} \\
    \end{pmatrix}^{T}\, .
\end{align}
The optimal bias vector $\mathbf{b}$ can then be solved analytically. It admits the following solution:
\begin{align}\label{eq: bestfit bv analytic}
    \hat{\mathbf{b}}_v &= \left(\mathbf{L}^T\text{Cov}^{-1}\left(\mathbf{C}^{vg}\right)\mathbf{L}\right)^{-1}\left(\mathbf{L}^T\text{Cov}^{-1}\left(\mathbf{C}^{vg}\right)\hat{\mathbf{C}^{vg}}\right)\, ,
\end{align}
and 
\begin{equation}\label{eq: bestfit bv error analytic}
    \text{Cov}^{-1}(\hat{\mathbf{b}}_v) = \mathbf{L}^T\text{Cov}^{-1}\left(\mathbf{C}^{vg}\right)\mathbf{L} \, .
\end{equation}
The inverse covariance matrix is just the QML Fisher matrix defined in Eq.~\ref{eq:QML Fisher matrix}. For the cross-power-purified case it only depends on the auto-powers and the noise models:
\begin{align}
    \text{Cov}^{-1}\left(\mathbf{C}\right) &= \mathbf{F}\left(\mathbf{C}^{gg}, \mathbf{C}^{vv}, \mathbf{N}^{gg}, \mathbf{N}^{vv}\right) \, .
\end{align}

In the case of a one-parameter amplitude fit, the estimator for the overall amplitude is
\begin{align}
\label{eq:amplitude}
    \hat{A} = \frac{\mathbf{b}^{\text{T}}_{v}\text{Cov}^{-1}(\hat{\mathbf{b}}_v)\hat{\mathbf{b}}_v}{\mathbf{b}^{\text{T}}_{v}\text{Cov}^{-1}(\hat{\mathbf{b}}_v)\mathbf{b}_v} \, ,
\end{align}
and the SNR is given by
\begin{align}\label{eq:SNR definition A}
    \text{SNR} &= \frac{\hat{A}}{\sigma_A} = \frac{\mathbf{b}^{\text{T}}_{v}\text{Cov}^{-1}(\hat{\mathbf{b}}_v)\hat{\mathbf{b}}_v}{\sqrt{\mathbf{b}^{\text{T}}_{v}\text{Cov}^{-1}(\hat{\mathbf{b}}_v)\mathbf{b}_v}}  \, ,
\end{align}
where here $\mathbf{b}=\mathbf{1}$ is the fiducial value. This is the definition of SNR we use as our main result. Alternatively, one could define SNR assuming no knowledge of the redshift dependence of $b_v$, i.e. all $b_{v,i}$ are independent variables and we exclude the null hypothesis of all $b_{v,i}$ being zero. In that case the SNR would be defined as
\begin{equation}\label{eq:SNR definition bv}
    SNR = \sqrt{\hat{\mathbf{b}}^{\rm T}_v\text{Cov}^{-1}(\mathbf{\hat{b}})\hat{\mathbf{b}}_v} \, .
\end{equation} 
We refer to App.~\ref{app:snr} for an in-depth discussion of these definitions.

The SNR in Eq. \eqref{eq:SNR definition bv} is an upper limit on the SNR in Eq. \eqref{eq:SNR definition A}. We note that to convert the SNR in Eq. \ref{eq:SNR definition bv} to a p-value for the detection, one would use a $\chi^2$ distribution with the appropriate degrees of freedom. If the SNR is interpreted as sigmas of a Gaussian distribution, the appropriate definition is Eq. \ref{eq:SNR definition A}. We will use Eq. \ref{eq:SNR definition A} as our default definition. 

\section{Results: kSZ Velocity Reconstruction and Cross-Correlation}
\label{sec:results}
We present the main results of this paper: QML estimates for the velocity-galaxy cross-powers, the estimated kSZ amplitude, and the SNR for the detection of the cross-power in this section.

\subsection{kSZ velocity field reconstruction with ACT and DESI-LS}\label{subsec:kSZ reconstruction results}
\begin{figure}[h!]
    \centering
    \includegraphics[width=0.4\columnwidth]{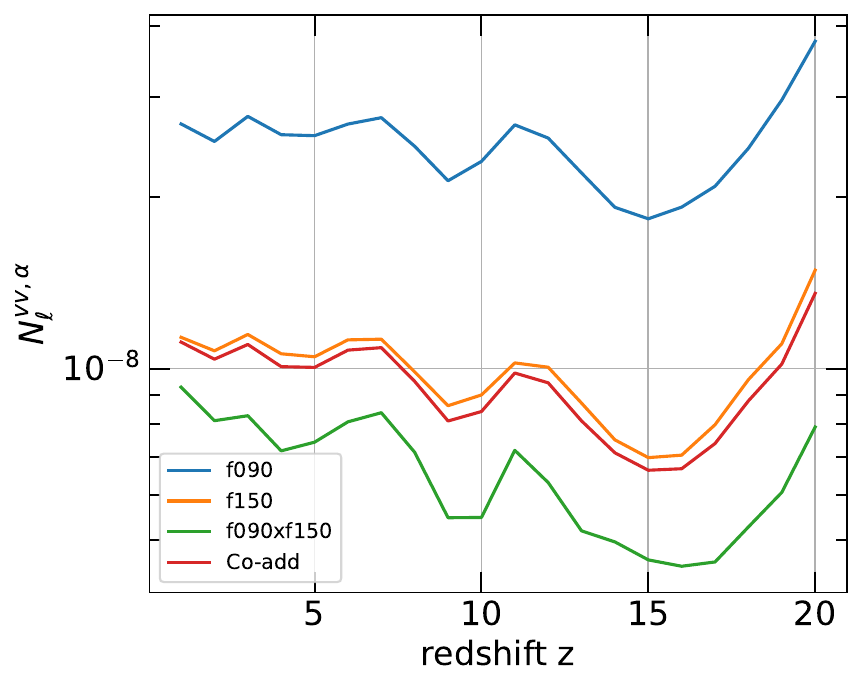}
    \caption{A summary of the best-fit kSZ reconstruction noises $N^{vv, \alpha}_{\ell}$ used in this analysis. As discussed in Sec.~\ref{subsec:velocity field QE}, the reconstruction noises are assumed to be uncorrelated between different redshift bins and constant across the multipoles $\ell$. The blue and orange curves give the noise for the velocity maps individually reconstructed from the 90 GHz and 150 GHz CMB maps. The green curve shows best-fit estimates of the cross-channel kSZ noise. The co-added noise in the red curve is first computed in pixel space and converted back to harmonic space.}
    \label{fig:Nvv_alpha}
\end{figure}
We apply the quadratic kSZ velocity estimator from Sec. \ref{subsec:velocity field QE} to the ACT and DESI-LS data described in Sec. \ref{subsubsec: DESI LS data}. Some example sky maps of the reconstructed radial velocity field are shown in Fig.~\ref{fig:galaxy_recon_compare} and further discussed below.

\paragraph{Reconstruction noise.} Fig.~\ref{fig:Nvv_alpha} shows the redshift dependence of various kSZ reconstruction noises $N^{vv, \alpha}_{\ell}$ involved in the analysis. As discussed in Sec. \ref{subsec:velocity field QE}, we assume that the noises are uncorrelated in the pixel space and are close to a constant value on large scales. To avoid biased estimation on the kSZ reconstruction noise from the theory as discussed in Sec.~\ref{subsec:velocity field QE}, we use the best-fit values for the 90 GHz, 150 GHz, and 90 GHz $\times$150 GHz frequency channels obtained from the data. The single frequency channel noises and the cross channel noise are used as input to compute the co-added velocity maps and the associated co-added reconstruction noises as discussed in Sec.~\ref{subsec:vmap coadd}.

\begin{figure}[h!]
    \centering
    \includegraphics[width=1.0\columnwidth]{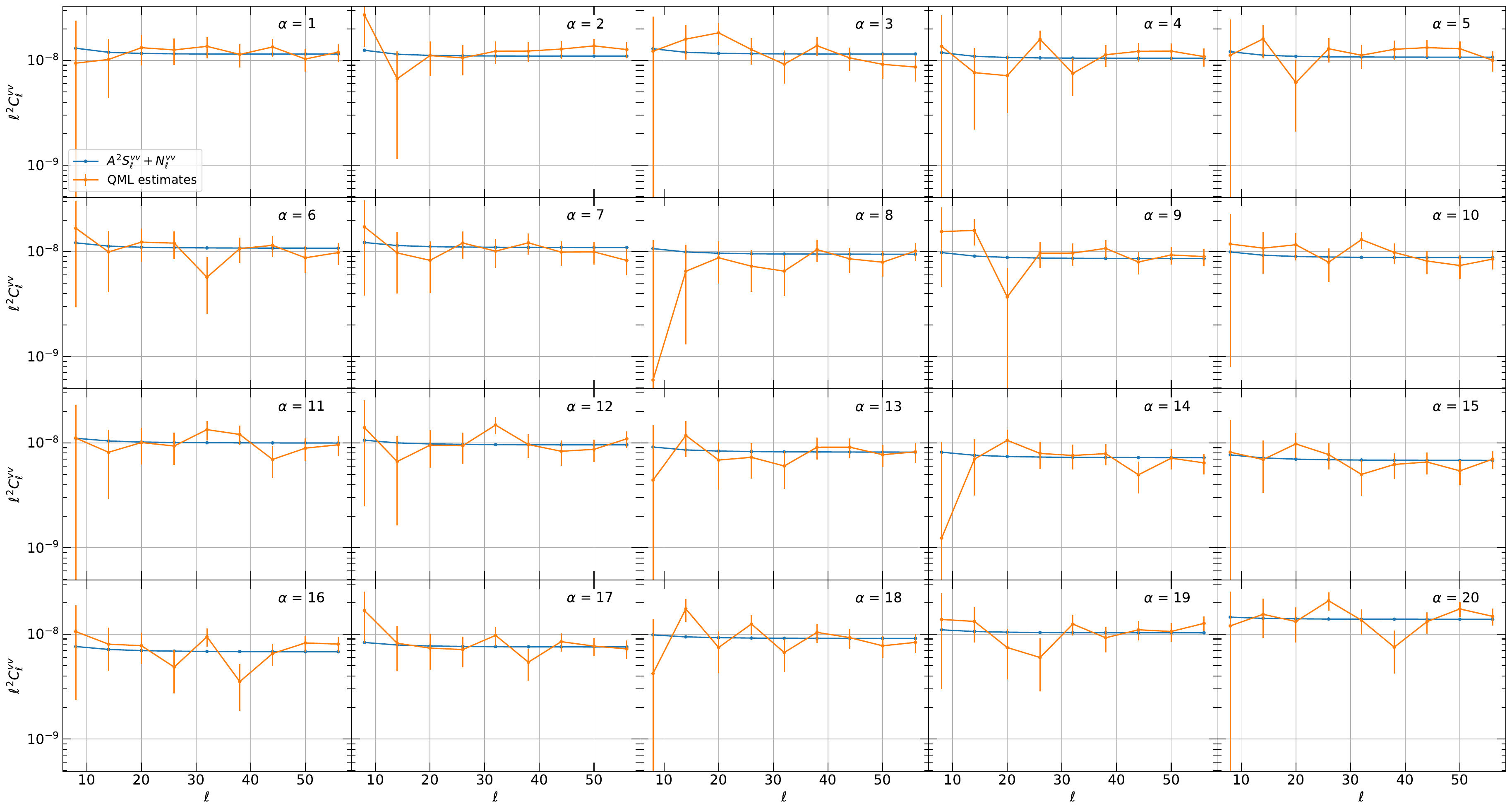}
    \caption{QML estimates of the velocity auto-powers $C^{vv, \alpha}_{\ell}$ computed from the kSZ reconstructed velocity maps. The data points and errorbars cover modes between $\ell = 5$ to $\ell = 60$ in bandpowers of width 6. Data points mark the center of each bandpower. The fiducial signal + best-fit reconstruction noise kSZ power $A^2 S^{vv}_{\ell}+N^{vv}_{\ell}$ with amplitude $A = 0.421$ (tSZ-masked, second iteration) are plotted as the blue curves. We note that the velocity auto-powers are noise-dominated in individual redshift bins and quickly flatten out at the level of the reconstruction noise.}
    \label{fig:Cvv_QML_estimation_binned}
\end{figure}

\paragraph{KSZ Velocity Estimator Auto-Power Spectrum.} We present the full-sky velocity auto-powers $C^{vv, \alpha}_{\ell}$ for the 20 redshift bins recovered from the QML estimator in Fig.~\ref{fig:Cvv_QML_estimation_binned}. Although the QML method provides a reconstruction for each individual mode, we bin the velocity auto-powers in each redshift bin in bandpowers of 6 modes using a uniform, normalized window function for the purpose of illustration. We compare them to the modelled total velocity auto-powers $A^2S^{vv,\alpha}_{\ell}+N^{vv,\alpha}_{\ell}$, which are obtained by adding the best-fit kSZ reconstruction noises to the fiducial signals rescaled by the squared best-fit kSZ amplitude $A^2$. The fiducial powers are convolved with the same bandpowering window function.
The data points are consistent with the fiducial powers within the 1-sigma interval and above zero. The reduced chi-square value is 1.01 for $N_{\rm dof} = 11969$ before bandpowering, suggesting a good fit to the data by the fiducial theory. We note that the auto-powers of the kSZ reconstruction are noise dominated, except for the lowest $\ell$ bins, which are most contaminated by systematics. For this reason we require a cross-correlation measurement to establish a detection.

\subsection{QML Velocity-Galaxy Cross-Power Spectrum Measurement}\label{subsec:Cvg result}

We now make the main measurement of this work, the cross-power SNR between the kSZ velocity reconstruction and the galaxy field on large scales, using our QML power spectrum estimator~\cite{Kvasiuk:qmlpaper}. A different option would be to first estimate a velocity map from the galaxy map and then cross-correlate this velocity-from-galaxy map with the velocity-from-kSZ map. However, for a linear velocity reconstruction, the resulting cross-correlation SNR would be identical to the one from a direct cross-correlation with the galaxy field in our dense matrix approach. For visualization purposes we also perform a velocity-from-galaxy reconstruction in Sec. \ref{sec:linearrec}, but we do not use it to establish the cross-correlation SNR. We now explain our kSZ velocity to galaxy density cross-correlation measurement.

\begin{figure}[h!]
    \centering
    \includegraphics[width=0.8\columnwidth]{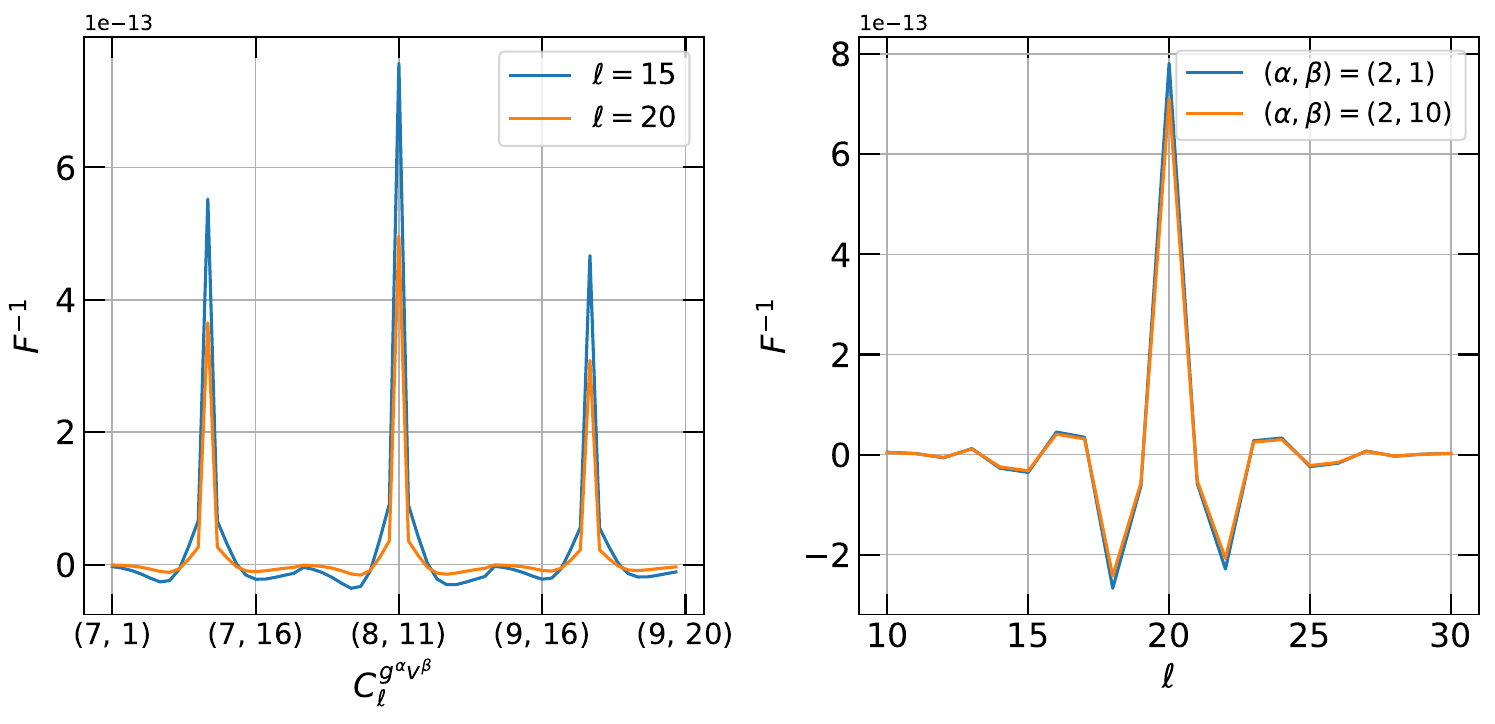}
    \caption{Left: Covariance of the cross-powers between the redshift bin combination $(\alpha, \beta) = (8, 11)$ and other redshift combinations centered around it for two different values of multipoles $\ell=15$ and $\ell = 20$. %The self-covariance is the largest, and the covariance becomes less correlated as the redshift bins gradually move away from $\alpha, \beta = (8, 11)$. 
    Right: Similar to the left plot but for two fixed redshift bin configurations $(\alpha, \beta) = (2,1), (2,10)$ and varying $\ell$'s around the mode $\ell = 20$.%. Again the covariance is the largest when coupling to itself and damps as the modes go further away from the center.
    }
    \label{fig:QML coupling matrix example}
\end{figure}

\paragraph{Evaluating the covariance matrix.} We first evaluate the covariance matrix from the inverse of Eq. \eqref{eq:QML Fisher matrix}. The left panel in Fig.~\ref{fig:QML coupling matrix example} shows the cross-power covariance between $C^{g^{8}, v^{11}}_{\ell}$ and other redshift bin configurations $(\alpha, \beta)$ at two fixed modes $\ell = 15$ and $\ell = 20$. The x-axis follows an ordering of $(\alpha, \beta)\,,\,(\alpha, \beta+1)...\,,\,(\alpha, \beta+20)\,,\,(\alpha+1, \beta)$. The covariance reaches maximum for autocorrelation and decays as it goes off-diagonal (0.73 at $(7, 11)$, 0.67 at $(9, 11)$). Another feature about the covariance matrix is that the correlation structure evolves slowly along the galaxy index $\alpha$ and rapidly along the velocity index $\beta$. This can be understood in the full-sky limit with Gaussian random fields, in which the covariance of the nearby ($\alpha$, $\beta$) contains either a term $C^{g^\alpha, g^{\alpha+1}}_\ell$ or $C^{v^{\beta+1}, v^{\beta}}_\ell$. As $C^{g^\alpha, g^{\alpha+1}}_\ell$ decays much slower than $C^{v^{\beta+1}, v^{\beta}}_\ell$, it will also be reflected in the cross-power covariance matrix.

\begin{figure}[h!]
    \centering
    \includegraphics[width=0.9\columnwidth]{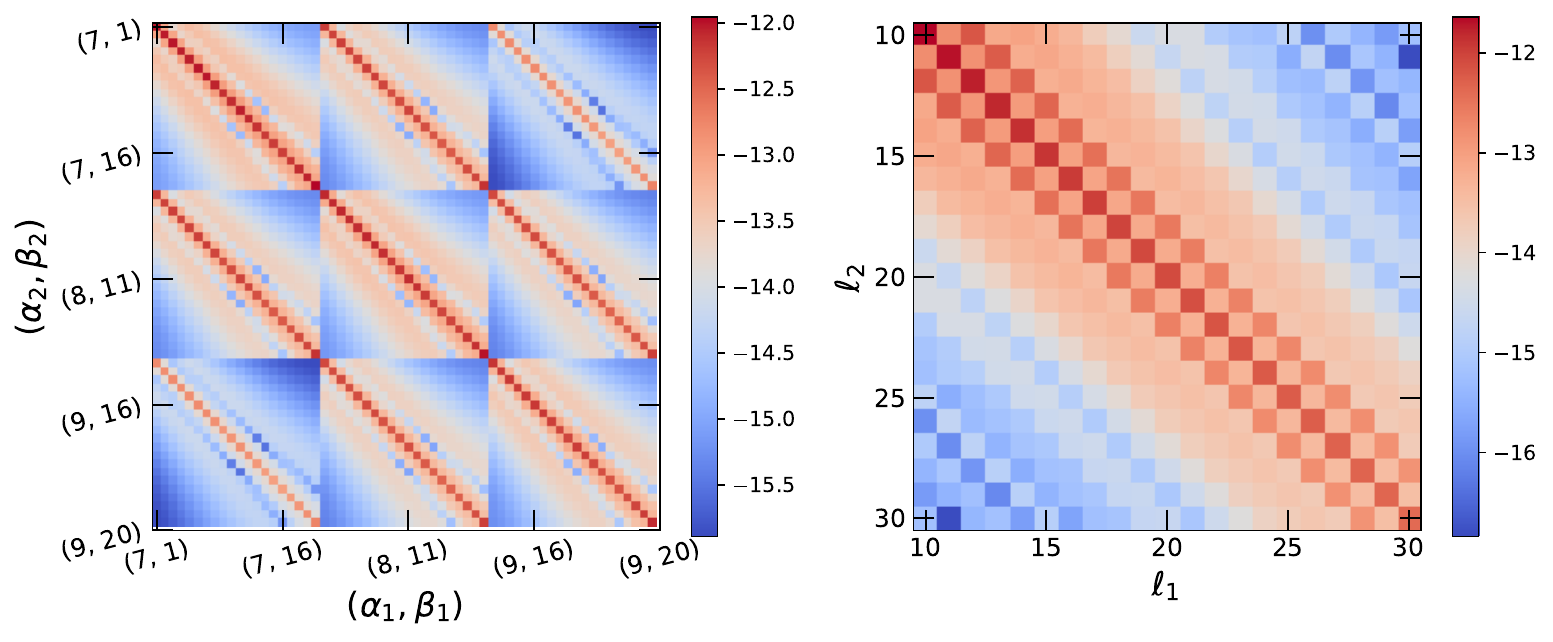}
    \caption{Left: Sub-blocks of the analytic covariance matrix between the first cross power in redshift bin $(\alpha_1, \beta_1)$ and the second cross power in redshift bin $(\alpha_2, \beta_2)$ for one mode $\ell = 15$. Right: Similar to left but for $(\alpha_1, \beta_1) = (\alpha_2, \beta_2) = (2,10)$ and varies across the modes $\ell_1, \ell_2$. The color shows the log-absolute magnitude of the covariance in base 10.}
    \label{fig:QML covariance matrix example}
\end{figure}

We also present in the right panel of Fig.~\ref{fig:QML coupling matrix example} the cross-power covariance matrix at two fixed redshift configurations $(\alpha, \beta) = (2,1), (2,10)$. We focus on the covariances between $\ell = 20$ and modes around it. They highlight a universal behavior where the two modes become less correlated if they are further separated. We provide an example sub-block of the full covariance matrix at a fixed multipole $\ell$ in the left panel of Fig.~\ref{fig:QML covariance matrix example} and at a fixed redshift-bin pairs on the right of Fig.~\ref{fig:QML covariance matrix example}.

\paragraph{Galaxy-Velocity Cross-Power Spectrum Measurement.} We then measure the cross-powers using the QML estimator in Eq. \eqref{eq:qmlmaster}. The full set of galaxy-velocity cross-powers is of size 20 galaxy bins $\times$ 20 velocity bins $= 400$ cross-powers and is plotted in App. \ref{sec:full Cgv plot}. We present a co-added galaxy-velocity cross-power in Fig.~\ref{fig:Cgv_QML_estimation_binned} which is reduced from all the cross-powers included in the SNR analysis. These cross-powers have most of the correlations residing in nearby redshift bins and contain both positive and negative fiducial values, so we project all measured cross-power spectra to a reference power spectrum in Fig.~\ref{fig:Cgv_QML_estimation_binned}. We describe this co-adding scheme in App.~\ref{app:coadding scheme}. We also bandpower the co-added cross-power from $\ell = 5$ to $\ell = 60$ in intervals of width 6. The bandpowering scheme is detailed in App.~\ref{app:bandpowering scheme}. While the co-adding plot is for visualization and does not represent the full SNR recovered from the effective cross-powers, one can still see that most of the SNR comes from the large scales. The data points align well with the best-fit theory within the 1-$\sigma$ error bars, illustrating a high SNR detection that rejects the null hypothesis ($C^{gv}_{\ell} = 0$). We also evaluate the goodness-of-fit by computing the reduced chi-square value with respect to the fiducial theory rescaled by the best-fit amplitude ($A_{\rm init}$ for the first iteration, $A_{\rm fit}$ for the second) after selecting the limited redshift bin range described below in Sec.~\ref{subsec: SNR result}. The reduced chi-square values are evaluated with a mode-by-mode covariance matrix. We find a value of 1.02 for our main result, which corresponds to the $A_{\rm fit} = 0.421$, tSZ masked configuration (Table~\ref{Tb:QML summary}, fourth row). We also present the reduced chi-square and the p-value for other configurations in Table~\ref{Tb:QML summary}.

\subsection{Total Signal-To-Noise of the QML Cross-Power Spectrum Measurement}\label{subsec: SNR result}

To establish the total SNR of our cross-power spectrum measurement, we select a limited range of cross-redshift bins to avoid being biased by systematics, as we now explain.\footnote{The choice of redshift range for the cross-powers and the investigation of the goodness-of-fit (also see App.~\ref{app:outlier_removal}) is newly introduced in version 2 of this paper.}

\paragraph{Removing equal redshift bins.} We notice from the full QML estimates that there are consistent outliers when the galaxy redshift bin and the velocity redshift bin are equal ($\alpha = \beta$). This is also observed in the diagonal plots provided in App.~\ref{sec:full Cgv plot}. We note that this is also found in~\cite{McCarthy:2025brx} in their cross-power analysis using a similar dataset. Following the discussion in~\cite{McCarthy:2025brx, EmbilVillagra:2026fnh}, the emergence of the equal-redshift outliers can be associated with the fact that the cross-power estimator $\hat{C}^{g^{\alpha}v^{\alpha}}_{\ell}$ is inherently sensitive to any tSZ and CIB foregrounds through the three-point correlation $\left<g^{\alpha}_{L}g^{\alpha}_{S}T_{S}\right>$. The subscripts `L' and `S' refer to the large and the small scale. This type of foreground can also contaminate the first off-diagonal cross-power estimates through $\left<g^{\alpha}_{L}g^{\alpha\pm 1}_{S}T_{S}\right>$ in our pipeline, mainly due to an even narrower redshift bin definition (20 bins against 16 bins in~\cite{McCarthy:2025brx}). We thus remove the diagonal and first off-diagonal redshift bins from our set of cross-power spectra when calculating the SNR.

\paragraph{Removing far-separated redshift bins.}
There are also outliers from the far-separated redshift bins, i.e., $|\alpha - \beta| > 10$. These cross-powers usually contain either a kSZ reconstruction velocity field or a galaxy overdensity field from a high-redshift bin. Therefore, the outliers can be a result of mismodelling the redshift window function $W^{\alpha}(z)$ at large redshifts, where the number density of galaxies is low and the actual photo-z distribution departs from a Gaussian. We thus decide to also drop these cross-power spectra when calculating the SNR, since they contain virtually no expected signal and are contaminated by galaxies with catastrophic photo-z errors.

\paragraph{Effect on SNR and goodness of fit.} We explore the effect of excluding the outlying cross-powers on the QML pipeline in App.~\ref{app:outlier_removal}, where we compute the inferred kSZ amplitude $A$, the goodness-of-fit and the SNR for different setups. We find that removing the cross-powers from far-separated and equal/first off-diagonal redshift bins greatly improves the goodness-of-fit. The kSZ amplitude $A$ and the SNR are more sensitive to the removal of cross-powers from nearby redshift bins, where most of the signal is concentrated. To achieve a good fit, we exclude any cross-power that satisfies $|\alpha - \beta| < 2$ or $|\alpha - \beta| > 10$ in the final SNR analysis presented below (grey-scale cross-powers in App.~\ref{sec:full Cgv plot}). This reduces the number of effective cross-powers from $400$ to $252$. The selected cross-powers are highlighted as colored plots in App.~\ref{sec:full Cgv plot}.

\begin{figure}[h!]
    \centering
    \includegraphics[width=0.65\columnwidth]{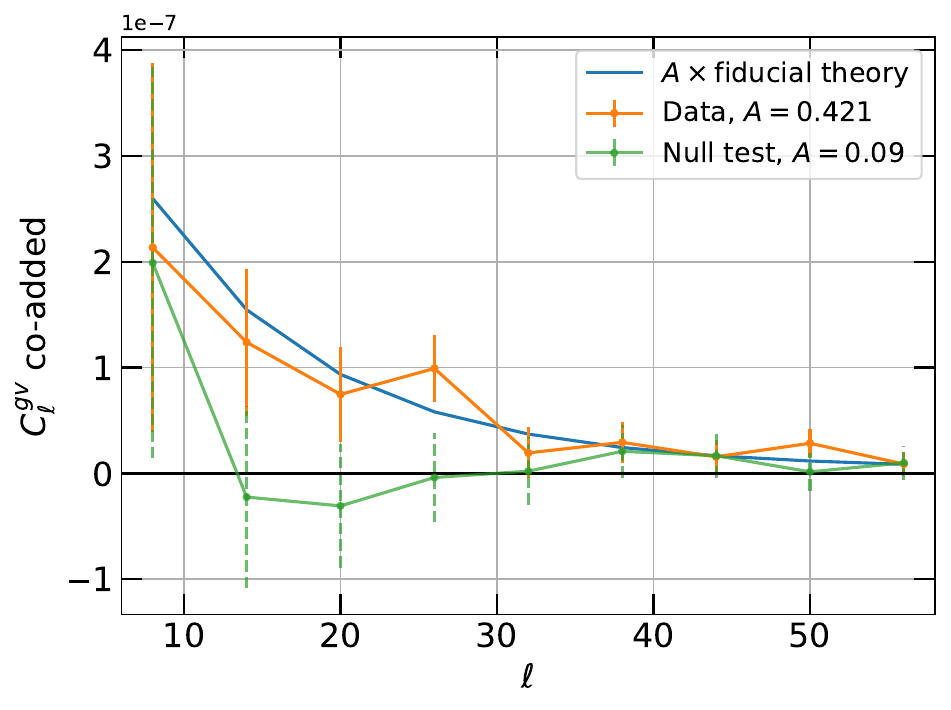}
    \caption{QML estimates of the galaxy-velocity cross-powers co-added from the SZ-cluster masked, $252$ redshift bin configurations following the co-adding scheme in App.~\ref{app:coadding scheme} and using $C^{g^1 v^2}_\ell$ as the pivot. The best-fit fiducial power spectrum, denoted by $A\times$fiducial theory, with $A=0.421$ obtained from the tSZ-masked, iterated QML analysis, is shown by the blue curve. The co-added data and the associated error bars, obtained from analytic QML covariance matrix, are shown by the orange curve. The null test cross-power, computed from the f090$-$f150 velocity maps, is shown by the green curve. We quantify the goodness-of-fit for the full spectra as the reduced chi-square value and the p-value in the last row of Table~\ref{Tb:QML summary}. Details of the null test are provided in App.~\ref{app:null test}. The co-added data shown in this figure are for illustration only and are not used in the SNR determination.}
    \label{fig:Cgv_QML_estimation_binned}
\end{figure}

\begin{figure}[h!]
    \centering
    \includegraphics[width=0.6\columnwidth]{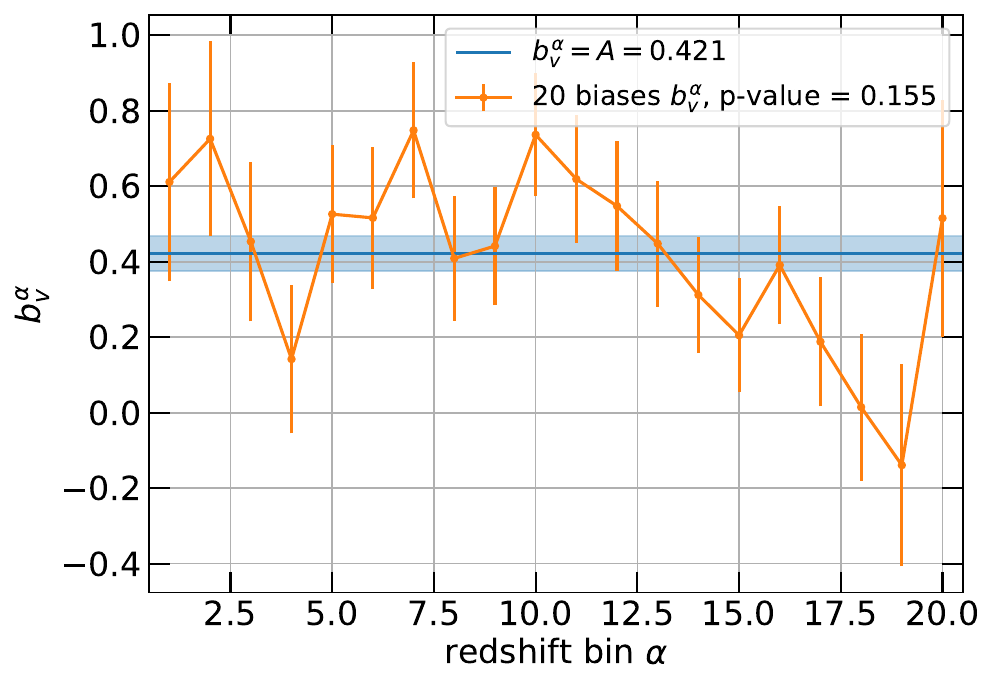}
    \caption{The best-fit velocity biases $\mathbf{b}^\alpha_v$ obtained from analytic solution Eq.~\ref{eq: bestfit bv analytic} over the 20 redshift bins. The errorbars are retrieved from the diagonal terms of the covariance matrix for the velocity biases. The biases are defined with respect to a halo model calculation of $C^{\tau g}$ using the Battaglia profile, as explained in Sec. \ref{subsec: fiducial powers}. The best-fit velocity bias amplitude $A=0.421$ (tSZ-masked configuration, second iteration) and the associated $1$-$\sigma$ uncertainty are also highlighted as blue for reference. The p-value 0.155 comes from the compressed 20-parameter $\mathbf{b}^\alpha_v$ covariance matrix and is different from Table.~\ref{Tb:QML summary}.
    }
    \label{fig:velocity biases}
\end{figure}

\paragraph{Best-fit velocity biases and overall amplitude.} We now examine the model fit using the selected cross-power spectra. Fig.~\ref{fig:velocity biases} shows the 20 velocity biases obtained from the QML velocity-galaxy cross-powers, using the linear model from Eq.~\ref{eq:linear bias model}. The velocity biases and the errorbars are solved analytically through Eq.~\ref{eq: bestfit bv analytic} and Eq.~\ref{eq: bestfit bv error analytic}. We also estimate the overall amplitude $A$ using Eq. \ref{eq:SNR definition A}. Most of the values are consistent with the estimated overall amplitude $A = 0.421$ relative to the Battaglia model. Our measurement of a relatively small A is consistent with a high feedback in massive halos (see \cite{Hadzhiyska:2024qsl}) and also consistent with the independent analysis which appeared in parallel to this paper in \cite{SelimKendrickPaper}.

\paragraph{Total Signal-To-Noise.} A summary on the SNRs with different setups is shown in Table~\ref{Tb:QML summary}, where we use the range $\ell=5$ to $\ell=60$ to avoid the largest modes which are most strongly affected by systematics. To increase the total SNR of the cross-correlation, we run two iterations of the QML, as described at the bottom of Sec. \ref{subsec:QML theory}. In the first iteration without the SZ mask applied, we measure $A=0.476 \pm 0.0479$ for the kSZ amplitude. Then we adjust $C^{\tau g}$ by this value in the second iteration of the QML (i.e. we learn in the first iteration that the halo model over-estimates the expected kSZ signal). The QML, unlike the pseudo-C$_\ell$, is not invariant under this rescaling, as we explain in more detail in App.~\ref{sec:ksz estimator rescaling}. This procedure gains an extra 9.7\% increase in the cross-correlation significance (tSZ-masked configuration). We define the final kSZ amplitude $A_{\rm fin}$ as $A_{\rm init} \times A_{\rm fit}$, where $A_{\rm init}$ and $A_{\rm fit}$ are the amplitudes inferred from the first and second iteration.

We also provide two definitions of the SNR: a 1-parameter model that represents the velocity-galaxy cross power $A$, which is the same across all modes and redshift bins (Eq.~\ref{eq:SNR definition A}) and a 20-parameter model where each parameter represents the velocity bias of a redshift bin $\alpha$ (Eq.~\ref{eq:SNR definition bv}). We provide the SNRs for both definitions in Table~\ref{Tb:QML summary}. As discussed above, if the SNR is interpreted as the significance of the rejection of the null hypothesis, the first definition as a 1-parameter model is more appropriate. In any case, the two definitions only differ by about one sigma. We thus choose the significance of the one-parameter fit in the second QML iteration as our final result. 

\paragraph{TSZ masking.} We investigate a potential remaining tSZ contamination by applying the ACT DR5 tSZ cluster mask~\cite{ACT:2020tSZ_mask} and find a slight drop in SNR, about $0.8$, when applying the mask. This is in line with the findings in~\cite{Lague:2024czc_7.2sigma}. However, we notice that the central value of the kSZ amplitude $A_{\rm fin}$ shifts from 0.480 to 0.421. The shift is 4.5$\sigma$ by the nested-dataset criterion of~\cite{Gratton:2019fru}. Therefore, we conclude that the galaxy-velocity cross-power from the CMB map without SZ masking is still significantly affected by the foregrounds. We report the tSZ-masked SNR 9.96 from the second iteration as our main result. Note that the p-value that corresponds to the first QML iteration is very close to 1 (Table~\ref{Tb:QML summary}, third row). This suggests that the error bars in this setup are slightly overestimated, which is a known behavior of the QML estimator when fiducial power is offset from the underlying truth. The p-value from the second iteration (0.077) results in an acceptable fit. To further illustrate the robustness of the SNR reported against potential correlated foregrounds, we perform a null test by running the QML pipeline on a map created by differencing the single-frequency maps to remove all kSZ signal. The null test is also plotted in Fig.~\ref{fig:Cgv_QML_estimation_binned} and is consistent with zero. Readers are referred to App.~\ref{app:null test} for details of the setup.

\paragraph{Comparison with pseudo-$C_\ell$.}  Finally, for an analysis with identical choices using the pseudo-C$_\ell$ formalism we refer to App.~\ref{sec:app_pseudocl}, where we find $8.3 \sigma$ cross-correlation significance and an amplitude of $A_{\rm fin}=0.44$ using a covariance matrix estimated from $10^5$ simulations (we also show convergence of kSZ amplitude and 1-sigma error with this choice of number of simulations in App.~\ref{sec:app_pseudocl}). The central value from pseudo-C$_\ell$ is consistent with that from the QML pipeline. The improvement from the QML pipeline is about $20\%$ compared to the pseudo-C$_\ell$ pipeline and mainly comes from a reduction in the error bar ($\sigma_A = 0.0423$ compared to $0.053$ for pseudo-$C_{\ell}$, see App.~\ref{sec:app_pseudocl}), demonstrating the power of the approach.

\vspace{1.5cm}

\begin{table}[h!]
\centering
\begin{tabular}{|l|r|r|r|r|r|r|}
\hline
\textbf{Configuration} & \textbf{Amplitude $A$} & \textbf{Error $\sigma_A$} & \textbf{SNR ($b^\alpha$)} & \textbf{SNR ($A$/$\sigma_A$)} & \textbf{$\chi^2$/dof} & \textbf{p-value}\\
\hline
$A_{\mathrm{init}}$  & 0.476 & 0.0479  &  11.1 & 9.93 & 0.983 & 0.914\\
$A_{\rm fin}$ & 0.480 & 0.0443 & 12.3 & 10.8 & 1.04 & $5.19 \times 10^{-4}$\\
$A_{\mathrm{init}}$, tSZ masked & 0.436 & 0.0480& 10.4 & 9.08 & 0.961 & 0.9996\\
$A_{\rm fin}$, tSZ masked &  \textbf{0.421} & \textbf{0.0423} & \textbf{11.2}  &  \textbf{9.96}& \textbf{1.02} & \textbf{0.0769}\\

\hline
\end{tabular}

\caption{Summary of the estimated velocity-galaxy cross-power amplitude, 1$\sigma$ uncertainty and the SNR ratio from the first and second QML iteration. All quantities are computed from the QML estimates between $\ell = 5$ and $\ell = 60$ using a mode-by-mode Fisher matrix with the outlying cross-powers excluded. Our key results are the amplitude with respect to the halo model $A_{\rm fin}=0.421$ from a tSZ masked configuration, and the total SNR of the cross-correlation in the second QML iteration of $10 \sigma$. We also quantify the goodness-of-fit with the reduced chi-square and the p-value. They are computed from pre-bandpowering estimates and with respect to the best-fit fiducial cross-powers. The bolded result (second QML iteration, tSZ masked) is the final measurement quoted in abstract and conclusion.}
\label{Tb:QML summary}
\end{table}

\begin{figure}[h!]
    \centering
    \includegraphics[width=\columnwidth]{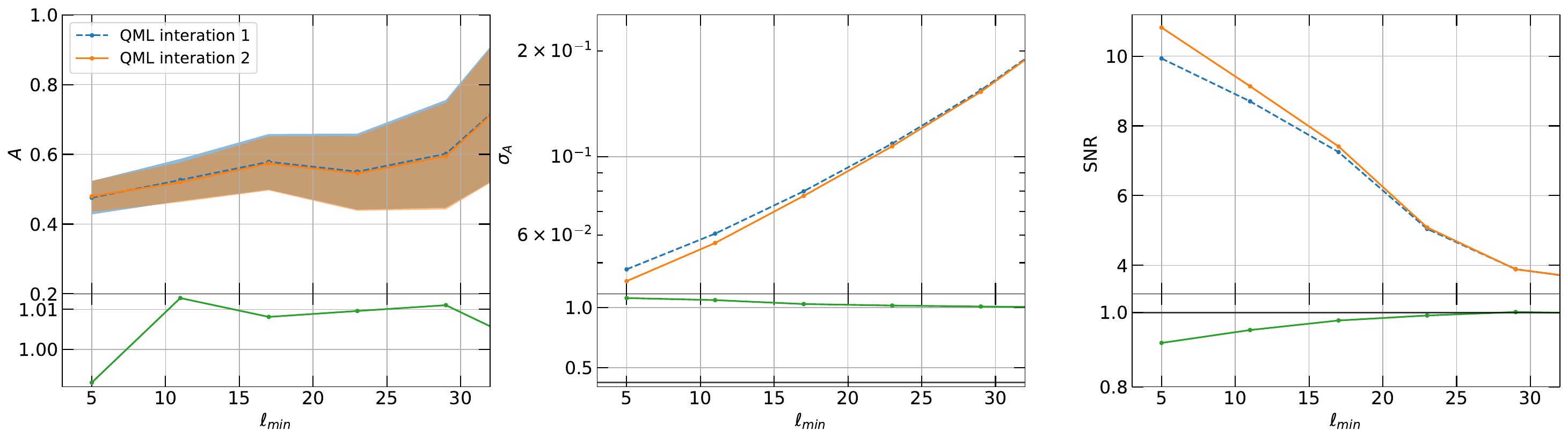}
    \caption{Dependence of the QML galaxy-velocity cross-power analysis on $\ell_{min}$. Blue and orange curves give the results in the first and second QML iteration. The quantities are plotted against different values of $\ell_{\text{min}}$, the lowest mode used for SNR calculation, up to 35. Left: the velocity-galaxy cross-power amplitude $A$, such that $A = A_{\rm init}$ for the first iteration and $A = A_{\rm fin}$ for the second iteration. The 1-$\sigma$ regions are also highlighted as the shaded region. Middle: The 1-$\sigma$ values of the kSZ amplitude $A$ computed from the fiducial power and covariance. The curves are monotonically increasing with fewer modes involved in the SNR calculation. Right: The total SNR $A/\sigma_A$. Bottom row: The ratio of the corresponding quantities with the first QML iteration as the numerator and the second QML iteration as the denominator. While this analysis was performed without SZ cluster mask, similar qualitative features are expected with SZ cluster mask.
    }
    \label{fig:QML_summary}
\end{figure}

\paragraph{Dependence on $\ell_{min}$.} We demonstrate the robustness of the QML estimator with respect to the choice of $\ell_{min}$ in Fig. \ref{fig:QML_summary}. This analysis is performed without the tSZ mask.
The left plot shows the cross power amplitude $A$, which falls around $0.48$ at the first QML iteration and is similar to the second QML iteration. The value of the cross-power amplitude does not depend strongly on $\ell_{min}$, and our conclusion of a low amplitude with respect to the fiducial theory is thus robust with respect to this cut. In the middle plot we show how the uncertainty on the amplitude scales with $\ell_{min}$, showing that most SNR comes from the lower $\ell$. The plot on the right gives the SNR as a function of $\ell_{min}$. An interesting aspect of this plot is that the second QML iteration only improves the SNR in the regime where the analysis is strongly signal dominated. This is in line with the expectation from our analysis in App.~\ref{sec:ksz estimator rescaling}, where we show that in the noise dominated regime the QML becomes invariant under a rescaling of the fiducial model, in the same way as the pseudo-C$_\ell$.

\paragraph{Optimality and future improvements.} Our analysis, in terms of data representation and power spectrum estimator, should be close to optimal. In App. \ref{app: fisher forecast qml} we explore redshift binning and $\ell$ range using a Fisher forecast. This Fisher forecast uses the full pixel-wise covariance matrix, so it takes into account the survey geometry. By reducing from 20 bins to 10 bins, we only find a small loss in SNR, and we again find convergence in $\ell_{max}$ at around 60, which is well covered with our $N_{\rm side}$ of 32. On the other hand, other aspects of our analysis are likely sub-optimal. For example, we do not consider anisotropic noise (even though this is technically easy to do in our dense matrix representation). We will improve these choices in future work.  

\section{Results: Linear Velocity Map Reconstruction from Galaxy Overdensity maps}
\label{sec:linearrec}

In Sec.~\ref{subsec:kSZ reconstruction results} we present the results of the velocity fields that are reconstructed from the kSZ QE. Alternatively, one can utilize the linear relation between the overdensity fields and the velocity fields to reconstruct the large-scale velocity field from the galaxy density field (rather than the kSZ). In addition, one can cross-correlate the velocity field from the galaxy reconstruction to the kSZ reconstruction, forming a cross-power $C^{v_{\text{gal}}v_{\text{ksz}}}$ between the two fields. This method allows the SNR to be evaluated by comparing $C^{v_{\text{gal}}v_{\text{ksz}}}$ to the fiducial velocity auto-power. Mathematically, the SNR obtained from this method is equivalent to that obtained from cross-correlating the galaxy overdensity field with the kSZ reconstruction (Sec.~\ref{subsec:Cvg result}), because a linear transformation does not change the SNR in the cross-correlation. However, the former provides a natural way to visualize the correlation at the map level, since both maps are velocity fields from different reconstruction methods. We therefore develop a linear velocity reconstruction method based on the redshift-binned 2-dimensional galaxy overdensity.

\subsection{Linear reconstruction}

We formulate a pixel-space, linear estimator for the velocity map from the galaxy overdensity map as follows:
\begin{align}\label{eq:v estimator from g}
    \hat{v}_{i,g} &= M_{ij}\delta^j_g\, ,
\end{align}
where the index $i,j$ runs over all the unmasked pixels and redshift bin index $\alpha$ to account for the potential coupling between pixels due to incomplete sky coverage and coupling between redshift bins. We use the subscripted variable $\hat{v}_{i,g}$ to differentiate it from the kSZ reconstruction. We look for an optimal estimator by minimizing the variance $\left<(\hat{v}_{i,g} - v_{i})(\hat{v}_{i,g} - v_{i})^T\right>$ with respect to $M_{ij}$. This gives:
\begin{align}\label{eq:Mij}
    \mathbf{M} &= \mathbf{C}^{vg}\left(\mathbf{C}^{gg}\right)^{-1} \, ,
\end{align}
where $\mathbf{C}^{vg} = \left<\mathbf{v}\mathbf{g}^T\right>$ and $\mathbf{C}^{gg} = \left<\mathbf{g}\mathbf{g}^T\right> = \mathbf{S}^{gg} + \mathbf{N}^{gg}$ are the fiducial pixel-space covariance matrices for the velocity-galaxy cross-power and galaxy auto-power. They are identical to the components defined in Eq.~\ref{eq: pixel covariance matrix explicit} for the signal covariance matrix. Note that the estimator Eq.~\ref{eq:Mij} can also be derived from maximizing the joint probability distribution between the observed galaxy overdensity maps and the underlying velocity maps that are assumed to be described by the fiducial velocity power spectrum. We refer interested readers to App.~\ref{sec:galaxy reconstruction details} for more details.

To compare the galaxy-reconstructed velocity fields with the kSZ reconstructed ones, we compute the following correlation coefficient $R_{\ell}$:
\begin{align}\label{eq: correlation coefficient}
    R_{\ell} &= \frac{C_{\ell}^{\hat{v}_g \hat{v}}}{\sqrt{C_{\ell}^{\hat{v}_g \hat{v}_g}C_{\ell}^{\hat{v} \hat{v}}}} \, ,
\end{align}
where $\hat{v}$ refers to the velocity field from kSZ. We use $R_{\ell}$ to evaluate the performance of galaxy reconstruction and to probe the agreement between both reconstructions.

\subsection{Result on DESI-LS}

\begin{figure}[h!]
    \centering
    \includegraphics[width=0.9\columnwidth]{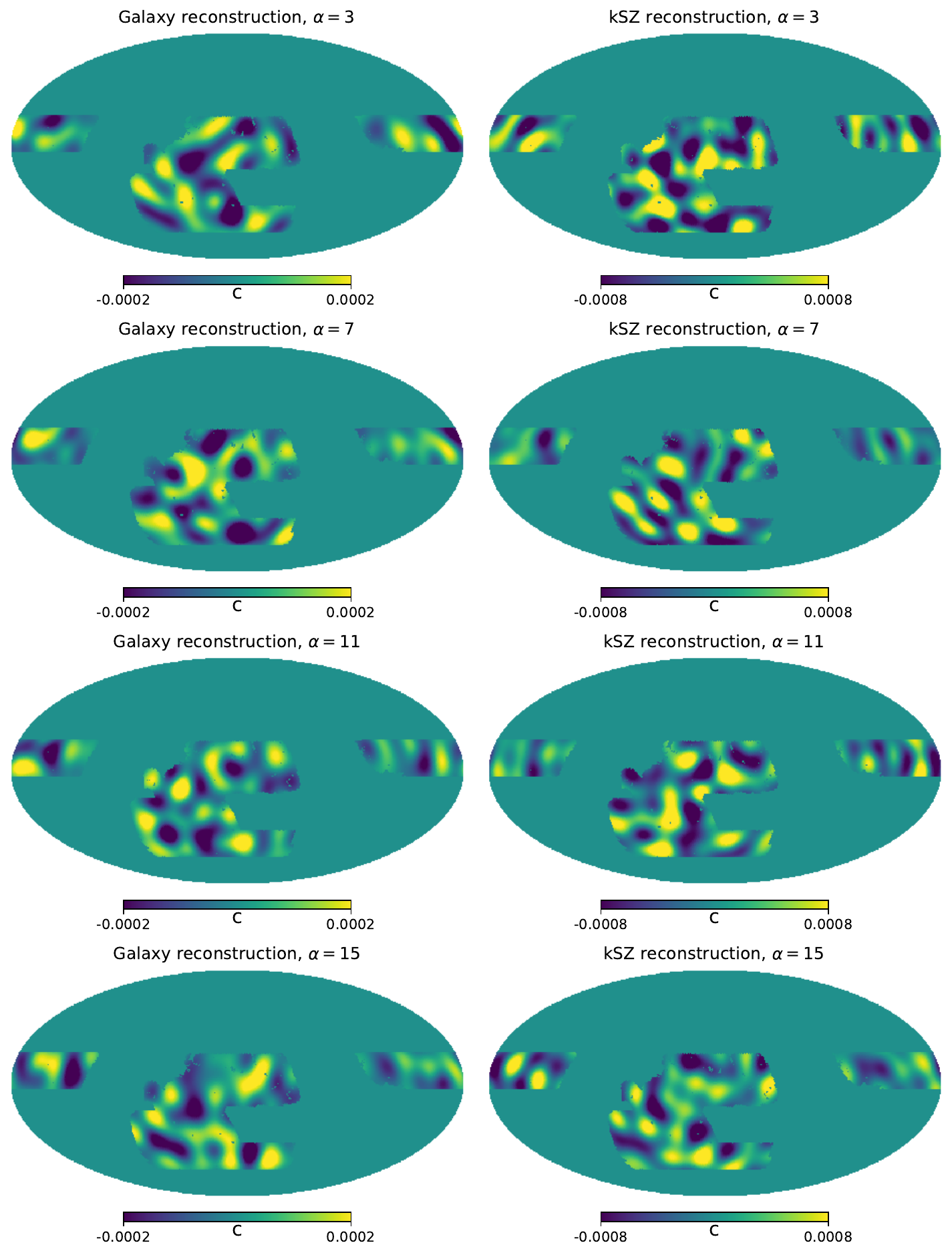}
    \caption{A side-by-side comparison between the kSZ reconstructed velocity field and the galaxy reconstructed velocity field. Modes below $\ell = 5$ and above $\ell = 20$ are removed to stress the correlation between the two fields within the range of multipoles that are signal-dominated. A cutoff on the maximum and minimum reconstructed velocities is applied to the maps, with $v \in [-2\times10^{-4}, 2\times10^{-4}]$ for the galaxy reconstruction and $v \in [-8\times10^{-4}, 8\times10^{-4}]$ for the kSZ reconstruction.}
    \label{fig:galaxy_recon_compare}
\end{figure}

We apply the estimator Eq.~\ref{eq:v estimator from g} to the DESI-LS LRG data described in Sec.~\ref{subsubsec: DESI LS data} to reconstruct the velocity fields. Results for four redshift bins $(\alpha = 3, 7, 11, 15)$ are presented in the left column in Fig.~\ref{fig:galaxy_recon_compare}. To stress the correlation between the velocity maps from different reconstruction methods in the signal-dominated regime, we apply a filter to the modes, such that only $5 \leq\ell \leq 20$ are preserved in the maps. We compare the galaxy reconstruction to those from the kSZ reconstruction, which are given in the right column of Fig.~\ref{fig:galaxy_recon_compare}. The kSZ velocity fields shown are mean-subtracted. One can find some similarities between the galaxy reconstructions and the kSZ reconstructions by eye, but given that the $10\sigma$ SNR is split over the correlation of 20 $\times$ 20 bins, no very strong visual correspondence is expected. Further, the excess in the galaxy auto-powers on large scales may lead to a bias in the galaxy-based velocity reconstruction (see Fig.~\ref{fig:desils_dr9_extlrg_Cgg}, at $\ell < 10$). We defer a more detailed investigation of systematics in this velocity estimate to future work.

\begin{figure}[h!]
    \centering
    \includegraphics[width=0.5\columnwidth]{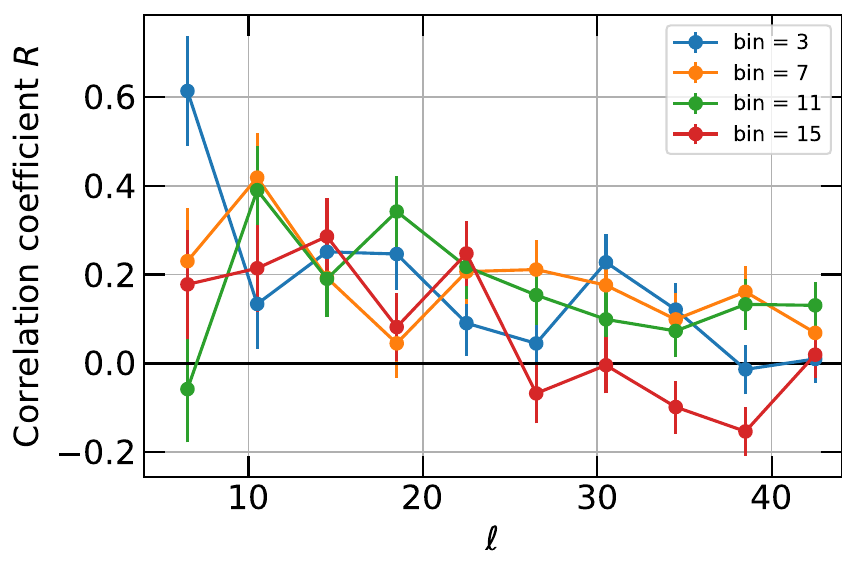}
    \caption{The correlation coefficient $R_\ell$ between the velocity maps from galaxy reconstruction and kSZ reconstruction for 4 redshift bins. The data points cover modes from $\ell = 5$ to $\ell=44$ in bandpowers of 4. It can be seen that most data points lie above the black horizontal line ($R_\ell$ = 0), suggesting a nonzero, positive correlation between velocity maps from the two reconstruction methods.}
    \label{fig:vgal_vksz_corr}
\end{figure}

The similarities in the velocity structure between the two maps are quantified in Fig.~\ref{fig:vgal_vksz_corr} in terms of the cross-correlation coefficient. The cross-correlation coefficient fluctuates within a range from -0.1 to 0.6 on large scales, showing a strong preference for a positive correlation. It is also expected that the correlation between the galaxy reconstruction and kSZ reconstruction is washed out on smaller scales, where the kSZ velocity auto-powers become noise-dominated. We find that most of the redshift bins give a correlation coefficient between 0 and 0.4 within the signal-dominated regime $0 < \ell <20$. We also analyze a different setup by re-binning the DESI LRG catalog into 4-bins (not shown in this paper) and find a lower significance of the correlation coefficient, since the linear velocity estimator relies mostly on the galaxy overdensity information from nearby redshift bins, while a setup with large redshift bins tends to remove the correlation between them. 

We also investigate the expected correlation coefficient between galaxy and kSZ reconstruction with Gaussian simulations. These simulation results are presented in App.~\ref{sec:galaxy reconstruction test on simulation}. 

\section{Conclusion}
\label{sec:conclusion}

In this paper, we apply kSZ velocity reconstruction to data from ACT DR6 and DESI-LS DR9. We establish the kSZ signal in the cross-correlation with a significance of $10\,\sigma$, which is significantly higher than the SNR reported in \cite{McCarthy:2024_ACT_DESI_ksz} and almost equal to the SNR reported in \cite{SelimKendrickPaper} (published at the same time). We find a kSZ amplitude of $A \simeq 0.42$ with respect to the halo model prediction, again in a close match to \cite{SelimKendrickPaper}, which performed their analysis independently from us. This value of $A$ is consistent with strong feedback in massive halos, which was also found in other studies \cite{Hadzhiyska:2024qsl}. To compare our novel QML power spectrum estimator with a more traditional pipeline, we perform a pseudo-$C_{\ell}$ analysis with equivalent data input, finding an SNR of 8.3 with a velocity-galaxy cross-power amplitude of 0.44. We note that while the covariance matrix we used for pseudo-$C_{\ell}$ does not fully converge in $10^5$ simulations for weakly-correlated redshift bin pairs, it has negligible effect on the kSZ amplitude and the 1-sigma error (see App.~\ref{sec:app_pseudocl} for details). We therefore conclude that our QML pipeline improves the pseudo-$C_{\ell}$ method by about 20\%. Finally, we also provide a linear estimator for redshift-binned velocity reconstruction from a photometric galaxy catalog.

To achieve our cross-correlation measurement, we develop a novel implementation of the optimal QML power spectrum estimator, adapted to redshift-binned scalar fields. We find that this estimator substantially outperforms the traditional pseudo-C$_\ell$ formalism. Our estimator is somewhat unusual in that we calculate and invert the full dense covariance matrix, which is tractable at the resolution of $2 \times 20$ bins and $\ell_{max} = 60$ after some computational and algorithmic optimizations. For most analyses in cosmology, this approach would be computationally prohibitive, but for kSZ velocity reconstruction, where almost all the SNR comes from the largest scales, it is appropriate and convenient. This work is the first that demonstrates the capability of the QML pipeline to handle a large number of correlated binned fields. We describe our QML approach and code in more detail in a companion paper~\cite{Kvasiuk:qmlpaper}. The code is publicly released and can be found in~\url{https://github.com/ykvasiuk/qmlfast}.

In future work, we will extend our analysis to include a cosmological parameter inference, especially of primordial non-Gaussianity $f_{\rm NL}$, as proposed in \cite{Munchmeyer:2018eey}. Our pipeline is convenient for this purpose, as again the signal comes from the largest scales. We expect that the dense matrix approach will be convenient to include anisotropic noise or to down-weight calibration contaminated modes. $f_{\rm NL}$ measurements with kSZ reconstruction have already been made in \cite{Smith:2018bpn,Lague:2024czc_7.2sigma}, which demonstrated the feasibility and promise of the method. 

The kSZ velocity reconstruction has a bright future, with SNR predicted in the hundreds for the upcoming generation of CMB and galaxy surveys \cite{Smith:2018bpn}. The present pipeline provides a powerful starting point for future kSZ velocity reconstruction-based cosmology analysis with these experiments.

\section*{Acknowledgements}

We thank Fiona McCarthy and Selim Hotinli for discussions and in particular Matthew Johnson and Kendrick Smith for comments on our draft that improved our analysis. M.M., Y.K. and A.L. acknowledge the support of the U.S. Department of Energy, Office of Science, Office of High Energy Physics under Award Number DE-SC0017647, the support of the National Science Foundation (NSF) under Grant Number 2307109 and
the Wisconsin Alumni Research Foundation (WARF).

\bibliographystyle{unsrt}
\bibliography{ref}

\clearpage

\appendix

\section{Imaging Calibration for the DESI LRG Extended Sample}
\label{app:calibration}

\begin{figure}[h!]
    \centering
    \includegraphics[width=0.7\columnwidth]{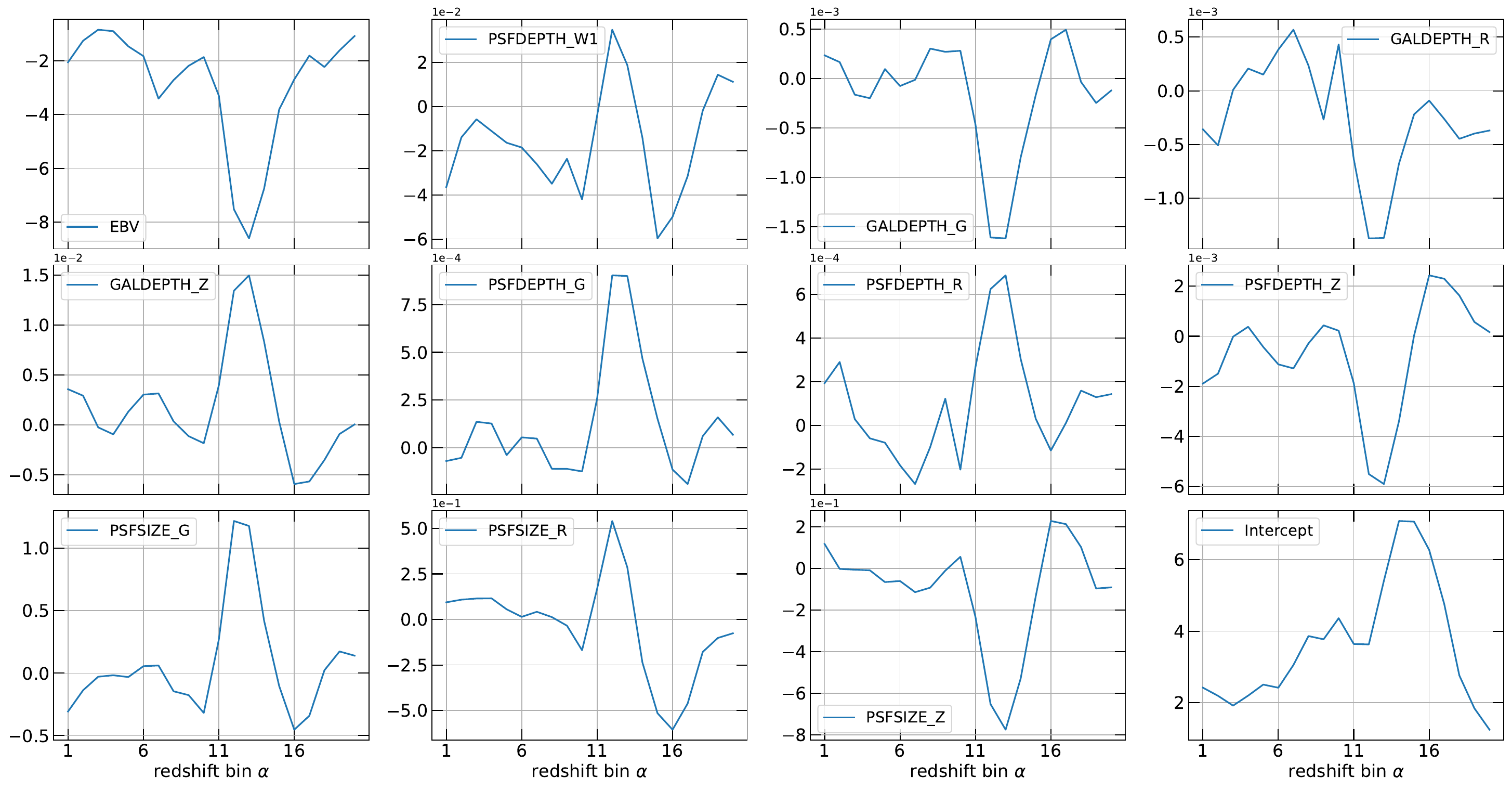}
    \caption{The imaging properties used for constructing the linear regression model and their best-fit values. The imaging maps associated with these properties are available from the data product in~\cite{DESI:202DESILS_Cggsample1}. The unit of the modeling parameters is 1/(imaging property units). The 'Intercept' panel models a constant term in addition to the 11 imaging properties in order to account for potential constant bias from the data.}
    \label{fig:imaging model parameters}
\end{figure}

As described in Sec.~\ref{subsubsec: DESI LS data}, the DESI extended LRG sample suffers an excess in the galaxy auto-powers on large scales. This can partly be removed by modeling the imaging properties of the objects at different redshift bins using the observed data. We provide the 11 imaging properties + intercept and their corresponding best-fit modeling parameter in Fig.~\ref{fig:imaging model parameters}. Note that the parameters depend on the redshift bin, such that one has to recompute the regression model and generate a new set of coefficients for each redshift bin.

\begin{figure}[h!]
    \centering
    \includegraphics[width=0.7\columnwidth]{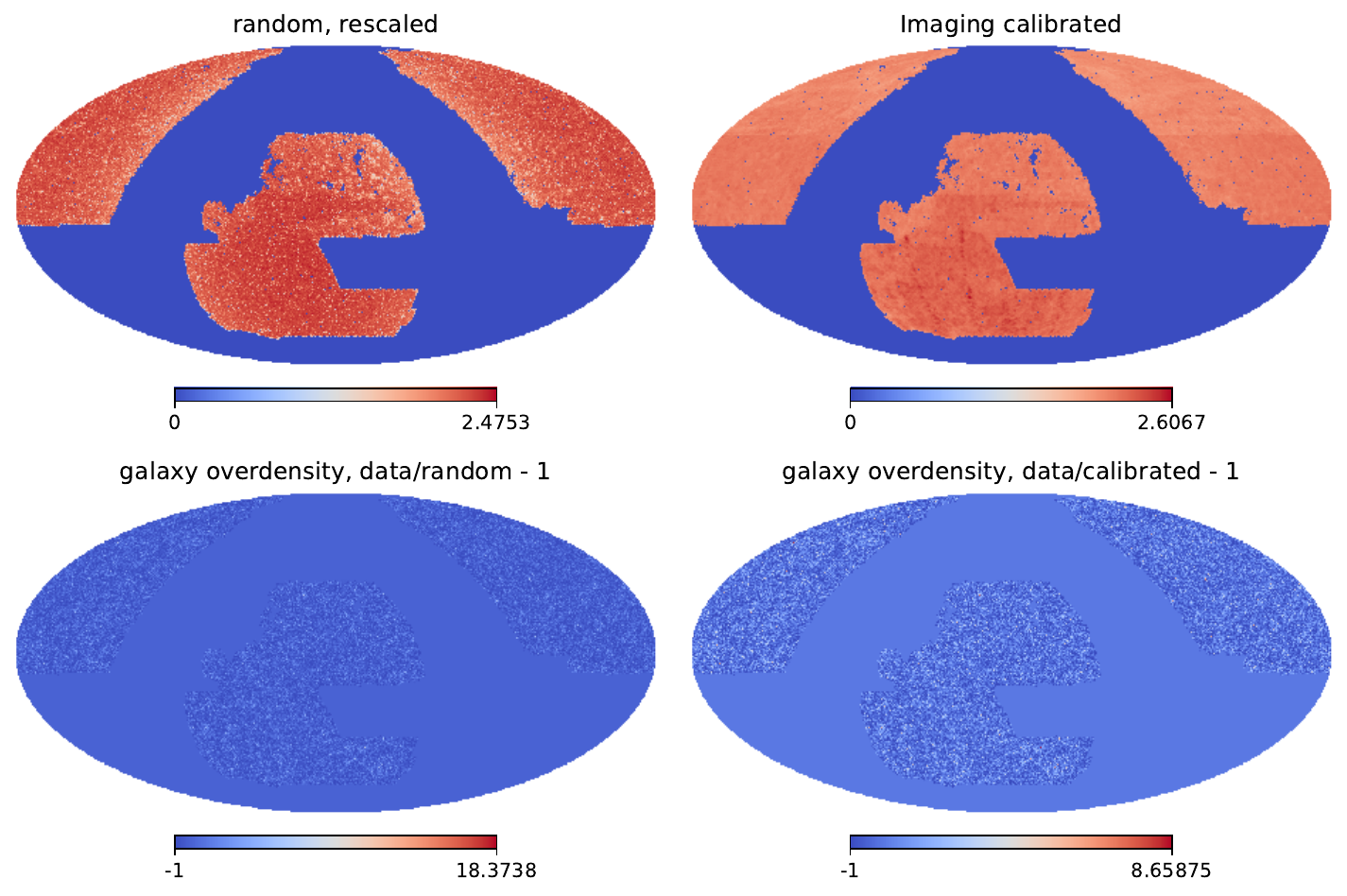}
    \caption{Top row: Random maps obtained by projecting the random catalog points onto Healpix maps. The total count is normalized to match the total number of the galaxies. Left map shows the one before imaging calibration. Right map shows the calibrated random map. Bottom row: The galaxy overdensity maps computed from dividing the galaxy density map with the random map (Left) and the calibrated random map (Right).}
    \label{fig:imaging maps}
\end{figure}

Fig.~\ref{fig:imaging maps} illustrates the difference in the galaxy overdensity map before and after the imaging calibration using the second redshift bin as an example. The original random map (top left) only accounts for the biasing in galaxy counts due to the survey footprint. Therefore, it is inadequate to describe the actual survey condition due to lack of imaging properties. The calibrated random map (top right), however, includes the effect from imaging systematics that can account for photometry at a particular redshift bin. For example, one can see that the calibrated random map tends to overcount the number of galaxies around zero declination. The effect of using a calibrated random map for overdensity computation is shown in the bottom row, where anomalies such as unphysical, overdense regions vanish from the overdensity maps.

\section{Fisher Forecast for the QML Estimator}
\label{app: fisher forecast qml}
It is known that the QML estimator is optimal for Gaussian fields, i.e., it saturates the Cramer-Rao lower bound, and has the smallest possible error bars for the power spectrum. In this section, we make this statement more explicit for our application to velocity-galaxy cross-power by showing that the Fisher information obtained from the QML estimator is equivalent to the one directly obtained from the pixel space covariance matrix. We then use the Fisher matrix to study whether our redshift binning and $\ell_{max}$ exhaust the available SNR, using the full pixel-wise covariance matrix rather than the typical but imprecise $f_{sky}$ approximation.

Starting from the definition of the fiducial pixel covariance matrix, $\mathbf{C}(A)$:
\begin{align}
    \mathbf{C}(A) &= 
    \begin{pmatrix}
        \mathbf{S}^{gg}+\mathbf{N}^{gg}, &A\mathbf{C}^{gv} \\
        A\mathbf{C}^{vg}, &\mathbf{S}^{vv}+\mathbf{N}^{vv}
    \end{pmatrix} \, ,
\end{align}
here we consider a redshift-independent parameter $A$. We remind the reader that the cross-power amplitude $A$ is included as the only parameter to be constrained in this problem. Also note that $A = 0$ recovers the cross-power-purified pixel space covariance matrix Eq.~\ref{eq: pixel covariance matrix explicit}. The Fisher information with respect to A is given by, according to the definition:
\begin{align}
    F_{AA} &= \frac{1}{2}\text{Tr}\left.\left(\mathbf{C}^{-1}\frac{\partial \mathbf{C}}{\partial A}\mathbf{C}^{-1}\frac{\partial \mathbf{C}}{\partial A}\right)\right|_{A=0}\ ,
\end{align}
as we measure the number of sigmas away from $A = 0$. Now we can write $\frac{\partial \mathbf{C}}{\partial A}$ in terms of Legendre matrix $\mathbf{P}_{\ell}$, such that:
\begin{align}
    \frac{\partial \mathbf{C}}{\partial A} = \sum_{\ell \in \mathbf{C}^{vg}} C^{vg}_{\ell}\mathbf{P_{\ell}} \, ,
\end{align}
rewriting the Fisher information with this definition gives:
\begin{align}
    F_{AA} &= \sum_{\ell' \ell}\frac{1}{2}\text{Tr}\left.\left(\mathbf{C}^{-1}C^{vg}_{\ell}\mathbf{P_{\ell}}\mathbf{C}^{-1}C^{vg}_{\ell'}\mathbf{P_{\ell'}}\right)\right|_{A=0}\ , \nonumber \\
     &= \sum_{\ell' \ell} C^{vg}_{\ell}C^{vg}_{\ell'} \frac{1}{2}\text{Tr}\left.\left(\mathbf{C}^{-1}\mathbf{P_{\ell}}\mathbf{C}^{-1}\mathbf{P_{\ell'}}\right)\right|_{A=0} \, , \nonumber \\
     &= \sum_{\ell' \ell} C^{vg}_{\ell} C^{vg}_{\ell'} \mathbf{F}^{\text{QML}}_{\ell\ell'} \, , \nonumber \\
     &= (\mathbf{C}^{vg})^{\mathbf{T}}\mathbf{F}^{\text{QML}}\mathbf{C}^{vg} \, , 
\end{align}
where from second to third row we use the definition of the QML Fisher matrix Eq.~\ref{eq:QML Fisher matrix} and we label it with a superscript 'QML' to distinguish it from the Fisher information of the cross-power amplitude. We have also replaced the summation over the cross-power modes $\ell, \ell'$ with a matrix multiplication of the vectorized cross-powers $\mathbf{C}^{vg}$. Note that the last line is equivalent to the definition of the error in cross-power amplitude $\sigma_A$ using a cross-power-purified covariance matrix, Eq.~\ref{eq: equations for A and sigmaA}. This completes the proof.

\begin{figure}[h!]
    \centering
    \includegraphics[width=0.5\columnwidth]{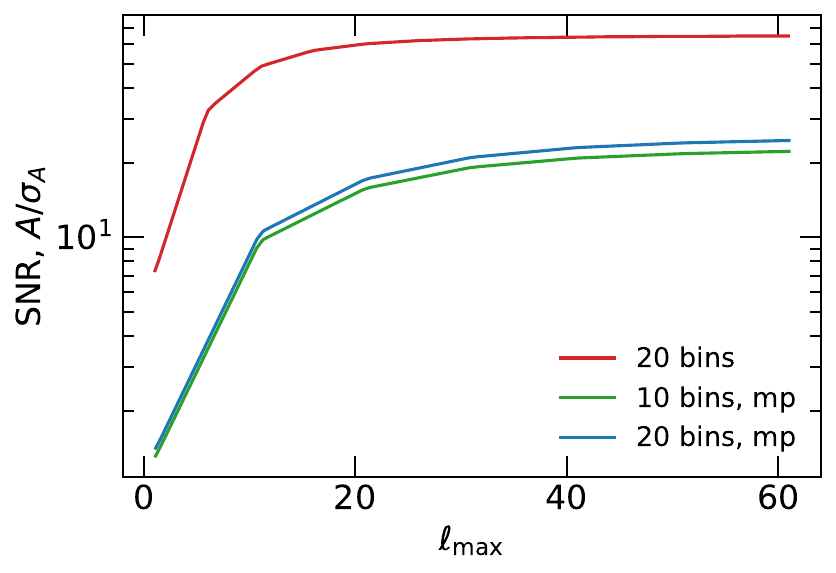}
    \caption{A figure that illustrates the Fisher forecast for various configurations. The covariance matrix is computed using the prescribed fiducial powers in Sec.~\ref{subsec: fiducial powers}. 'mp' stands for a cross-power-purified covariance matrix that is computed from the cross-power $C^{gv}$ contribution only (with $C^{gg}$ and $C^{vv}$ decoupled from the SNR evaluation). The red curve shows the 20-bin SNR computed without cross-power-purification. The SNR is defined as $A/\sigma_A$ with the assumption of $C^{gv}$ amplitude parameter $A = 1$.}
    \label{fig:fisher_forecast}
\end{figure}

We present the Fisher forecast for three different covariance matrix setups in Fig.~\ref{fig:fisher_forecast}. The x-axis shows the maximum number of multipoles $\ell_{\rm max}$ preserved in the Fisher forecast. The red curve with the highest SNR corresponds to a 20-bin analysis without cross-power purification, where both the galaxy and velocity auto-powers are included in the SNR evaluation pipeline. We also show the blue and the green curve, which correspond to cross-power-purified 20-bin and 10-bin configurations. They are evaluated from just the galaxy-velocity cross-powers. All three configurations assume an underlying cross-power amplitude $A = 1$. From Fig.~\ref{fig:fisher_forecast}, we see that the SNR starts to saturate at $\ell = 50$ for all configurations. Our choice of $\ell_{\rm max} = 60$ in the analysis therefore captures all of the available SNR. The setup with 20 bins and without cross-power purification (red curve) has the highest SNR among the 3 curves presented and is three times larger than the cross-power-purified configuration (blue curve).

From Fig.~\ref{fig:fisher_forecast} it is also clear that for a given estimated cross-power amplitude, most of the SNR is located in the first 20 modes. For the case of 20 redshift bins with cross-power purification (blue curve), the SNR increases by 3 in the first 10 modes and 1.5 in the next 10 modes, accounting for 75\% of the total SNR. Therefore, the SNR is mainly affected by the quality of the low $\ell$ cross-powers reconstructed from the data.

\section{Definition of the SNR as a Significance of the Rejection of the Null Hypothesis}
\label{app:snr}
Let us define broadly the hypothesis $H_1: \{S \neq 0\}$ that we detect some "signal", which is defined precisely later. The null hypothesis is $H_0:\{S=0\}$. We define SNR in all cases as 
\begin{equation}
    SNR = \sqrt{-2\Delta\ln\mathcal{L}\bigg|_{H_0}} \, .
\end{equation}
It has a nice meaning of the significance in "number of sigmas" of the rejection of $H_0$. We immediately note that in the simplest possible case of one Gaussian variable, this definition implies the usual $SNR = \frac{\hat{\mu}}{\sigma}$.
Let us define $\bf{\hat{x}}$ - an output of the QML pipeline for cross powers. Then by construction
\begin{equation}
    -2\ln\mathcal{L}(\bf{x}|\bf{\hat{x}}) = (\bf{\hat{x}} - \bf{x})^T\bf{F}\bigg|_x(\bf{\hat{x}} - \bf{x}) \, .
\end{equation}
Hence, for the null hypothesis that $\bf{\hat{x}}$ is consistent with noise (that is equivalent to $\bf{x=0}$), we get
\begin{equation}
    SNR = \sqrt{\bf{\hat{x}^T}\bf{F}\hat{x}}\, .
\end{equation}
Because not all the parameters are independent parameters of our theory model, we usually define the reduced set of parameters that we are interested in: e.g. $\bf{b}$: $\dim \mathbf{b} = N_b, \dim \mathbf{x} = N$, and $N_b \ll N$. Let's also define a grouping matrix $\mathbf{G}$, $\dim \mathbf{G} = (N,N_b)$ and diagonal matrix $\mathbf{X}, \dim \mathbf{X} = (N,N)$ with $\mathbf{x}$ on the diagonal. The grouping matrix is a general way to group estimates with the same property. For example, one can group all the multipoles in the same pair of redshifts and define a bias coefficient for the corresponding template.
Then we have
\begin{equation}
    -2\ln\mathcal{L}(\bf{b}|\bf{\hat{x}}) = (\bf{\hat{x}}^T - b^TG^T\bf{X})\bf{F}\bigg|_x(\bf{\hat{x}}-XGb)\, .
\end{equation}
This likelihood is maximized with
\begin{equation}
    \mathbf{\hat{b}} = (\bf{G^T XFXG})^{-1}\mathbf{G^T X^T F\hat{x}}\, .
\end{equation}
Then
\begin{equation}
        -2\ln\mathcal{L}(\bf{b}|\bf{\hat{b}}) = (\bf{\hat{b}}^T - b^T)\bf{C}_b^{-1}(b-\bf{\hat{b}})\, .
\end{equation}
With 
$\mathbf{C_b} = (\bf{G^T XFXG})^{-1}$.
It follows that SNR in this case is (under $H_0:\bf{b}=0$)
\begin{equation}\label{eq:SNR null hypothesis}
    SNR = \sqrt{\mathbf{\hat{b}^T C_b^{-1}\hat{b}}}\, .
\end{equation}
We also note that in the extreme case of one parameter $\mathbf{b}=A$, we have
\begin{equation}\label{eq: equations for A and sigmaA}
    \hat{A} = \frac{\mathbf{x^T F\hat{x}}}{\mathbf{x^T Fx}},\quad  \sigma^2_{A} = \frac{1}{\mathbf{x^T Fx}},\quad  SNR_A=\sqrt{\hat{A}\sigma_A^{-2}\hat{A}} = \frac{\mathbf{x^T F\hat{x}}}{\sqrt{\mathbf{x^T Fx}}}\, .
\end{equation}
Dependent on how we define signal $S$: $\bf{x}$, $\bf{b}$, or $A$, we can use either of the expressions as a consistent definition of SNR. We adopted the definition of the SNR via $\bf{b}$ in the main text, where we group according to redshift binning, so that $\bf{b}$ represents the amplitude of the cross spectrum in a given bin. The SNR in Eq. \eqref{eq:SNR definition bv} is an upper limit on the SNR in Eq. \eqref{eq:SNR definition A}: since covariance and its inverse are positive definite, the first definition of the SNR is just a norm with respect to the inner product $(\cdot,\cdot )_{cov^{-1}}$. The inequality that relates two definitions of the SNR follows from the fact that the Cauchy-Schwarz inequality tells us that $(\bf\hat{b},b)_{cov^{-1}}\leq\sqrt{(\bf\hat{b},\hat{b})_{cov^{-1}}(\bf b,b)_{cov^{-1}}}$. We also note that even though $SNR_A \leq SNR_{\bf{b_v}}$, the detection significance should be interpreted via p-value. For the case of a single variable A, $SNR^2_A = \hat{A}\sigma^{-2}\hat{A} \sim \chi^2_1(SNR^2_A)$ and $p_A = 1-F_{\chi_1^2}(SNR^2_A)$, while for $\bf{b_v}$, $p_{\bf{b_v}} = 1-F_{\chi^2_n}(SNR^2_{\bf{b_v}})$. Here, $F_{\chi_n^2}$ is the cumulative density function (CDF) of the $\chi^2$ distribution with $n$ degrees of freedom.
\paragraph{Comment on the alternative definition of the SNR.}
Having estimated $\hat{\mathbf{b}}_v$, one could've defined the SNR as
\begin{align}
    \text{SNR} &= \frac{\hat{\mathbf{b}}^{\text{T}}_{v}\text{Cov}^{-1}(\hat{\mathbf{b}}_v)\mathbf{b}_v}{\sqrt{\mathbf{b}^{\text{T}}_{v}\text{Cov}^{-1}(\hat{\mathbf{b}}_v)\mathbf{b}_v}} \, ,
\end{align}
where $\mathbf{b}_v = \mathbf{1}$ as a fiducial value. Looking more carefully at this expression and recalling that $\text{Cov}(\hat{\mathbf{b}}_v)^{-1} = \mathbf{L}^T\mathbf{F}\mathbf{L}$, we can notice that it's the same as the expression of the SNR for the single amplitude. First, we note that $\mathbf{L} = \mathbf{X}\mathbf{G}$. Then
\begin{align}
\mathbf{G^T XFXG1} = \mathbf{G^T XFx}\, .
\end{align}
Since $\mathbf{G^T XFXG}$ is invertible, we can write  
\begin{equation}
    \mathbf{1} = (\mathbf{G^T XFXG})^{-1}\mathbf{G^T XFx}\, .
\end{equation}
Then under the root we have:
\begin{equation}
\bf{x^TFL(L^TFL)^{-1}L^TFx} = \texttt{Tr}[FL(L^TFL)^{-1}L^TFxx^T]\, .    
\end{equation}
Then we notice that $\bf{xx^T} = X11^TX$, use cyclic identity of trace and the fact that 
\begin{equation}
    \texttt{Tr}[G^TAG] = \texttt{Tr}A \, ,
\end{equation}
because of the property of the grouping matrix. We simplify
\begin{align}
    &\texttt{Tr}[\bf{FL(L^TFL)^{-1}L^TFxx^T}] \, ,\\
    & =\texttt{Tr}[\bf{G^TXFL(L^TFL)^{-1}L^TFXG11^T}] \, ,\\
    & = \bf{x^TFx}\, .
\end{align}
The numerator can be similarly simplified, so that indeed
\begin{align}
    \text{SNR} &= \frac{\hat{\mathbf{b}}^{\text{T}}_{v}\text{Cov}^{-1}(\hat{\mathbf{b}}_v)\mathbf{b}_v}{\sqrt{\mathbf{b}^{\text{T}}_{v}\text{Cov}^{-1}(\hat{\mathbf{b}}_v)\mathbf{b}_v}} = \frac{\bf{x^TF\hat{x}}}{\sqrt{\bf{x^TFx}}} \, .
\end{align}

\section{Rescaling of the kSZ Estimator under Fiducial Power $C^{\tau g}$}\label{sec:ksz estimator rescaling}
To understand the effect to the SNR from either the QML and pseudo-$C_{\ell}$ due to potential mismodelling, we consider the case of the overall galaxy-velocity cross-power amplitude $A$.
Consider a rescaling of fiducial $C^{\tau g}_{\ell}$ by a factor of b due to a mismodelling, such that $\bar{C}^{\tau g} = bC^{\tau g}$. Now we ask the question how does the kSZ estimator change with it, assuming $b$ is not scale dependent, i.e., it is not a function of $\ell$, then according to the definition Eq.~\ref{eq:v estim result}:
\begin{align}\label{eq:ksz estim rescaling relation}
    \bar{N}^{vv, \alpha}_{\ell} &= b^{-2}N^{vv, \alpha}_{\ell} \, , \nonumber \\
    \bar{\hat{v}}_{\ell m} &= b^{-1}\hat{v}_{\ell m} \, ,
\end{align}
then one can ask how the SNR for velocity-galaxy cross-power,
\begin{align}\label{eq:SNR}
\frac{A}{\sigma_A} &= \frac{(\hat{\mathbf{C}}^{vg})^T\text{Cov}^{-1}\mathbf{C}^{vg}}{\sqrt{(\mathbf{C}^{vg})^T\text{Cov}^{-1}\mathbf{C}^{vg}}} \, , \nonumber \\
&= \frac{(\hat{\mathbf{C}}^{vg})^T\,F\,\mathbf{C}^{vg}}{\sqrt{(\mathbf{C}^{vg})^T\,F\,\mathbf{C}^{vg}}} \, .
\end{align}
changes with respect to the scaling, both for pseudo-$C_{\ell}$ and QML \, .

For pseudo-$C_{\ell}$, since the mask decoupling is a linear operation and it does not depend on the choice of fiducial powers, one can safely propagate this scaling to the reconstructed full-sky velocity auto-powers and cross-powers:
\begin{align}\label{eq:pcl ps rescaling relation}
    \bar{C}^{\hat{v}\hat{v}}_{\ell, \text{pcl}} &= b^{-2}C^{\hat{v}\hat{v}}_{\ell, \text{pcl}} \, , \nonumber \\
    \bar{C}^{g\hat{v}}_{\ell, \text{pcl}} &= b^{-1}C^{g\hat{v}}_{\ell, \text{pcl}} \, , \nonumber \\
    \bar{C}^{gv}_{\ell} &= C^{gv}_{\ell} \, .
\end{align}
Note that the fiducial powers do not scale with the mismodelling factor, since the cross-power does not contain the reconstruction noise. Moreover, the covariance of the cross-powers, which is evaluated from rescaled pcl power spectrum, is also corrected by a factor of $b^{-2}$:
\begin{align}
    \overline{\text{Cov}}^{-1} &= b^{2}\text{Cov}^{-1} \, ,
\end{align}
Then the SNR of the rescaled velocity field, Eq.~\ref{eq:SNR} is given by:
\begin{align}
    \left(\frac{A}{\sigma_A}\right)_{\text{pcl}} &= \frac{(b^{-1})(b^{2})}{\sqrt{(b^{2})}}\frac{A}{\sigma_A} \, , \nonumber \\
    &= \frac{A}{\sigma_A} \, ,
\end{align}
i.e. the significance of the cross-correlation estimated with pseudo-C$\ell$ does not change under such a rescaling.

\begin{figure}[h!]
    \centering
    \includegraphics[width=0.9\textwidth]{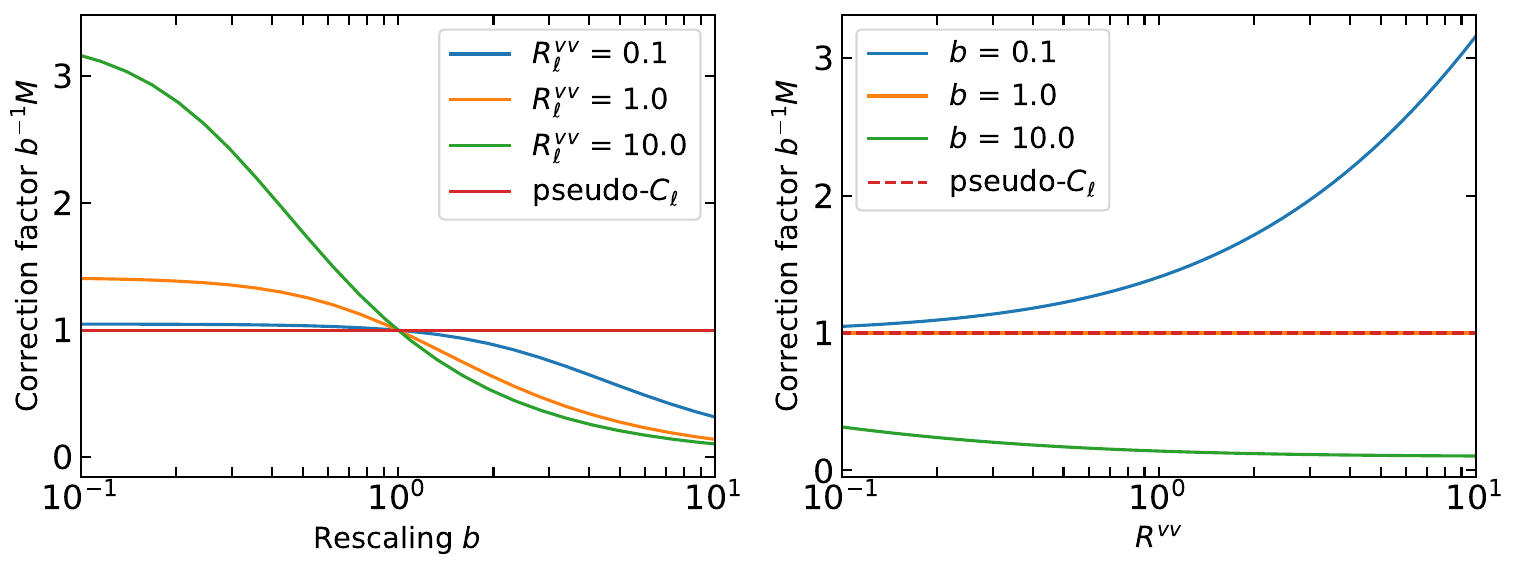}     
    \caption{Illustrations of the QML SNR correction factor Eq.~\ref{eq:SNR correction factor simplify} with varying rescaling $b$ and velocity auto-power SNR $R^{vv}$. Left: varying $b$ for three different values of $R^{vv} = 0.1, 1$ and $10$, corresponding to the noise-dominated, signal-noise-comparable and signal-dominated regimes. Right: varying $R^{vv}$ for three different values of $b = 0.1, 1$ and $10$, corresponding to the underestimating, exact-modelling and overestimating of the observed velocity map. The correction factor $b^{-1}M =1$ for pseudo-$C_{\ell}$ is also plotted as the red line (dashed line for the right plot to avoid overlapping with the $b = 1$ in QML) for reference.}
    \label{fig:mismodelling}
\end{figure}

On the other hand, consider the QML estimator. Eq.~\ref{eq:ksz estim rescaling relation}-Eq.~\ref{eq:pcl ps rescaling relation} still hold (since QML is unbiased over the power spectrum by construction). However, the inverse covariance matrix/Fisher matrix is now computed from fiducial powers. We instead have the following relations:
\begin{align}
    \bar{F} &= \bar{F}\left(S^{vv}_{\ell}+\bar{N}^{vv}_{\ell}\right) \, , \nonumber \\
    &= \bar{F}\left(S^{vv}_{\ell}+b^{-2}N^{vv}_{\ell}\right) \, .
\end{align}
The Fisher matrix for QML estimator depends on the fiducial velocity auto-power $C^{vv}_{\ell}$, which contains the signal $S^{vv}_{\ell}$ that does not scale with the mismodelling, and the reconstruction noise $\bar{N}^{vv}_{\ell}$ that is obtained from Eq.~\ref{eq:ksz estim rescaling relation}. Therefore, the QML inverse covariance matrix does not follow a simple rescaling relation as in pseudo-$C_{\ell}$, and we arrive at the expression for the new SNR:
\begin{align}
    \left(\frac{A}{\sigma_A}\right)_{\text{QML}} &= b^{-1}\frac{(\hat{\mathbf{C}}^{vg})^T\,\bar{F}\,\mathbf{C}^{vg}}{\sqrt{(\mathbf{C}^{vg})^T\,\bar{F}\,\mathbf{C}^{vg}}} \, , \nonumber \\
    &= b^{-1}M\left(b\right)\frac{(\hat{\mathbf{C}}^{vg})^T\,F\,\mathbf{C}^{vg}}{\sqrt{(\mathbf{C}^{vg})^T\,F\,\mathbf{C}^{vg}}} \, , \nonumber \\
    &= b^{-1}M(b)\frac{A}{\sigma_A} \, ,
\end{align}
where we have defined the function:
\begin{align}
    M(b) &= \sqrt{\frac{(\mathbf{C}^{vg})^T\,F\,\mathbf{C}^{vg}}{(\mathbf{C}^{vg})^T\,\bar{F}\,\mathbf{C}^{vg}}}\frac{(\hat{\mathbf{C}}^{vg})^T\,\bar{F}\,\mathbf{C}^{vg}}{(\hat{\mathbf{C}}^{vg})^T\,F\,\mathbf{C}^{vg}}\, ,
\end{align}
note that $M \rightarrow 1$ as $b \rightarrow 1$. To illustrate the effect of the correction factor $b^{-1} M$ we restrict ourselves to one redshift bin and one mode, such that both $C^{vg}$ and $F$ are numbers. In this limit, $F$ can be easily determined from the Gaussian covariance:
\begin{align}
    F^{-1} &= C^{vv}C^{gg} \, , \nonumber \\
    &=(S^{vv}+N^{vv})(S^{gg} + N^{gg}) \, .
\end{align}
We have omitted the subscript $\ell$ for simplicity and $C^{vg} = C^{gv} = 0$ for a cross-power-purified estimator. Apparently, $\bar{F}^{-1} \neq b^{2}F^{-1}$, and the correction factor is given by:
\begin{align}\label{eq:SNR correction factor simplify}
    b^{-1}M &= b^{-1} \sqrt{\frac{S^{vv}+N^{vv}}{S^{vv}+b^{-2}N^{vv}}} \, , \nonumber \\
    &= b^{-1}\sqrt{\frac{R^{vv}+1}{R^{vv}+b^{-2}}} \, .
\end{align}
The kSZ SNR ratio $R^{vv}$ is defined as the ratio between the fiducial velocity auto-power and kSZ reconstruction noise $S^{vv}/N^{vv}$. We display the function Eq.~\ref{eq:SNR correction factor simplify} in Fig.~\ref{fig:mismodelling} for different values of mismodelling factor $b$ and kSZ SNR $R^{vv}$. From the left panel one can see that in signal-dominated regime $R^{vv} > 1$, any mismodelling in the electron-galaxy power spectrum ($b \neq 1$) will induce a correction factor to the cross-power SNR. In particular, $b < 1$ gives a boost in SNR, while $b > 1$ gives a drop in SNR. The correction factor converges to 1 as the kSZ reconstruction enters the noise-dominated regime $R^{vv} < 1$. To summarize, the galaxy-electron cross-power SNR obtained from the QML pipeline can vary with fiducial power spectrum input for the Fisher matrix calculation. In the regime where $C^{\tau g}_{\ell}$ is overestimated and the kSZ reconstruction is signal-dominated, one can record a boost in the SNR through a linear correction factor $b$.

\section{Bandpowering Scheme for Velocity-Galaxy Cross-Powers}\label{app:bandpowering scheme}
Here we briefly mention the bandpowering scheme used for making the plots in Fig.~\ref{fig:Cgv_QML_estimation_binned}, which takes into account the non-zero coupling between different modes and redshift bins. Consider a bandpowered cross-power defined as follows:
\begin{align}\label{eq:bandpower scheme}
    \mathbf{C}^{gv}_{(i,j,L)} &= \mathbf{M}_{(i,j,L),(\alpha, \beta,\ell)}\mathbf{C}^{gv}_{(\alpha, \beta,\ell)}\, , \\
    &=\frac{1}{V}\sum_{\substack{\alpha \in i,\\ \beta\in j,\ell \in L}} C^{g^{\alpha}v^{\beta}}_{\ell} \, ,
\end{align}
here $\mathbf{C}^{gv}_{(i,j,L)}$ gives the bandpowered, vectorized cross-power with a new set of indexing $(i, j, L)$ that represents the first redshift interval, second redshift interval and the bandpower. For simplicity we weight each mode equally, such that $V$ is the normalization factor that gives the total number of modes within the interval $(i, j, L)$ and $\mathbf{M}$ is a rectangular matrix of dimension $(i\times j\times L)$ by $(\alpha\times \beta\times \ell)$, such that:
\begin{align}
    \mathbf{M}_{(i,j,L),(\alpha, \beta,\ell)} = \frac{1}{V} \,\,\,\, \text{if} \,\,\,\,(\alpha, \beta,\ell)\in(i,j,L) \, .
\end{align}
The bandpowered covariance matrix, with the diagonal terms as the squared errorbars, is compressed as:
\begin{align}
    \mathbf{F}^{-1}_{\text{bandpowered}} = \mathbf{M}\mathbf{F}^{-1}\mathbf{M}^T \, .
\end{align}
We note that this choice of uniform weighting across the modes is suboptimal; i.e., it leads to a loss in SNR compared to the un-binned cross-powers. This is also different from a direct 20 galaxy redshift bins $\times$ 20 velocity redshift bins QML analysis. The purpose of Eq.~\ref{eq:bandpower scheme} is to provide a simple averaging, so that qualitative features of the QML cross-powers can still be illustrated given that the complicated mixing between redshift bins and modes is minimally preserved.

\section{Linear Estimator of the Velocity Field from Galaxies}\label{sec:galaxy reconstruction details}

\subsection{Construction of the Velocity Estimator from Maximizing the Joint Probability $\mathcal{P}(v^{\text{true}}_{r}, \delta^{\text{obs}}_{g})$}
The true, mode-dependent radial velocity field at different slices of redshift shells $v^{\alpha, \text{true}}_{r}$ are correlated to each other. This is because matter in nearby regions should interact with each other and move in similar velocities. We wish to reconstruct the large-scale radial velocity field from a galaxy survey through linear theory. Therefore, to the lowest order, we assume the true radial velocity field can be related to observed matter tracer overdensity $\delta_g$, which will be the galaxy overdensity in this case. To start with, we construct the pixel-space radial velocity vector $\mathbf{v}^{\text{true}}_{r}$ and the galaxy overdensity vector $\mathbf{g}^{\text{obs}}$, defined as:
\begin{align}
    \mathbf{v}^{\text{true}}_{r} &= \left(v^{\alpha_1, \text{true}}_{r}, v^{\alpha_2, \text{true}}_{r}, ..., v^{\alpha_n, \text{true}}_{r}\right)^\text{T} \, ,\nonumber \\
    \mathbf{g}^{\text{obs}} &= \left(g^{\alpha_1, \text{obs}}, g^{\alpha_2, \text{obs}}, ..., g^{\alpha_n, \text{obs}}\right)^\text{T} \, ,
\end{align}
here $\alpha$ runs across a total number of redshift bins $n$. Each $v^{\alpha_n, \text{true}}_{r}$ and $g^{\alpha_n, \text{obs}}$ should be understood as the vectorized velocity and galaxy overdensity map. The radial velocity vector denotes the underlying true radial velocity distribution and the galaxy overdensity vector is based on the observed data, which is the Gaussian simulation in this case.

Armed with the definitions, we assume that $\mathbf{v}^{\text{true}}_{r}$ can be constructed from a linear model of $\mathbf{g}^{\text{obs}}$. That is:
\begin{align}\label{eq:linear model assumption}
    \mathbf{v}^{\text{true}}_{r} = \textbf{M}\,\mathbf{g}^{\text{obs}} \, .
\end{align}
The fact that Eq.~\ref{eq:linear model assumption} lies in pixel space makes it more adaptive to the partial-sky scenario, since masking over a certain region on the sky will only reduce the dimension of the data vector and matrix $\textbf{M}$ in Eq.~\ref{eq:linear model assumption}, while one has to account for the mode coupling between all the modes $(\ell, m)$ in the spherical harmonic space, such that the dimension of the matrix cannot be reduced. From the expression Eq.~\ref{eq:linear model assumption}, we construct a Gaussian likelihood joint probability distribution $\mathcal{P}\left(\mathbf{v}^{\text{true}}, \mathbf{g}^{\text{obs}}\right)$.  

\begin{align}\label{eq:joint prob likelihood}
    \mathcal{P}\left(\mathbf{v}^{\text{true}}_{r}, \mathbf{g}^{\text{obs}}\right) &= \text{exp}\left(-\frac{1}{2}\mathbf{u}^\text{T}\mathbf{C}^{-1}\mathbf{u}\right) \, ,
\end{align}
where $\mathbf{u}$ is an abstract notation of the composite vector $\mathbf{u} = \left(\mathbf{v}^{\text{true}}_{r}, \mathbf{g}^{\text{obs}}\right)^{\text{T}}$ and $\mathbf{C}$ is a large covariance matrix defined as:
\begin{align}\label{eq:Covariant block}
    \mathbf{C} = 
    \begin{pmatrix}
        C^{v_rv_r} & C^{v_rg} \\
        C^{gv_r} & C^{gg}
    \end{pmatrix} \, ,
\end{align}
each entry in Eq.~\ref{eq:Covariant block} represents an $n\times n$ block covariance matrix, for example:
\begin{align}
    C^{v_rv_r} &= 
    \begin{pmatrix}
        C^{v^1_r v^1_r} & ... &  C^{v^1_r v^{\alpha_n}_r}\\
        ... & ... & ... \\
        C^{v^{\alpha_n}_r v^1_r} & ... &  C^{v^{\alpha_n}_r v^{\alpha_n}_r}
    \end{pmatrix} \, ,
\end{align}
This covariance matrix is completely determined by the linear theory and can be directly calculated from the equations in last section. Assuming that $\left(C^{v_rv_r}\right)^{-1}$ and $\left(C^{gg}\right)^{-1}$ exist, then $\mathbf{C}^{-1}$ can be represented as follows:
\begin{align}\label{eq:covariant inverse}
    \mathbf{C}^{-1}= 
    \begin{pmatrix}
        \mathbf{Q}\left(C^{v_rv_r}, C^{gg}\right) & -\mathbf{Q}\left(C^{v_rv_r}, C^{gg}\right)\,C^{v_r g}\left(C^{gg}\right)^{-1} \\
        -\mathbf{Q}\left(C^{gg}, C^{v_r v_r}\right)C^{v_r g}\left(C^{v_r v_r}\right)^{-1} & \mathbf{Q}\left(C^{gg}, C^{v_r v_r}\right)
    \end{pmatrix} \, ,
\end{align}
where the matrix $\mathbf{Q}\left(C^{v_rv_r}, C^{gg}\right)$ is given as follows:
\begin{align}
    \mathbf{Q}\left(C^{v_rv_r}, C^{gg}\right) = \left(C^{v_rv_r} - C^{v_r g}\left(C^{gg}\right)^{-1}C^{v_r g}\right)^{-1} \, ,
\end{align}
notice that we repeatedly used the symmetric property $C^{v_r g} = (C^{g v_r})^T$.

Inserting Eq.~\ref{eq:covariant inverse} and the definition of $\mathbf{u}$ to the joint probability $\mathcal{P}\left(\mathbf{v}^{\text{true}}, \mathbf{g}^{\text{obs}}\right)$, we arrive at the following expression for the exponential term:
\begin{align}\label{eq:gaussian term}
    \mathbf{u}^\text{T}\mathbf{C}^{-1}\mathbf{u} &=
    \begin{pmatrix}
        \left(\mathbf{v}^{\text{true}}\right)^{\text{T}} & \left(\mathbf{g}^{\text{obs}}\right)^{\text{T}}
    \end{pmatrix}
    \begin{pmatrix}
        \mathbf{Q}\left(C^{v_rv_r}, C^{gg}\right) & -\mathbf{Q}\left(C^{v_rv_r}, C^{gg}\right)\,C^{v_r g}\left(C^{gg}\right)^{-1} \\
        -\mathbf{Q}\left(C^{gg}, C^{v_r v_r}\right)C^{v_r g}\left(C^{v_r v_r}\right)^{-1} & \mathbf{Q}\left(C^{gg}, C^{v_r v_r}\right)
    \end{pmatrix}
    \begin{pmatrix}
        \mathbf{v}^{\text{true}} \\ \mathbf{g}^{\text{obs}}
    \end{pmatrix} \, \nonumber \\
    &= \left(\mathbf{v}^{\text{true}}\right)^{\text{T}} \mathbf{Q}\left(C^{v_rv_r}, C^{gg}\right)\mathbf{v}^{\text{true}} + \left(\mathbf{v}^{\text{true}}\right)^{\text{T}} \left[-\mathbf{Q}\left(C^{v_rv_r}, C^{gg}\right)\,C^{v_r g}\left(C^{gg}\right)^{-1}\right]\mathbf{g}^{\text{obs}} + ... \, ,
\end{align}
other terms are abbreviated since they are not involved in the minimization. Recall that we want to obtain a maximum likelihood estimation of the underlying radial velocity field $\mathbf{v}^{\text{true}}_{r}$. This can be achieved by minimizing the expression Eq.~\ref{eq:gaussian term} with respect to $\left(\mathbf{v}^{\text{true}}_{r}\right)^{\text{T}}$. All abbreviated terms vanish upon the derivative with respect to $\left(\mathbf{v}^{\text{true}}_{r}\right)^{\text{T}}$, leaving:
\begin{align}
     \frac{\partial\left(\mathbf{u}^\text{T}\mathbf{C}^{-1}\mathbf{u}\right)}{\partial\left(\mathbf{v}^{\text{true}}\right)^{\text{T}}} &= \mathbf{Q}\left(C^{v_rv_r}, C^{gg}\right)\,\left[\mathbf{v}^{\text{true}} - C^{v_r g}\left(C^{gg}\right)^{-1}\mathbf{g}^{\text{obs}}\right] \, .
\end{align}
Setting this to zero, we immediately obtain a reconstruction of the radial velocity field:
\begin{align}\label{eq:radial velocity reconstruction from linear density}
    \mathbf{v}^{\text{true}} &= C^{v_r g}\left(C^{gg}\right)^{-1}\mathbf{g}^{\text{obs}} \, .
\end{align}
This result is also expected from finding $\textbf{M}$ through the minimization of the variance 
\begin{align}
\left<\left(\hat{\mathbf{v}} - \mathbf{v}^{\text{true}}\right)\left(\hat{\mathbf{v}} - \mathbf{v}^{\text{true}}\right)^{\text{T}}\right>.
\end{align}

\subsection{Validating the Velocity Reconstruction on Simulations}\label{sec:galaxy reconstruction test on simulation}

To explore the validity of the linear estimator Eq.~\ref{eq:v estimator from g}, we construct the pixel covariance matrices $\mathbf{C}^{gg}$, $\mathbf{C}^{vg}$ and $\mathbf{C}^{vv}$ according to the prescription in Sec.~\ref{subsec:QML theory}. We include the galaxy shot noise for the galaxy auto-powers, such that $\mathbf{C}^{gg} = \mathbf{S}^{gg}+\mathbf{N}^{gg}$ while keeping only the signal for the velocity covariance $\mathbf{C}^{vv} = \mathbf{S}^{vv}$. The pixel covariance matrix spans the unmasked footprint of the DESI-LS survey on $N_{\text{side}}=16$, accounting for a sky fraction $\approx44\%$. We simulate full-sky, Gaussian velocity maps and galaxy maps from the same power spectrum that are used to generate the pixel covariance matrices above. The survey mask is then applied to maps to reproduce the observation, such that the velocity fields are reconstructed from the masked galaxy overdensity maps following Eq.~\ref{eq:v estimator from g}.
\begin{figure}[h!]
    \centering
    \includegraphics[width=0.9\columnwidth]{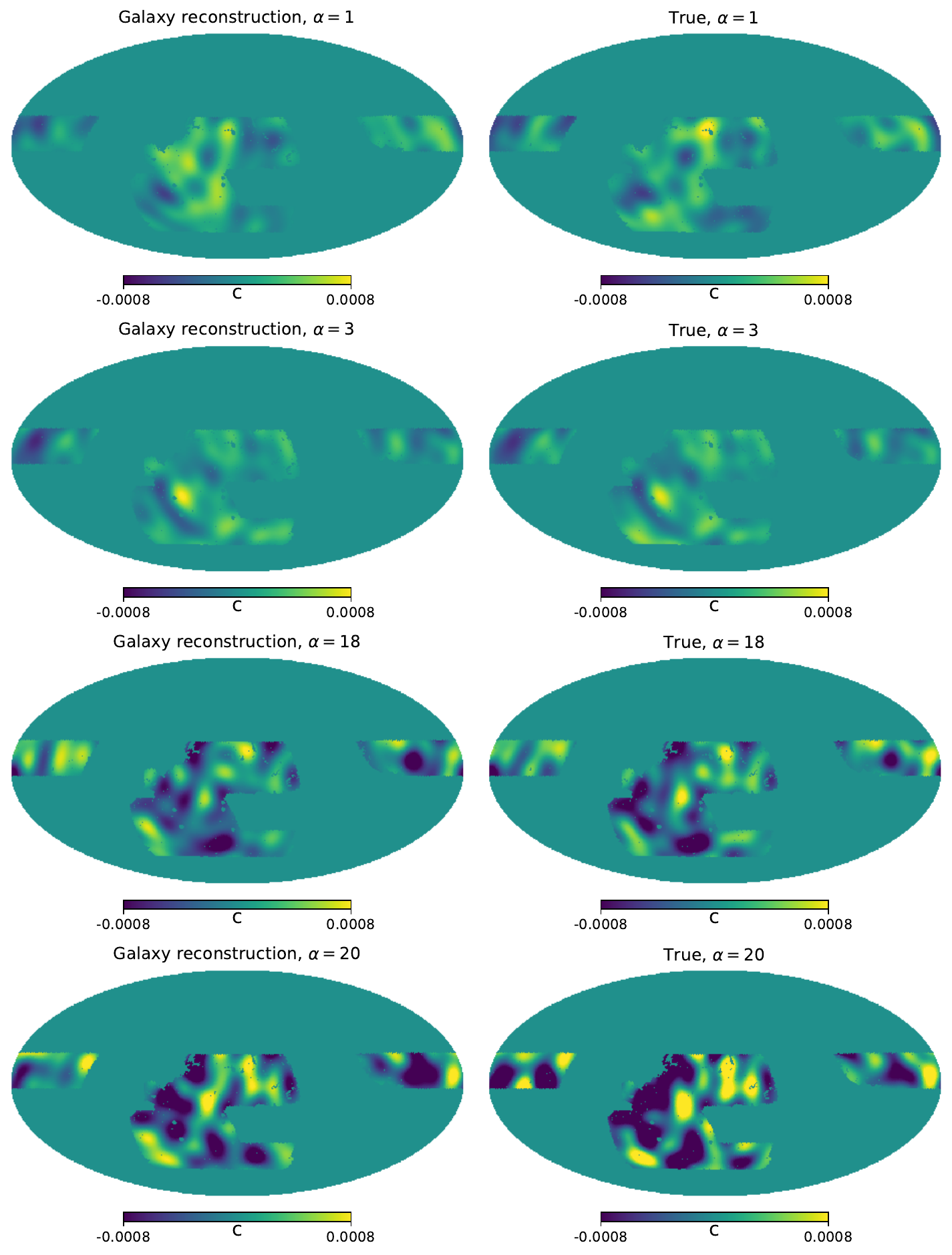}
    \caption{The velocity fields reconstructed from the masked galaxy overdensity maps (left column) compared to the true velocity fields (right column). Four redshift bins ($\alpha = 1,3,18,20$) near the edges of the interval are selected for illustration. The maps are first obtained on $N_{\text{side}} = 16$ and then projected to $N_{\text{side}} = 128$. Modes with $\ell < 7$ and $\ell > 20$ are removed for better visualization.}
    \label{fig:vgal_vksz_map_compare_sim} 
\end{figure}

We provide a side-by-side comparison between the galaxy reconstruction and the underlying velocity fields directly simulated from the fiducial velocity power spectrum in Fig.~\ref{fig:vgal_vksz_map_compare_sim}. The reconstruction quality is high, as one can see most of the structure between the scale $\ell = 7$ and $\ell = 20$ is recovered. This is partly due to the large sky coverage of the DESI-LS footprint. In general, the level of galaxy shot noise in each redshift bin increases with the number of redshift bins over a fixed interval. However, more redshift bins also imply more tomographic structure, which is important for reconstruction methods that rely heavily on the information from nearby redshifts.

\begin{figure}[h!]
    \centering
    \includegraphics[width=0.9\columnwidth]{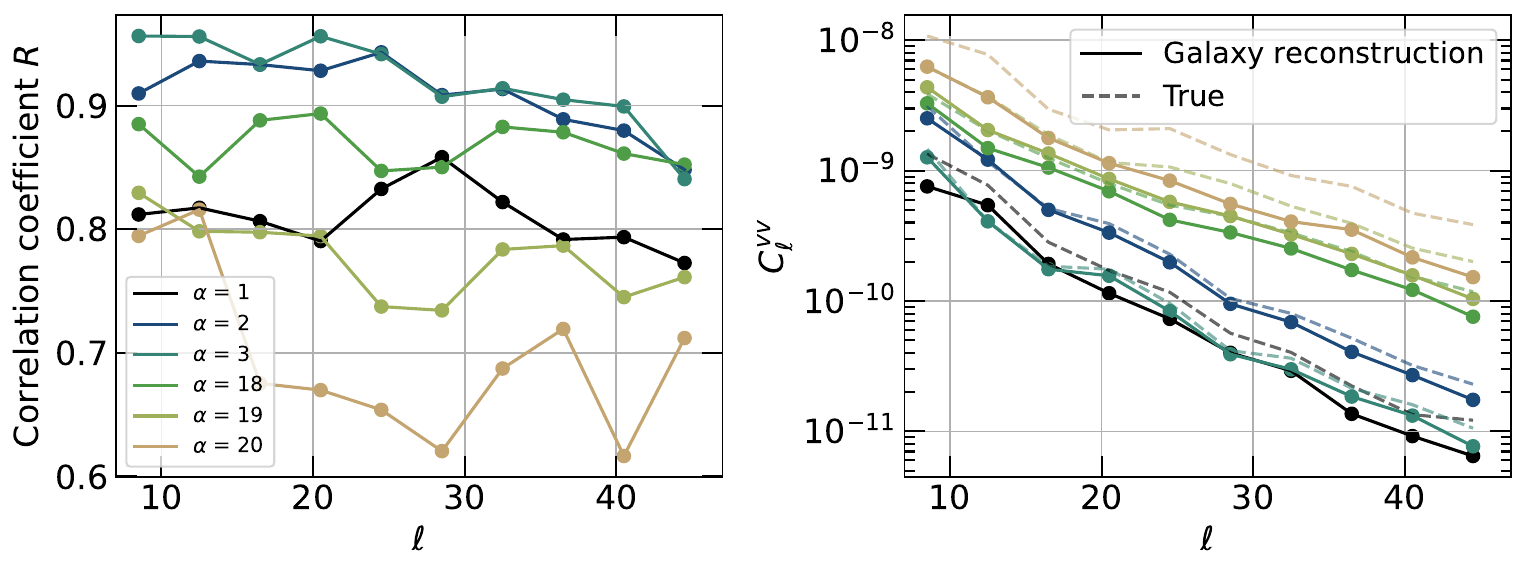}
    \caption{Left: The correlation coefficient $R_{\ell}$ between the reconstructed velocity fields and true velocity fields for redshift bins around the edge of the interval. Right: The velocity auto-powers directly computed from the unmasked pixels, neither $f_{\text{sky}}$ approximation nor pseudo-$C_{\ell}$ was applied since the coupling between masks and modes is the same for both the reconstructed field and the fiducial field. A bandpowering between $\ell = 7$ to $\ell = 48$ is applied to both plots in 4 modes per bandpower.}
    \label{fig:vgal_vksz_summary_sim}
\end{figure}

The quality of the reconstruction is described more quantitatively in the left panel of Fig.~\ref{fig:vgal_vksz_summary_sim} in terms of the correlation coefficient $R_{\ell}$. For redshift bins that are away from the two edges of the interval ($\alpha = 3$, $\alpha = 18$), $R_{\ell}$ ranges from 0.85 to 0.95, suggesting a strong correlation between the galaxy-reconstructed velocity fields and the true velocity field. However, the correlation coefficient $R_{\ell}$ begins to fall for redshift bins that are closer to the edge of the interval ($\alpha = 1$, $\alpha = 20$). These redshift bins suffer more from not getting information beyond the interval than those away from the edge. This is because the velocity-galaxy cross-power peaks when the velocity field and galaxy fields come from nearby but not exactly the same redshift bins. In general, there is a slight drop in correlation as $\ell$ increases, but the effect is not obvious. To conclude, the linear estimator proposed in Sec.~\ref{sec:linearrec} is an effective way to reconstruct 2D velocity fields from 2D galaxy overdensity fields as long as the correlations between different redshift bins are properly included in the construction of the covariance matrix. The fact that the estimator is performed on pixel space also suggests that it is highly compatible with sky masks in arbitrary shapes.

\section{A Comparison with Pseudo-$C_\ell$}\label{sec:app_pseudocl}
To evaluate the improvement factor from adopting the QML pipeline, we perform an SNR analysis on the velocity-galaxy cross-power amplitude $A$ using the pseudo-$C_{\ell}$ estimates. The pseudo-$C_{\ell}$ analysis is carried out with the \texttt{PyMaster} package. The input velocity maps correspond to the tSZ-masked $N_{\text{side}}=32$ maps. Both the input velocity maps and the galaxy overdensity maps are apodized on a scale of 4 degrees with a type `C1' apodization scheme using the \texttt{pymaster.mask\_apodization} function. The sky fraction $f_{\rm sky}$ changes from 23\% before apodization to 21.8\% after apodization. We create bandpowers for every 6 modes from $\ell = 5$ to $\ell = 60$. We note that, because pseudo-$C_{\ell}$ requires apodization, it is possible that increasing $N_{\text{side}}$ would be more beneficial to the pseudo-$C_{\ell}$ than the QML, and somewhat reduce the improvement. 

\begin{figure}[h!]
    \centering
    \includegraphics[width=0.8\columnwidth]{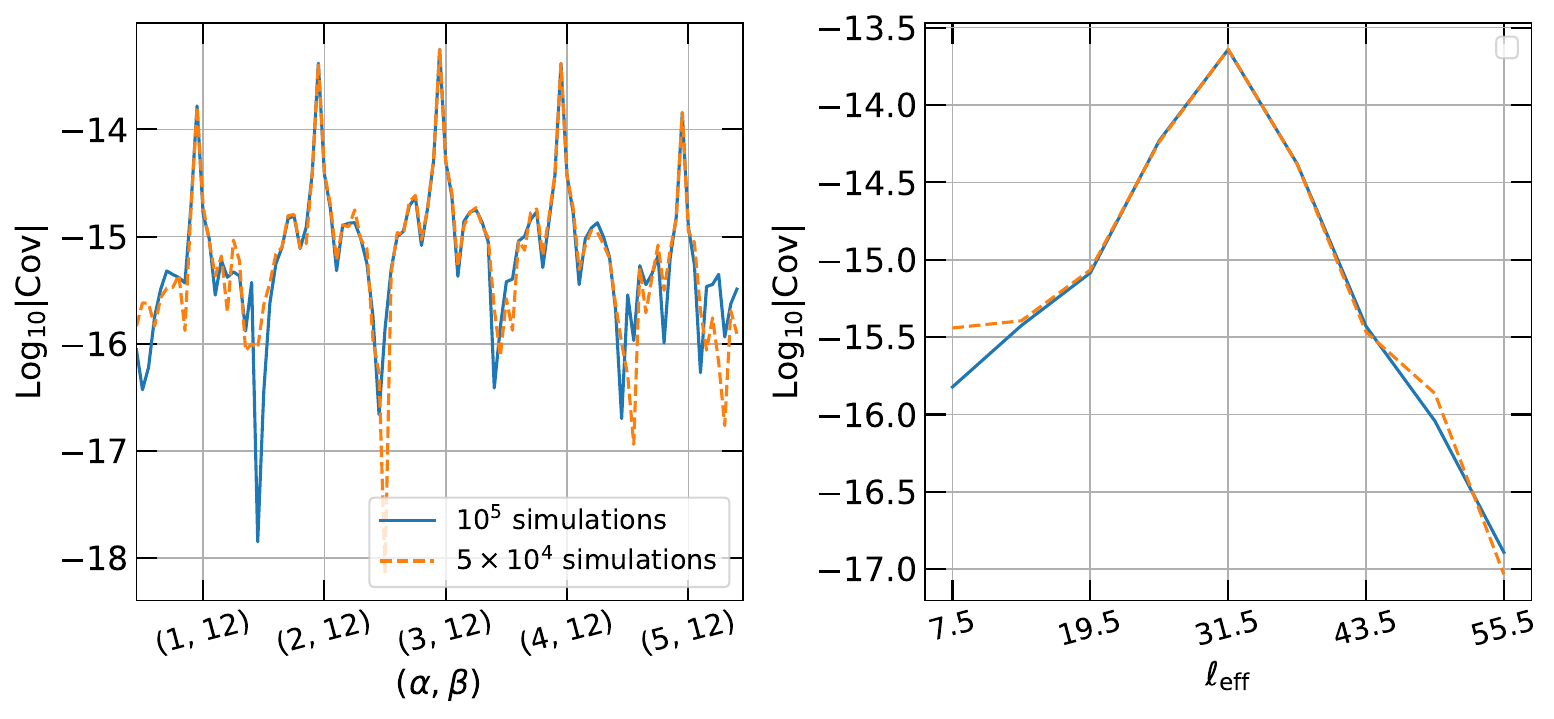}
    \caption{Left: The covariance of the velocity-galaxy cross-power between ($\alpha = 3$, $\beta = 12$) and other redshift bins centering at ($\alpha = 3$, $\beta = 12$) for the third bandpower. Log-absolute is taken for the covariances for the purpose of illustration. The solid line is estimated from $10^5$ simulations; the dashed line is estimated from $5\times10^4$ simulations. Highly-correlated redshift bins with large covariances are converging for both setups, while less-correlated redshift bins with smaller covariances (1-2 orders of magnitude lower) remain numerically unstable. Right: The log-absolute covariances between the fifth bandpower and other bandpowers in ($\alpha = 7$, $\beta = 8$). The x-axis shows the effective $\ell$ of each bandpower.}
    \label{fig:pcl_covaraiance_comparison}
\end{figure}

\begin{figure}[h!]
    \centering
    \includegraphics[width=1.0\columnwidth]{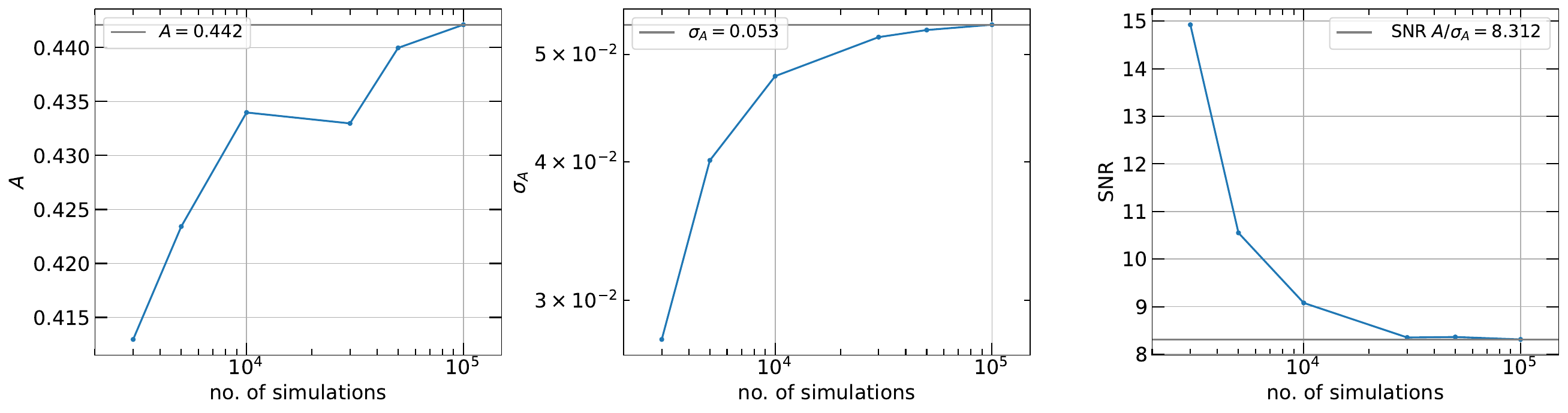}
    \caption{Pseudo-$C_{\ell}$ convergence test. We show the kSZ amplitude (left panel), the 1-$\sigma$ error (middle panel) and the SNR (right panel) from the pseudo-$C_{\ell}$ over a range of number of simulations. Outlier cuts (see App.~\ref{app:outlier_removal} and discussion in main text for more detail) are applied to all covariance matrices. We note that the 1-$\sigma$ error, which is completely determined by the simulation, starts converging from $3\times10^4$ simulations. The SNR is also insensitive to the fluctuations of the kSZ amplitude and the 1-$\sigma$ error beyond $3\times10^4$ simulations.}
    \label{fig:pcl_convergence_test}
\end{figure}

\begin{figure}[h!]
    \centering
    \includegraphics[width=1.0\columnwidth]{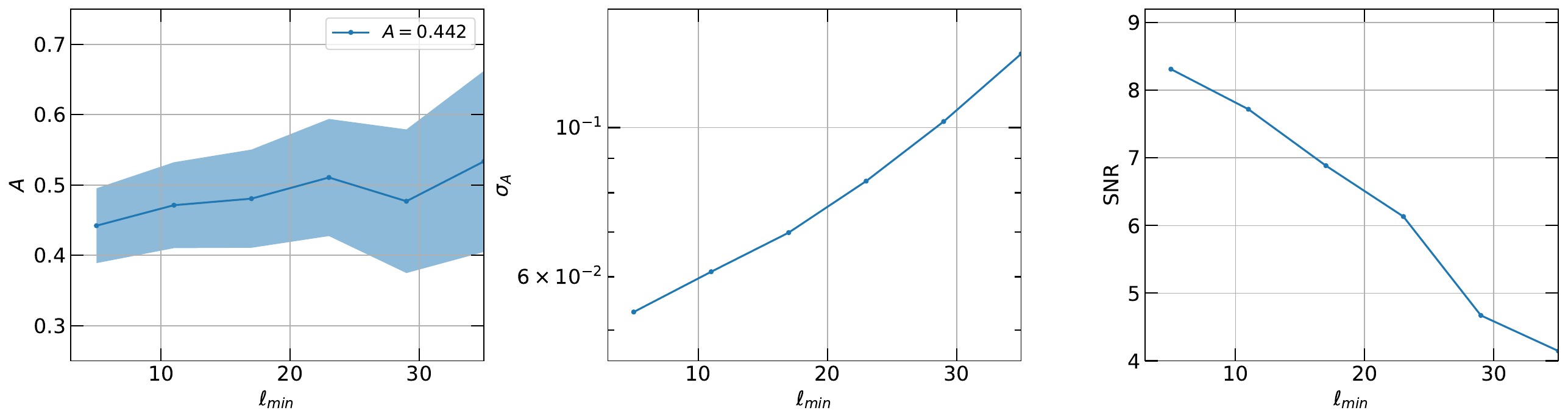}
    \caption{A summary of the results from the pseudo-$C_{\ell}$ pipeline using the covariance matrix evaluated from $10^5$ simulations. The definitions of the velocity-galaxy cross-power amplitude $A$ and 1-sigma uncertainty follow from Eq.~\ref{eq:SNR definition A}. The quantities are plotted against the lower cutoff of the bandpowers, with the first data point starting at $\ell_{\rm min} = 5$.}
    \label{fig:pcl_summary}
\end{figure}

To construct the covariance matrix for the pseudo-$C_{\ell}$ estimates, we generate $10^5$ full-sky, Gaussian simulations using the fiducial power spectrum $C^{v^\alpha v^\beta}_{\ell}, C^{g^\alpha g^\beta}_{\ell}$ and $C^{g^\alpha v^\beta}_{\ell}$ as outlined in Sec.~\ref{subsec: fiducial powers}. Maps from each simulation are then overlaid with the overlapping mask and processed with the PyMaster pipeline to obtain the pseudo-$C_{\ell}$ power spectrum. The covariance is computed from the pseudo-$C_{\ell}$ power spectrum. We present the covariance between the second bandpower of $C^{v^3 g^{11}}$ and the second bandpower of nearby redshift bins in Fig.~\ref{fig:pcl_covaraiance_comparison}. As a comparison, we also plot the covariance matrix obtained from $5\times10^4$ simulations. It can be observed that the covariance converges for both sets of simulations when the redshift-bin configurations $(\alpha, \beta)$ are highly correlated. However, we find that convergence is difficult to achieve for less correlated redshift bins. We investigate the overall convergence of the covariance matrix with respect to the number of simulations in Fig.~\ref{fig:pcl_convergence_test} and confirm that the offset in the covariance matrix between $5\times10^4$ and $10^5$ simulations is negligible. We include a summary from the pseudo-$C_{\ell}$ pipeline in Fig.~\ref{fig:pcl_summary}, using the covariance matrix estimated from $10^5$ simulations and applying same redshift-bin pair selection as in QML pipeline. The velocity-galaxy cross-power amplitude $A$ is estimated to be 0.442, with a 1-sigma uncertainty at $\sigma_A = 0.053$, resulting in an SNR of 8.3. The p-value with respect to the best-fit fiducial theory is 0.081 using a bandpowered covariance matrix.

\section{Null Tests}\label{app:null test}
\begin{figure}[h!]
    \centering
    \includegraphics[width=0.5\columnwidth]{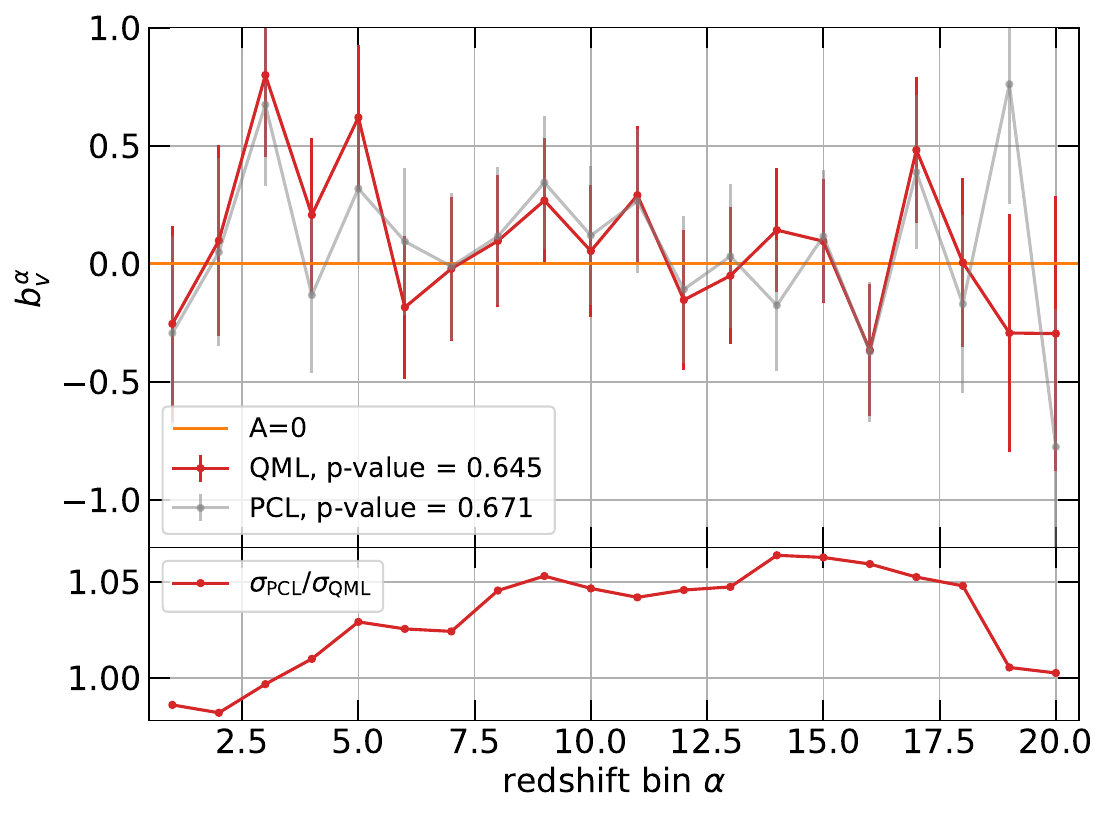}
    \caption{Null test, Top panel: The best-fit velocity bias $b^{\alpha}_v$ obtained from the QML estimates (red curve) and pseudo-$C_{\ell}$ estimates (grey curve) for each redshift bin. The error bars signify the 1-sigma values, which are obtained from the analytic covariance matrix for QML and from injecting $10^5$ simulations to \texttt{PyMaster} for pseudo-$C_{\ell}$. The orange line represents $b^\alpha_v = A= 0$. Bottom panel: The ratio of the 1-$\sigma$ pseudo-$C_{\ell}$ error bars to that of QML is consistently above 1, suggesting the robustness of QML as an optimal estimator. p-values in the legend are from the compressed 20-parameters covariance. We give the p-value from a full-spectra calculation in the main text.}
    \label{fig:bz_null}
\end{figure}

To check whether there are significant unaccounted foregrounds that correlate with the kSZ signal and bias our result, we perform a null test by constructing a set of $N_{\rm side} = 32$ null maps, which are defined by the difference of the two single-frequency, mean-subtracted velocity maps.
We proceed with the QML pipeline using the null maps as input and the following fiducial pixel-space covariance:
\begin{align}
    \mathbf{C}_{\rm null} &= 
    \begin{pmatrix}
        \mathbf{S}^{gg}+\mathbf{N}^{gg}, & 0 \\
        0, &0+\mathbf{N}^{vv}_{\rm null}
    \end{pmatrix} \, ,
\end{align}
such that the galaxy-velocity cross-powers and the signal part of the velocity auto-powers are set to zero due to the vanishing of kSZ signal. Reconstruction noise $N_{\rm null}$ is obtained as in Sec.~\ref{subsec:kSZ reconstruction results} with the null maps as input instead of the velocity maps from individual frequencies. To illustrate the robustness of the null test results from the QML pipeline, we also compute the velocity biases from the pseudo-$C_{\ell}$ galaxy-velocity cross-powers using the same set of null velocity maps. We apply the same outlier definition in Sec.~\ref{subsec: SNR result} and only include the 252 cross-powers in both pipelines. For QML, we find an overall cross-power amplitude of $A = 0.0904$ with a 1$\sigma$ error of $\sigma_A = 0.0619$ (the co-added cross-power is shown as the green curve in Fig.~\ref{fig:Cgv_QML_estimation_binned}). For pseudo-$C_{\ell}$, we find $A = 0.0754$ and $\sigma_A = 0.0838$. The p-value, which is now defined with respect to the null hypothesis ($A= 0$), is 0.388 for the QML pipeline and 0.671 for the pseudo-$C_{\ell}$ pipeline. The output of both pipelines is consistent with $A = 0$ within the 1.5-$\sigma$ region, suggesting that our main result is not significantly biased by foregrounds or systematics. We include the best-fit velocity biases and the p-value from the covariance matrix compressed with 20-parameters $b^{\alpha}_{\ell}$ from the null test in the top panel of Fig.~\ref{fig:bz_null}, which are consistent with zero (orange curve) for most of the redshift bins $\alpha$. In addition, QML estimates (red curves) have a narrower spread around zero with smaller 1-$\sigma$ error bars compared to pseudo-$C_{\ell}$ estimates, illustrating the optimal nature of QML as a power spectrum estimator.

\section{$C^{vg}_{\ell}$ co-adding}\label{app:coadding scheme}

In this appendix, we describe the procedure we use to plot the co-added cross-power spectrum measurement $\hat{C}_\ell^{vg}$ that unifies the individual redshift bin measurements $\hat{C}_{(\alpha,\beta,\ell)}^{vg}$. To obtain a single co-added cross-power spectrum, we first choose a pivot power spectrum $C^{\text{pivot}}_{\ell}$ by picking the redshift bin combination $\alpha$,$\beta$ with the highest fiducial $C_{(\alpha,\beta,\ell)}^{vg}$ amplitude. We then define the linear rescaling operator $\tilde{T}^{(\alpha, \beta, \ell)}$ by the ratio of the pivot power spectrum to the fiducial power spectra of the redshift bins:
\begin{align}\label{eq: transfer matrix}
    \tilde{T}^{(\alpha, \beta, \ell)} &= \frac{C_{\ell}^{\rm{pivot}}}{ C^{vg}_{(\alpha,\beta,\ell)}} \, .
\end{align}
We use this measurement-independent rescaling operator to bring all measured cross-power spectra  $\hat{C}_{(\alpha,\beta,\ell)}^{vg}$ to the same pivot form. We then construct a minimum variance estimator for the co-added power spectrum $\hat{C}^{\text{pivot}}_{\ell}$ from these rescaled cross-power measurements, taking into account their covariance.

To do so, we construct a minimum-variance estimator for $\hat{C}^{\text{pivot}}_{\ell}$ as follows. Let $\mathcal{T}^{(\alpha,\beta,\ell)}_{(\alpha',\beta',\ell')}$ be a diagonal matrix with $\tilde{T}^{(\alpha,\beta,\ell)}$ on the diagonal. Then 
\begin{equation}
    \hat{C}^{\text{pivot}}_{\ell} = w_{(\alpha, \beta)}\mathcal{T}^{(\alpha,\beta,\ell)}_{(\alpha',\beta',\ell')}\hat{C}_{(\alpha,\beta,\ell')} \, ,
\end{equation}
where we sum over repeated indices, and where $w_{(\alpha, \beta)}$ are the weights to be determined. We choose to neglect the $\ell\ell'$ correlations for simplicity, making the estimated pivot power spectrum not strictly optimal. By requiring the estimator to be unbiased and minimum variance, we obtain the weights
\begin{equation}
    \mathbf{w} = \left(\mathbf{i}^{\rm T}\mathbf{M}^{-1}\mathbf{i}\right)^{-1}\mathbf{M}^{-1}\mathbf{i} \, ,
\end{equation}
where the matrix $\mathbf{M}$ takes the following explicit form
\begin{equation}
    M^{(\gamma,\delta)}_{\ell} = \left[ \mathcal{T}^{(\alpha,\beta,\ell)}_{(\alpha',\beta',\ell')}(F^{-1})^{(\alpha',\beta',\ell')(\gamma',\delta',\ell'')}\mathcal{T}^{(\gamma,\delta, \ell)}_{(\gamma',\delta',\ell'')}\right] \, ,
\end{equation}
and $\mathbf{i}$ is a column vector of ones. $\hat{C}^{\text{pivot}}_{\ell}$ is what we refer to as the co-added cross-power spectrum and plot in Fig. \ref{fig:Cgv_QML_estimation_binned}.

\section{Effects of Removing Outlying Cross-powers}\label{app:outlier_removal}
\begin{figure}[h!]
    \centering
    \includegraphics[width=\columnwidth]{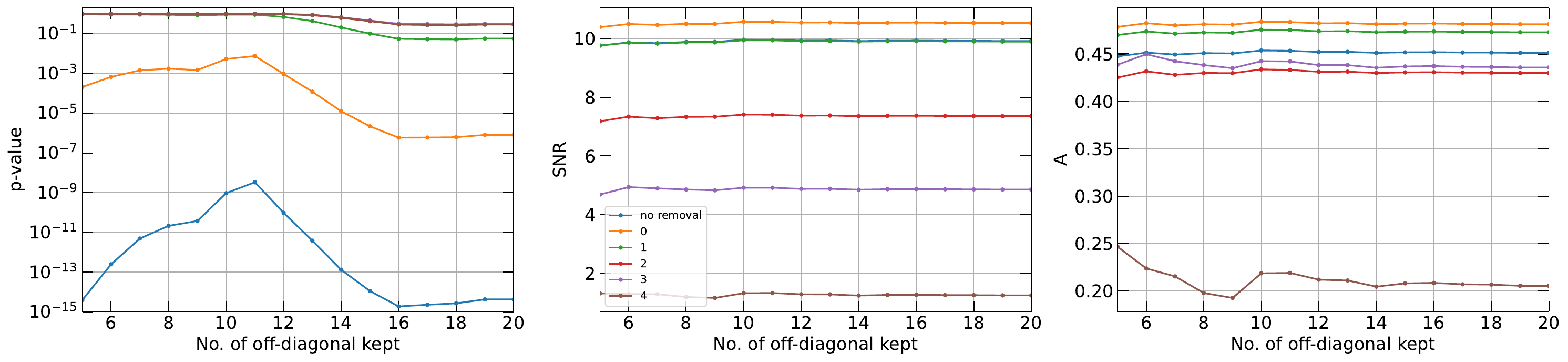}
    \caption{Left: The p-value computed from the total chi-square value and the degree of freedom. The x-axis shows the outermost off-diagonal used in the SNR analysis $i$, such that $|\alpha - \beta| > i$ is removed. Each curve corresponds to the number of innermost off-diagonals, j, that are excluded. For example, the green curve is evaluated with the diagonal + first off-diagonal removed. Middle: The signal-to-noise ratio SNR. Right: the kSZ amplitude $A$.}
    \label{fig:pvalue_notszmask}
\end{figure}

In Sec.~\ref{subsec: SNR result} we describe the effect of excluding $C^{g^{\alpha}v^{\beta}}_{\ell}$ outliers in the SNR pipeline. Here we show in Fig.~\ref{fig:pvalue_notszmask} the p-value, the SNR and the kSZ amplitude $A$ when different numbers of outermost off-diagonal redshift bins are removed ($|\alpha - \beta| > i$, $i$ is the x-axis value). Both $A$ and SNR are insensitive to the removal of far-separated redshift bins. On the contrary, we find an improvement in the p-value when we keep any cross-powers that satisfy $|\alpha - \beta| < 16$. The improvement saturates at $|\alpha - \beta| = 10$.

\begin{figure}[h!]
    \centering
    \includegraphics[width=\columnwidth]{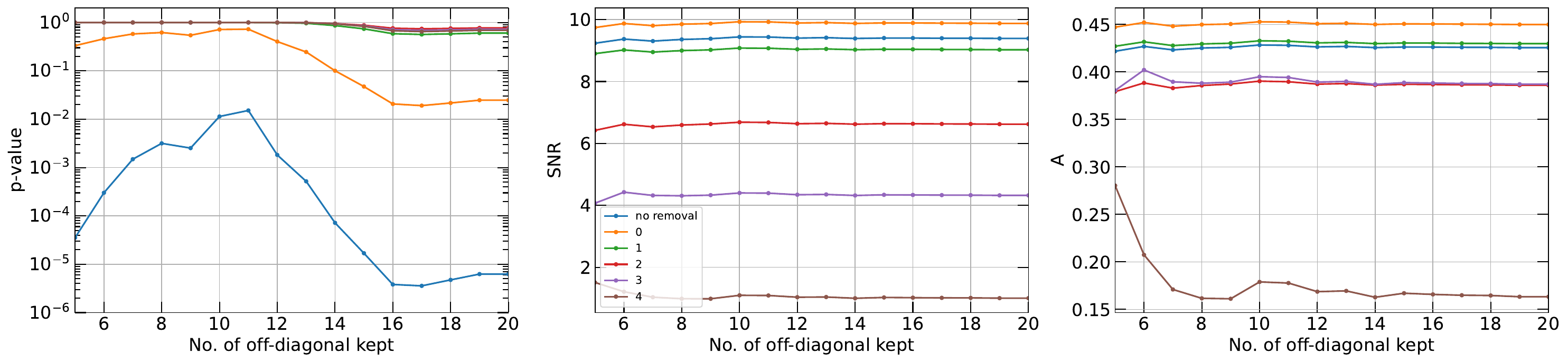}
    \caption{Similar to Fig.~\ref{fig:pvalue_notszmask}, but with the SZ cluster mask described in Sec.~\ref{sec:ACTDR6data}.}
    \label{fig:pvalue_withtszmask}
\end{figure}

We also vary the number of innermost off-diagonal redshift bins $j$, such that $|\alpha - \beta| \leq j$ is removed. They are presented in Fig.~\ref{fig:pvalue_notszmask} as curves of different colors. Note that the improvement to the p-value is huge by going from a full cross-power (blue curve, $\sim 10^{-15}$) to removing the diagonal and first off-diagonal (green curve, $\sim 10^{-2}$). This suggests that the goodness-of-fit is mainly affected by the residual, non-Gaussian CMB foreground that correlates to the small-scale galaxy overdensity (consistent with ~\cite{McCarthy:2025brx}, also see Sec.~\ref{subsec: SNR result}). We also present the same analysis on the SZ-cluster masked cross-powers in Fig.~\ref{fig:pvalue_withtszmask}. While the trend is similar to that in Fig.~\ref{fig:pvalue_notszmask}, we find an overall improvement in the goodness-of-fit. For example, the highest p-value for the orange curve moves from $\sim 10^{-2}$ to $\sim 0.7$. We therefore conclude that the employment of the tSZ mask and the removal of outlying cross-powers, while significantly improving the goodness-of-fit, do not lead to a large degradation of either the kSZ amplitude or the SNR. The results in the main text reported in Table~\ref{Tb:QML summary} correspond to the green curve (removing diagonal + first off-diagonal, and excluding any cross-powers with $|\alpha - \beta| > 10$).

\section{All Estimated 20 $\times$ 20 Bin Galaxy-Velocity Cross-Power Spectra $C^{gv}_{\ell}$}
\label{sec:full Cgv plot}
\begin{figure}[h!]
    \centering
    \includegraphics[width=\columnwidth]{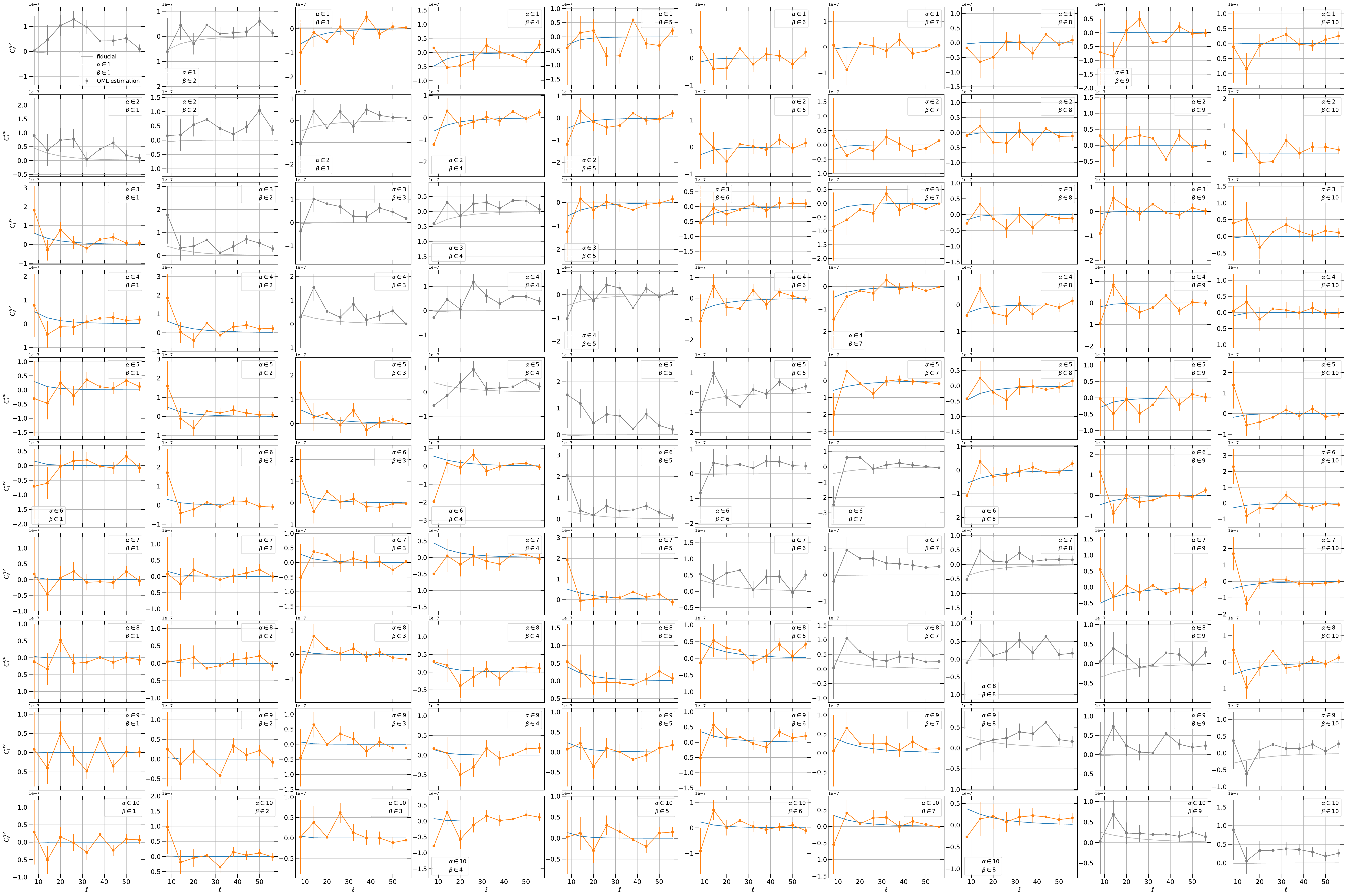}
    \caption{Part of the full 20-bins galaxy maps by 20-bins velocity maps cross-power from QML estimates. The modes are bandpowered in intervals of 6 using a uniform window function. Displayed is the $\alpha \in 1-10$, $\beta\in1-10$ block. The data points show the center of each bandpower. Plots in grey scales are excluded from the SNR analysis.}
    \label{fig:Cgv_QML_estimation_full1}
\end{figure}

\begin{figure}[h!]
    \centering
    \includegraphics[width=\columnwidth]{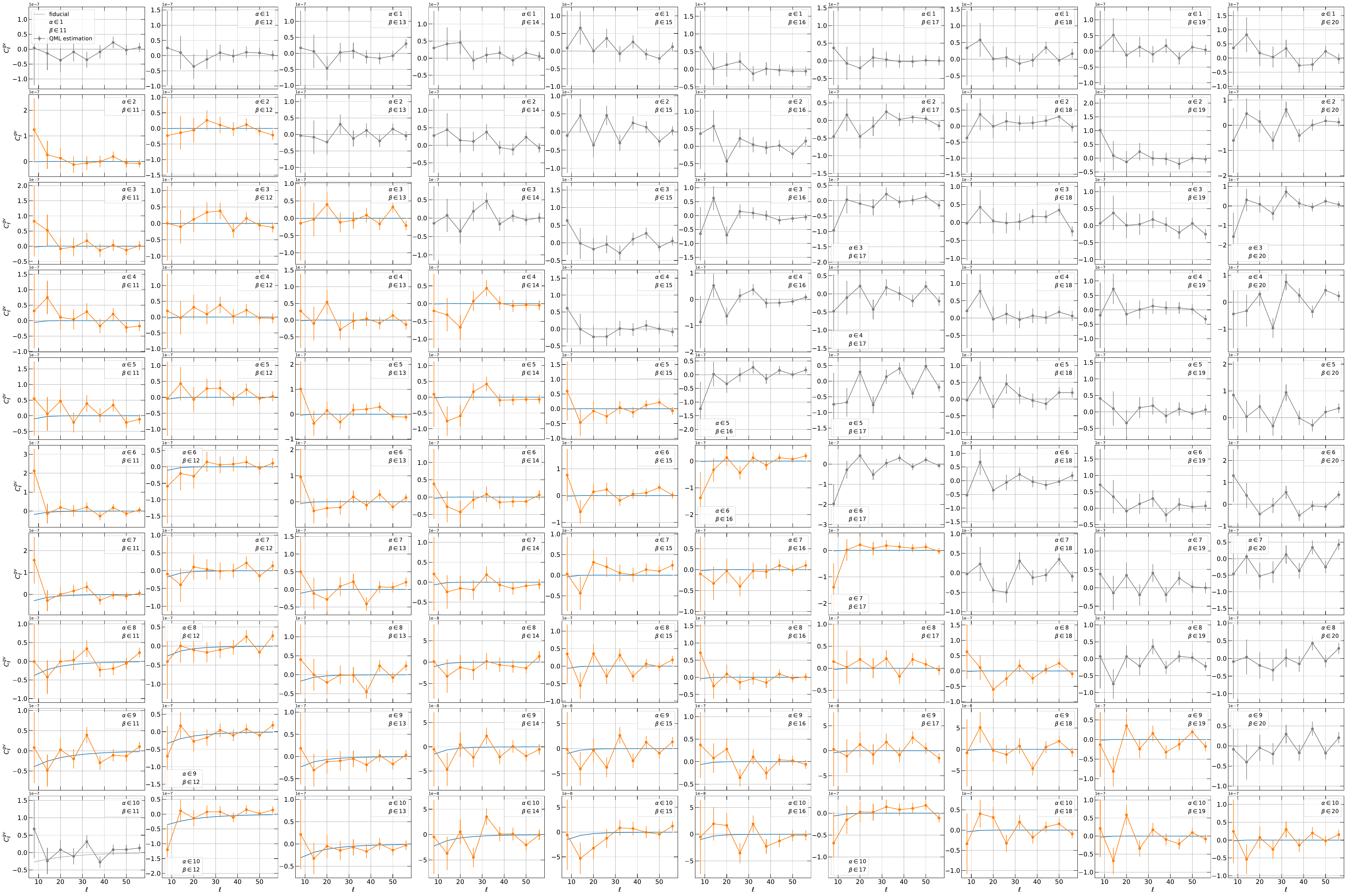}
    \caption{Same as Fig.~\ref{fig:Cgv_QML_estimation_full1} but with $\alpha \in 1-10$, $\beta\in 11-20$ block.}
    \label{fig:Cgv_QML_estimation_full2}
\end{figure}

\begin{figure}[h!]
    \centering
    \includegraphics[width=\columnwidth]{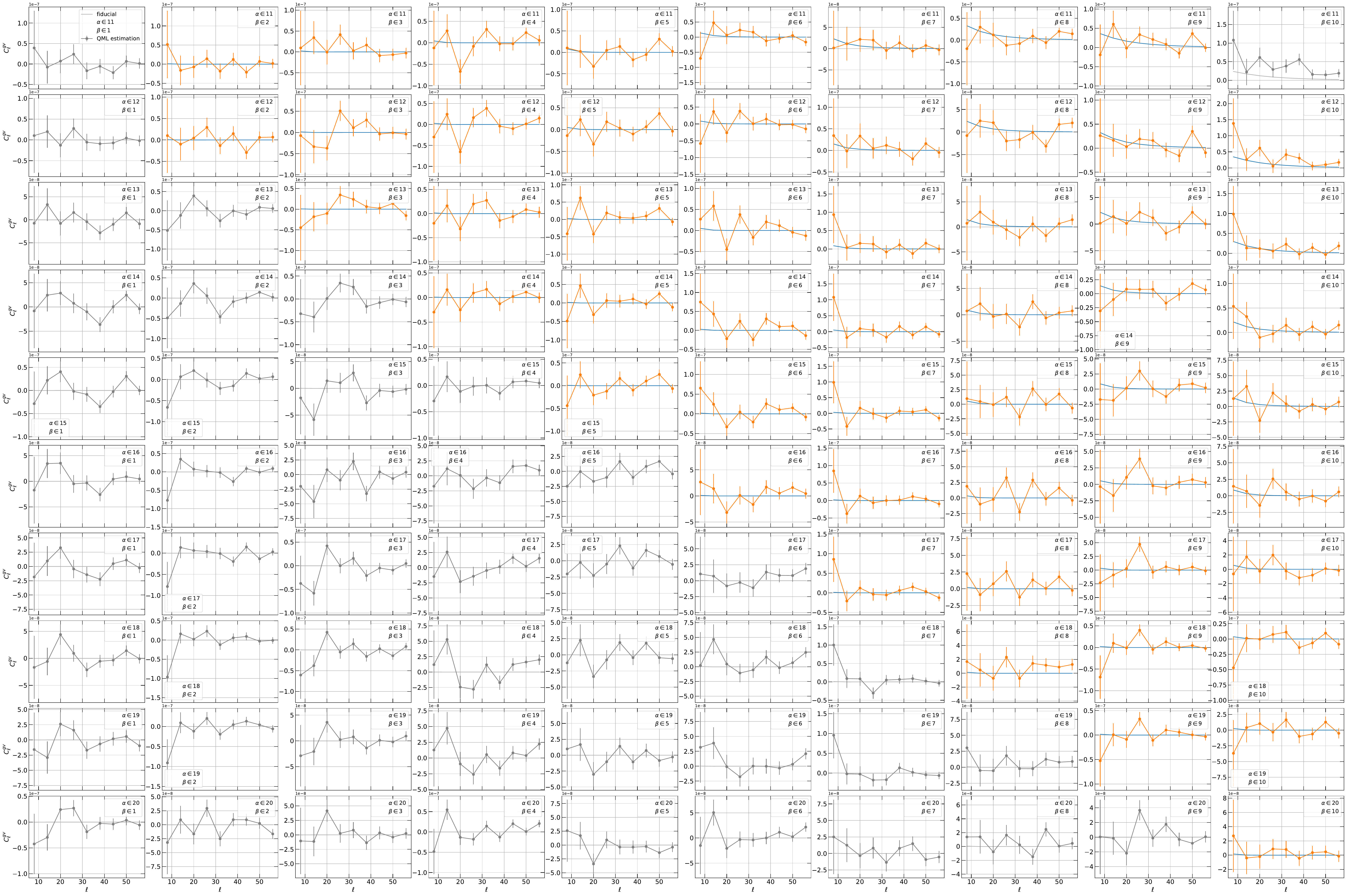}
    \caption{Same as Fig.~\ref{fig:Cgv_QML_estimation_full1} but with $\alpha \in 11-20$, $\beta\in 1-10$ block.}
    \label{fig:Cgv_QML_estimation_full3}
\end{figure}

\begin{figure}[h!]
    \centering
    \includegraphics[width=\columnwidth]{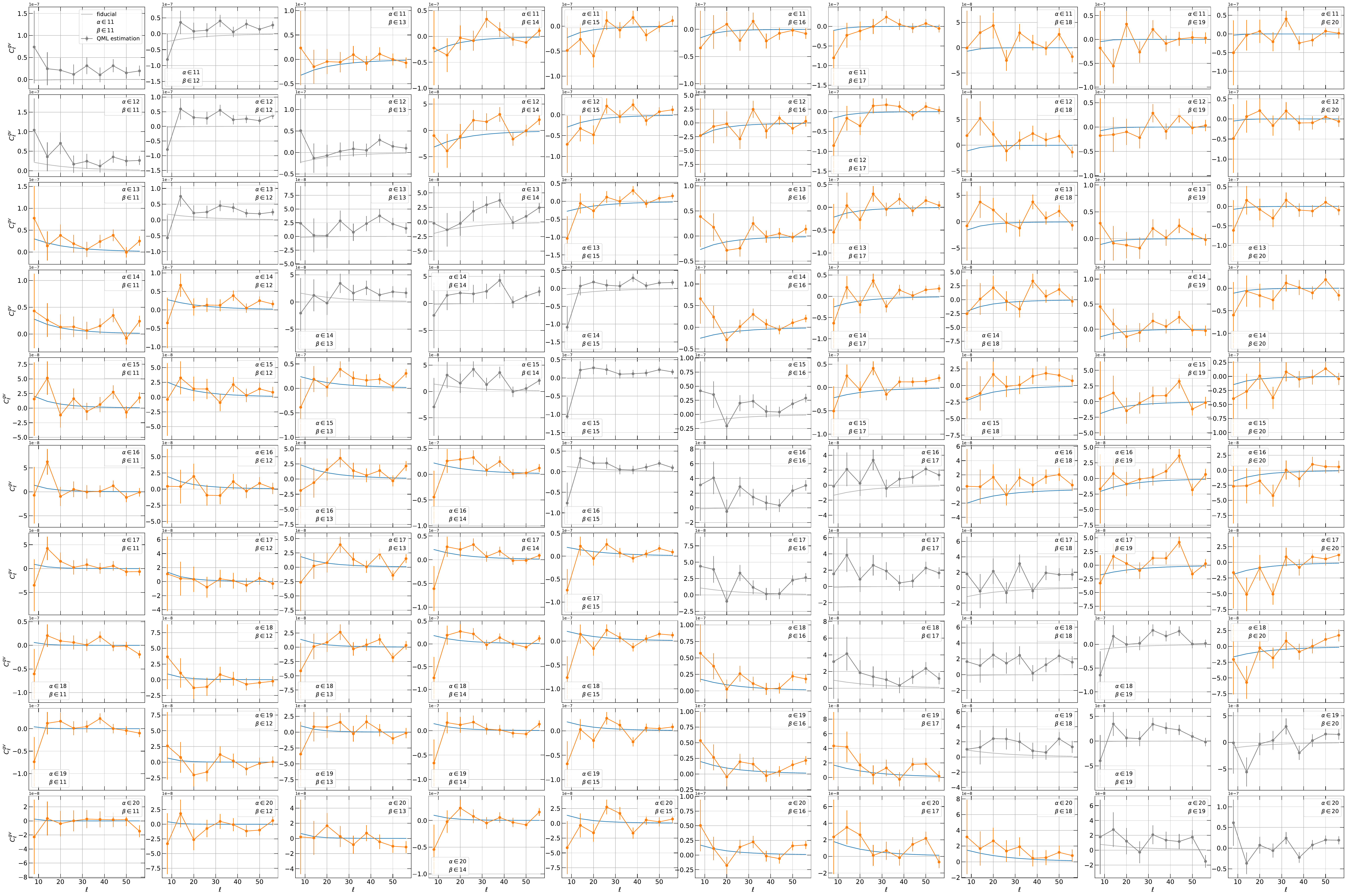}
    \caption{Same as Fig.~\ref{fig:Cgv_QML_estimation_full1} but with $\alpha \in 11-20$, $\beta\in11-20$ block.}
    \label{fig:Cgv_QML_estimation_full4}
\end{figure}

We provide a complete set of galaxy-velocity cross-power $C^{gv}_{\ell}$ without any averaging over redshift bins in the Figs.~\ref{fig:Cgv_QML_estimation_full1}-\ref{fig:Cgv_QML_estimation_full4}. Each plot contains $10 \times 10 = 100$ cross-powers that correspond to a block from the full $20 \times 20$ plot. While there is no averaging over redshift bins we band-power $\ell = 5$ to $\ell = 60$ in intervals of 6. However, the SNR calculation in Sec.~\ref{subsec: SNR result} uses the mode-by-mode covariance matrix.

\clearpage

\end{document}